\documentclass[longauth]{aa}  

\usepackage{graphicx}
\usepackage{txfonts}
\usepackage{lscape}
\usepackage{subfig}
\usepackage{amssymb,natbib,supertabular}
\usepackage{defs}
\usepackage{appendix}
\usepackage{color}
\usepackage{longtable}
\usepackage[breaklinks, colorlinks, citecolor=blue]{hyperref}
\usepackage{hyperref}
\begin{document}
\title{Spectroscopic observations of PHz\,G237.01$+$42.50: A galaxy protocluster at $z$=2.16 in the Cosmos field
    \thanks{Based on LBT/LUCI spectroscopic observations.}}
\titlerunning{Spectroscopic observations of PHz\,G237.01$+$42.5}
\subtitle{}

\author{M.~Polletta\inst{\ref{inst1},\ref{inst2}}
\and G.~Soucail\inst{\ref{inst2}}
\and H.~Dole\inst{\ref{inst3}} 
\and M.~D.~Lehnert\inst{\ref{inst4}}
\and E.~Pointecouteau\inst{\ref{inst2}}
\and G.~Vietri\inst{\ref{inst1}}
\and M.~Scodeggio\inst{\ref{inst1}}
\and L.~Montier\inst{\ref{inst2}}
\and Y.~Koyama\inst{\ref{inst5},\ref{inst6}}
\and G.~Lagache\inst{\ref{inst7}}
\and B.~L.~Frye\inst{\ref{inst8}}
\and F.~Cusano\inst{\ref{inst9}}
\and M.~Fumana\inst{\ref{inst1}}
}
  \offprints{M. Polletta\\ \email{maria.polletta@inaf.it}}
\institute{INAF - Istituto di Astrofisica Spaziale e Fisica cosmica (IASF) Milano, via A. Corti 12, 20133 Milan, Italy\label{inst1}
\and IRAP, Universit\'{e} de Toulouse, CNRS, CNES, UPS, (Toulouse), France\label{inst2}
\and Universit\'{e} Paris-Saclay, Institut d'Astrophysique Spatiale, CNRS, b\^at 121, 91400 Orsay, France\label{inst3}
\and Universit\'e Lyon 1, ENS de Lyon, CNRS UMR5574, Centre de Recherche Astrophysique
de Lyon, F-69230 Saint-Genis-Laval, France\label{inst4}
\and Subaru Telescope, National Astronomical Observatory of Japan, 650 North A'ohoku Place, Hilo, HI 96720, U.S.A.\label{inst5}
\and Graduate University for Advanced Studies (SOKENDAI), Osawa 2-21-1, Mitaka, Tokyo 181-8588, Japan\label{inst6}
\and Aix Marseille University, CNRS, CNES, LAM, Marseille, France\label{inst7}
\and Department of Astronomy/Steward Observatory, 933 North Cherry Avenue, University of Arizona, Tucson, AZ 85721, USA\label{inst8}
\and INAF - Osservatorio di Astrofisica e Scienza delle Spazio di Bologna, OAS, via Gobetti 93/3, I-40129 Bologna, Italy\label{inst9} 
}

\authorrunning{Polletta et~al.}

   \date{Received 19 February 2021 / Accepted 02 August 2021}
 
\abstract{The \planck\ satellite has identified more than 2,000 protocluster
candidates with extreme star formation rates (SFRs).  Here, we present the
spectroscopic identification of a \planck-selected protocluster located in
the Cosmos field, PHz\,G237.01$+$42.50.  PHz\,G237.01$+$42.50 contains a
galaxy overdensity of 31 spectroscopically identified galaxies at
$z{\simeq}$2.16 (significant at 5.4$\sigma$) in a
10\arcmin$\times$11\arcmin\ region.  The overdensity contains two
substructures or protoclusters at ${<}z{>}\simeq$2.16 and 2.195 with
estimated halo masses at $z$\,=\,0 of $\sim$5--6$\times$10$^{14}$\,\msun,
roughly consistent with Virgo-type clusters.  The overdensity total SFR,
$\sim$4000\,\msun\,yr$^{-1}$, is higher than predicted by simulations
but much smaller than the SFR derived from the \planck\ data (i.e.,
10,173\,\msun\,yr$^{-1}$).  The analysis of the \herschel\ data in the
field, in combination with the available ancillary data, shows that such a
difference is due to an effect of source alignment along the line of sight
that produces a 5$\sigma$ overdensity of red \herschel\ sources in the
field.  We analyze the members' ultraviolet (UV) spectra and
UV-far-infrared spectral energy distributions to derive their SFR, 
stellar mass, and metallicity.  Galaxy members include blue star-forming
galaxies and Active galactic nuclei (AGN) with SFRs and stellar masses
consistent with the main sequence.  Active galactic nuclei, identified through optical
spectroscopy or X-ray data, represent a significant fraction (20$\pm$10\%)
of all members of the protocluster at $z$\,=\,2.16, and they are powerful
enough to produce radiative feedback.  The core of this protocluster,
besides being denser, includes members that are, on average, more massive
and star-forming and contains a larger fraction of AGN and
\herschel-detected galaxies than the full sample, suggesting an
environmental effect on galaxy growth.  A comparison between
PHz\,G237.01$+$42.50 and other protoclusters in the literature at similar
redshifts reveals some common traits and differences that reflect both 
observational biases and a diversity in intrinsic properties that is not
yet fully understood.  }
   \keywords{large-scale structure of Universe --
		Galaxies: star formation --
		Galaxies: clusters: general --
		Galaxies: high-redshift --
		Submillimeter: galaxies}

\titlerunning{Spectroscopic observations of G237}
\authorrunning{Polletta et al.}

   \maketitle
%

\section{Introduction}

The study of galaxy cluster progenitors, so-called protoclusters, allows us
to investigate the growth of the most massive halos, the evolution of the
large-scale structure at high redshift, the history of star formation in
dense environments, and the link between galaxy evolution and
environment~\citep{kravtsov12,overzier16}.  Finding and studying high-$z$
growing structures is challenging as protoclusters are
rare~\citep{boylan09,overzier09}, necessitating either wide area surveys or
surveys that increase the volume probed by pushing to the highest redshifts.

In the past few years, the search and study of the earliest structure has
yielded hundreds of candidates and a dozen protoclusters studied in
detail~\citep[e.g.,
][]{pentericci00,matsuda05,chapman09,galametz10a,capak11,hatch11,tanaka11,koyama13,altieri14,cooke14,cucciati14,lemaux14b,diener15,chiang15,casey15,wang16,noirot16,noirot18}. 
These protoclusters have been discovered as overdensities of star-forming
galaxies (SFGs) identified through photometric~\citep[e.g., ][]{wang16} or
spectroscopic surveys~\citep[e.g., ][]{lemaux14b,steidel05,chiang14}, that
often target sources known to be tracers of dense environments, such as
high-$z$ radio galaxies (HzRGs), quasi stellar objects (QSOs), and
sub-millimeter (sub-mm) galaxies~\citep[e.g.,
][]{ivison13,dannerbauer14,cooke15,casey15,casey16,venemans07,kuiper12,hayashi16,noirot18}.

Different selection techniques have yielded a variety of protocluster
candidates whose members are often biased by the identification method. 
These biases hinder a proper comparison and obtaining a global picture of
their properties and of their galaxy members.

Ideally, we wish to have a homogeneous sample of protoclusters covering a
wide range of properties, such as redshift, predicted final cluster mass,
total star formation rate (SFR), and size.  For each protocluster it would
be desirable to have an unbiased census of their members by identifying them
in an unbiased way as well as to know their physical properties (stellar
mass, SFR, cold gas mass, morphology).  With such a sample it would then be
possible to investigate which processes play a dominant role in galaxy
growth and evolution in forming structures.

Many searches at $z{>}$1.5 have been carried out at sub-mm wavelengths as a
large fraction of member galaxies are expected to undergo a dust-obscured
star-formation
phase~\citep[e.g.,][]{dannerbauer14,clements16,kato16,wagner17,nantais17}. 
Indeed several sub-mm bright protoclusters have been discovered, at
small~\citep[$\sim$100\,kpc; ][]{ivison13,gomez19,wang16,miller18},
medium~\citep[$\sim$500\,kpc; ][]{oteo18,lacaille19}, and
large~\citep[several megaparsecs; ][]{dannerbauer14,casey15,kato16,hung16,umehata19}
scales.

\citet{negrello05} predict the detection by \planck~\footnote{\planck\
(http://www.esa.int/Planck) is a project of the European Space Agency (ESA)
with instruments provided by two scientific consortia funded by ESA member
states (in particular the lead countries France and Italy), with
contributions from NASA (USA), and telescope reflectors provided by a
collaboration between ESA and a scientific consortium led and funded by
Denmark.}~\citep{planck_mission} and the \herschel\ Space
Observatory~\footnote{\herschel\ is an ESA space observatory with science
instruments provided by the European-led Principal Investigator consortia and
with important participation from NASA.}~\citep{pilbratt10} sub-mm surveys
of unresolved intensity peaks made by the summed emission of dusty
star-forming high-$z$ protocluster members within the arcminute-sized beam
(arcminute scale).  For these reasons, we embarked on an ambitious program
to identify protoclusters based on their overall overdensity of ongoing star
formation.  We used the \planck\ satellite to search for high-$z$ structures
in a systematic way over the cleanest 26\% of the extragalactic sky, at an
angular resolution well matched to the expected size of these
structures~\citep[5\,arcmin, or $\sim$\,6\,Mpc in comoving scales at
$z{\simeq}$2--3; ][]{muldrew15}.  The \planck\ telescope, due to its
sensitivity at sub-mm wavelengths where SFGs at $z\gtrsim$2 are bright, has
provided a homogeneously selected sample of 2151 protocluster candidates at
redshifts $\simeq$2--4 -- the \planck\ high-$z$ (PHz)
sample~\citep{planck16}.  The sample selection and main properties are
summarized in Sect.~\ref{sec:phz_sample}.  The large sub-mm fluxes we see in
the PHz fields can be due to statistical fluctuations of the cosmic IR
background, single strongly lensed galaxies~\citep{canameras15},
overdensities of bright SFGs in the early Universe~\citep{planck15}, or
projections along the line of sight of bright galaxies or of
structures at different redshifts~\citep{negrello17}.  The most active
star-forming protoclusters must be among the brightest sub-mm sources of the
sky and are thus expected to be part of this sample.  Follow-up studies of
PHz protocluster candidates are relevant for a statistical characterization
of a large and homogeneous protocluster sample, as well as for the discovery
of extreme objects.

High-$z$ protoclusters offer also the possibility of investigating several
questions that are relevant in cosmology and galaxy evolution, for example
the processes that regulate the intense star formation observed in these
structures, the mechanisms responsible for their subsequent quenching, and
the role of the activity associated with active galactic nuclei (AGN).

In this work, we present the spectroscopic confirmation of the only PHz source
located in the Cosmos field~\citep{scoville07} for which spectroscopic
observations with the Large Binocular Telescope Near-infrared Spectroscopic Utility with Camera and Integral Field Unit for Extragalactic Research
(LUCI) spectrograph~\citep{seifert03,buschkamp12} on the Large Binocular
Telescope~\citep[LBT;][]{hill12} have been obtained, PHz\,G237.01$+$42.50
(G237).  The advantage of using this field is the availability of a wealth
of multiwavelength data, in particular of spectroscopic redshifts, that
allow us to identify and analyze the protocluster member galaxies.

Throughout this work we use magnitudes in the AB system, and adopt a
\citet{chabrier03} initial mass function (IMF).  We denote the stellar mass with
$\mathcal{M}$ and the characteristic stellar mass with $\mathcal{M^{\ast}}$. 
We assume a flat $\Lambda$ cold dark matter (CDM) model, with cosmological
parameters from the Planck 2018 release, 
$\Omega_{\Lambda}$\,=\,0.685, $\Omega_{\mathrm{m}}$\,=\,0.315,
H$_{\mathrm{0}}$\,=\,67.4\,\kmsMpc~\citep{planck_cosmo18}.  We express
physical and comoving distances in units of pkpc and ckpc, respectively. 
With the adopted cosmology, 1\arcsec\ corresponds to a scale of
$\sim$8.5\,pkpc, and 18.1\,ckpc at $z$\,=\,2.16.

\section{The \planck\ high-$z$ source sample}\label{sec:phz_sample}

The PHz catalog~\citep{planck16} was extracted from 26\% of the \planck\ all
sky maps.  The catalog includes all sources that were detected at more than
5$\sigma$ in the {\tt red-excess} (RX) map.  The RX map is obtained after
subtracting from the map at 550\um\ the image obtained by linearly
interpolating the signal in the 350\um, and 850\um\ maps.  The selection
requires a signal at more than 3$\sigma$ in the {\tt cleaned}\footnote{The
{\tt cleaned} maps are obtained after removing the signal from the cosmic
microwave background (CMB), and Galactic cirrus emission.} maps at 350, 550,
and 850\um, and a non-detection (S/N$<$3) in the 100\,GHz map.  In order to
favor only reliable high-$z$ candidates, all selected sources must also
satisfy a color selection criterion.  More specifically, a flux ratio
S$_{550}$/S$_{350}{>}$0.5 is required to reduce contamination from Galactic
cold sources, and S$_{850}$/S$_{550}{\lesssim}$0.9 to avoid contamination
from radio and Sunyaev-Zel'dovich sources~\citep[see Fig.~15
in][]{planck16}.  The PHz catalog contains 2151 sources.  The selection
procedure and the catalog are presented in~\citet{planck16}.  The color
selection criteria favor sources at 1.5${<}z{<}$4 assuming realistic values
of dust temperature and emissivity index.  The low angular resolution of the
$Planck$ beam (FWHM = 4.3\,arcmin at 550\um) and the 90\% completeness limit
of $\sim$500\,mJy in the sub-mm imply that the PHz sources correspond to
environments of typically 2\,Mpc in size with
L$_{IR}{\sim}$3$\times$10$^{13}$\,\lsun, for $z{\sim}$2--3.

Identifying and characterizing these sources require multiwavelength deep
photometric data over wide regions.  Several follow-up observations have
been carried out to obtain multiwavelength data for a representative
subsample. First, sub-mm observations with the Spectral and
Photometric Imaging REceiver~\citep[SPIRE; ][]{griffin10} instrument on the
\herschel\ Space Observatory have been obtained for 228 PHz. Of these, 217
targets (95\%) were found to correspond to overdensities of bright (S$_{\rm
350\mu m}{>}$37\,mJy) and red\footnote{A SPIRE source is defined
red if detected in all three SPIRE bands and with S$_{\mathrm{500}}$/S$_{\mathrm{350}}{>}$0.6 and
S$_{\mathrm{350}}$/S$_{\mathrm{250}}{>}$0.7~\citep[see ][]{planck15}.} SPIRE sources~\citep{planck15}. 
\spitzer/IRAC imaging at 3.6 and 4.5\,$\mu$m have been obtained for 92
sources out of the \herschel\ subsample~\citep{martinache18}.  Optical and
near-infrared (NIR) imaging data have been obtained for 42
PHz. Spectroscopic observations, necessary to confirm the protocluster
nature of these sources, have been obtained only for three PHz,
PHz\,G95.5$-$61.6, resulting in the identification of two potential
structures, one at $z\simeq$2 with six members, and a second one at
$z\simeq$1.7 with three members~\citep{flores16}, and PHz\,G237.01$+$42.50,
whose observations are presented here. A third PHz, PLCK\,G073.4$-$57.5,
has been observed with ALMA, yielding two CO line detections at the same
redshift, $z$\,=\,1.54~\citep{kneissl19}.  The paucity of spectroscopic
observations is due to the complication of identifying the counterpart to the
sub-mm sources, the need of data over wide enough areas to cover the
extension of the PHz, and the difficulty of obtaining redshifts for
galaxies at $z\sim$2, in particular, for dusty star-forming
galaxies~\citep[DSFGs; ][]{casey17}.

In addition to the data obtained through our follow-up campaign, we looked
for PHz sources located in public legacy fields, in particular
in the $>$50\,deg$^2$ HerMES fields~\citep{oliver12}, where a large set of multiwavelength
data, including \herschel\ SPIRE data, are available.  Based on the density
of the PHz sources (i.e., 0.2$\pm$0.004 sources/deg$^{2}$), we expect 
20$\pm$4 PHz in the HerMES fields (99.5\,\sqdeg).  There are 21 (1\% of the full sample)
PHz in HerMES, and one, PHz\,G237.01$+$42.50, in the 1.6\,\sqdeg\ Cosmos field, where,
in addition to multiwavelength photometric data, high-resolution imaging
and spectroscopic data are available.

\section{PHz\,G237.01$+$42.50: A \planck\ structure in the Cosmos field}\label{sec:G237}

PHz\,G237.01$+$42.50 (G237 hereinafter) is the only PHz source located in
the Cosmos field.  Its \planck\ position is $\alpha$\,=\,150.507\,deg, and
$\delta$\,=\,2.31204\,deg, and size is $\sim$12\arcmin$\times$5.5\arcmin. 
Its total \planck\ fluxes are $S_{\mathrm{550\mu m}}$\,=\,0.61$\pm$0.04\,Jy,
and $S_{\mathrm{350\mu m}}$\,=\,0.69$\pm$0.03\,Jy, and the signal-to-noise
ratios (S/Ns) at 850, 550 and 350\um, and in the RX 550\um\ map, are 5.2,
5.8, 3.5, and 6.0, respectively, making G237 among the 30\% most significant
detections in the PHz sample.  Based on these flux densities, and assuming a
modified black body model with temperature of 30\,K at $z$\,=\,2.2, and
the~\citet{kennicutt98a} relation between the IR luminosity and the SFR, the
estimated SFR of G237 is 18\,210\,\msun\,yr$^{-1}$~\citep[assuming a
\citet{salpeter55} IMF, and 10,173\,\msun\,yr$^{-1}$ assuming
a~\citet{chabrier03} IMF; ][]{planck16}.

\subsection{LBT spectroscopic observations}\label{sec:lbt}

\begin{figure} 
\centering
\includegraphics[width=9cm]{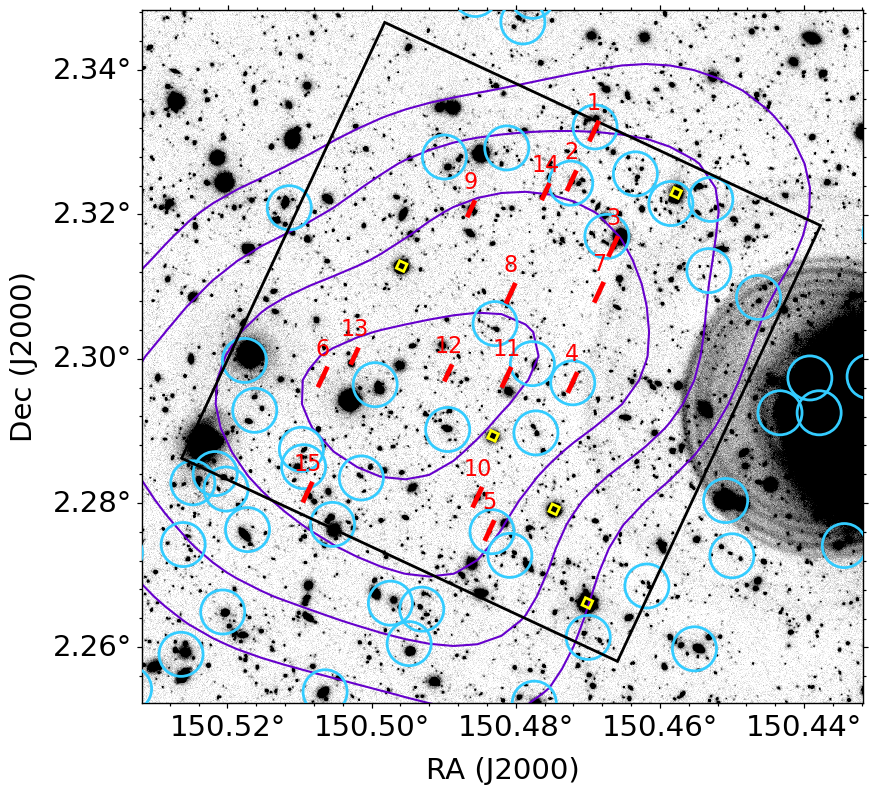}
\caption{{\small Layout of the LUCI mask (black square:
4\arcmin$\times$4\arcmin) overlaid on the UltraVISTA K$_{\mathrm s}$-band
6\arcmin$\times$6\arcmin\ image.  The slits assigned to the spectroscopic
targets are shown as red rectangles (1\arcsec$\times$8--10\arcsec) and
identified by the slit number, alignment stars are shown by yellow square. 
The SPIRE sources positions are indicated by blue circles (11\arcsec\ in
radius).  \planck\ {\tt red-excess} contours (at 50, 62.5, 75 and 87.5\% of
the highest value) are shown as purple lines. North is up and east to the left.}}
\label{fig:lbt_mask}
\end{figure}

As part of our follow up program, we obtained spectroscopic observations of
G237 (Italian Program 34; PI: Polletta) with
LUCI~\citep{seifert03,seifert10,ageorges10,buschkamp12} on the 8.4\,m
LBT~\citep{hill12} in monocular mode.  LUCI provides NIR multiobject
spectroscopy over a field of view of 4\arcmin$\times$4\arcmin.  We used the
N1.8 camera (0.249\,\arcsec/pixel) and the second order of the 200\,H+K
grating with a central wavelength of 1.93\,\micron, providing a resolving
power of R$\sim$1100.  The wavelength coverage is slightly different from
object to object (depending on the position of the slit in the mask) but
typically spans 1.50-2.30\,\micron.  We observed 15 targets with AB K-band
mag $\leq$22 selected from the ultra-VISTA 3$^{rd}$ data release
catalog~\citep{mccracken12}.  Five of them are within 11\,\arcsec\ of a
SPIRE source, our highest priority sources, the remaining ten are filler
sources selected because of their red optical-NIR colors.  The layout of the
LUCI mask overlaid on the UltraVISTA K$_{\mathrm s}$ image of G237 is shown
in Fig.~\ref{fig:lbt_mask}.  The observations were conducted on 2016 April
21--22.  Flats and arcs were taken on the same day of the
observations.  The observing conditions were good with mostly clear sky, an
airmass of 1.28, on average, and a seeing between 0.5 and 1.25\,\arcsec\
(0.9\,\arcsec, on average).  The observations were divided into five
observing blocks, with twenty 120s exposures each, for a total exposure time
of 3hr20min.  We used slit widths of 1\arcsec, and slit length of either 8
or 10\,\arcsec\ to sample enough sky background for sky subtraction.  The
nodding technique was applied to obtain a good quality sky subtraction
during the reduction phase.  A telluric star was observed at the end of each
night for the flux calibration and for correction of atmospheric absorption.

The data reduction procedure was carried out by the Italian LBT support team
using a dedicated pipeline developed at INAF
IASF-Milan~\citep{scodeggio05,garilli12}.  This includes flat-fielding, sky
subtraction, correction for distortions, and wavelength and flux
calibration.  The background was subtracted from frames in pairs obtained
from nodding along the slit.  Wavelength calibration was carried out using
OH sky lines, and flux calibration using a telluric standard star. 
Monodimensional spectra extraction was performed using the Horne extraction
algorithm~\citep{horne86}.  The final products include sky and noise
spectra, and flux calibrated bi- and mono-dimensional spectra.  The spectra
of the 15 observed objects are shown in Fig.~\ref{fig:lbt_spectra}. 
Redshifts were determined using the EZ code~\citep{garilli10}.

The complete list of spectroscopic targets, with coordinates,
estimated redshifts and associated quality flags, K-band magnitudes, and
IRAC colors from the Cosmos multiwavelength catalog~\citep[][; L16
hereinafter]{laigle16} are given in Table~\ref{tab:lbt_sample}. We were able
to obtain a reliable redshift ($z_{flg}>$1) estimate for five sources, of which
two are at the same redshift, $z{\simeq}$2.16.  One of these two
objects (ID SL01) exhibits a broad \halpha\ line
(FWHM\,=\,\,4004$\pm$327\,\kms)~\footnote{The width of the
broad \halpha\ line has been derived after fitting the rest-frame spectrum
in the 6400--6750\,\AA\ wavelength range with a power-law to model the
continuum, and four Gaussian components to model the narrow, and broad
\halpha\ lines, and the \nii\ doublet.} and it is thus
classified as broad line AGN~\citep{khachikian74}.

After analyzing these data, we discovered that six of our selected LUCI
targets had been previously observed by the Cosmos team and, for five of them,
a spectroscopic redshift was available from the zCosmos-Deep
survey~\citep{lilly09}.  These redshift are listed in
Table~~\ref{tab:lbt_sample}, and relative spectra are shown in
Fig.~\ref{fig:lbt_vimos_spectra}.  In Fig.~\ref{fig:lbt_spectra}, we show
the expected spectral features (see red vertical dashed line) in the LUCI
NIR spectra assuming the zCosmos-Deep redshift. In only one case we have a redshift estimate
from both LUCI and zCosmos, ID slit08. They differ significantly, the LUCI
estimate, $z$\,=\,1.52, is highly reliable ($z$ flag\,=\,4), while the zCosmos
one, $z$\,=\,0.86, is uncertain ($z$ flag\,=\,1). In the other four cases,
there are no strong features observable in the NIR, thus the spectra are
consistent with the zCosmos redshift estimates.

Since the chances of finding two sources at the same redshift out of such a
small sample are small, we investigate the hypothesis that G237 might
contain a structure at such a redshift.  To this end we searched for all the
available spectroscopic redshifts in a wide region centered on G237,
and analyzed its distribution, as described in the next section.

\subsection{Cosmos spectroscopic redshifts}\label{sec:cosmos}

\begin{figure} 
\centering
\includegraphics[width=9cm]{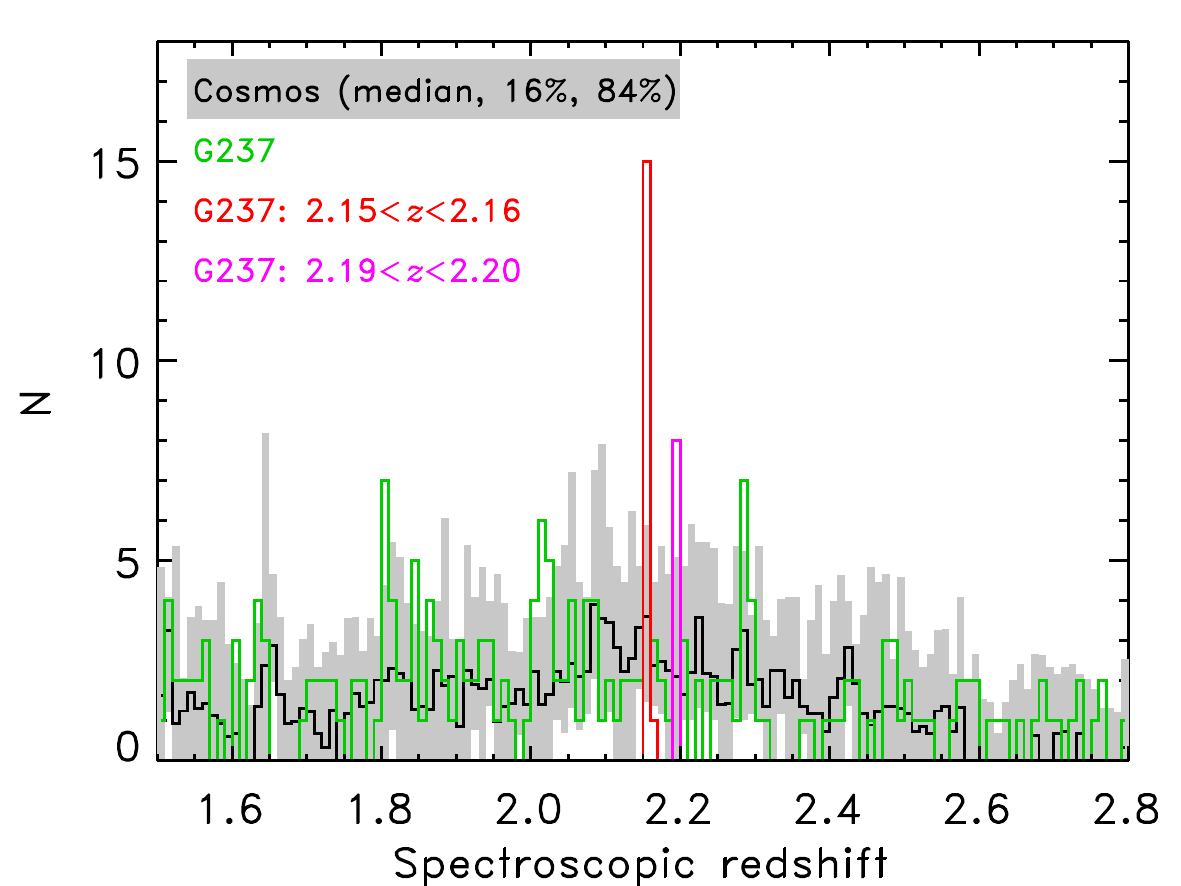}
\caption{{\small Spectroscopic redshift distribution of all
sources with known spectroscopic redshift from the Cosmos public redshift
release and the zCosmos-Deep catalog within a
10$^{\prime}\times12^{\prime}$ region centered at $\alpha$\,=\,150.465\deg, and
$\delta$\,=\,2.31\deg\ where G237 is located (green line). Sources with $z$\,=\,2.15--2.16
are shown with a red line, and those with $z$\,=\,2.19--2.20 with a magenta
line. The redshift distribution from the same spectroscopic redshift 
list obtained by taking the median values from 20 non overlapping 
10$^{\prime}\times12^{\prime}$ regions within the Cosmos field is shown as
black solid line, with associated 16$^{th}$ and 84$^{th}$ percentile values
(gray region).}}
\label{fig:specz_distribution}
\end{figure}

We collected a large number of spectroscopic redshifts in a region
$\sim$10\arcmin\ wide centered on G237 from various public surveys (zCosmos
Bright;~\citet{lilly07}; FMOS;~\citet{silverman15};
Magellan;~\citet{trump07}; \spitzer/IRS;~\citet{fu10};
DEIMOS;~\citet{hasinger19}), and from the zCosmos-Deep
survey~\citep{lilly09}. From this compilation, we obtained the redshift
distribution in the 10\arcmin$\times$12\arcmin\ region centered on G237 (see
Fig.~\ref{fig:specz_distribution}).  The distribution shows two narrow
peaks, one at $z_{\rm peak}\sim$\,2.155 with 15 members, and another at
$z_{\rm peak}\sim$\,2.195 with eight members where
$|z-z_{\mathrm{peak}}|/(1+z_{\mathrm{peak}})<0.0016$ (or
$|\Delta\varv|\leq$480\,\kms).  Six sources have intermediate redshifts,
2.16${<}z{<}$2.19.  To quantify the significance of the two redshift peaks,
we computed the median spectroscopic redshift distribution in a large region
of the Cosmos field where the density of sources with a spectroscopic
redshift is similar to the G237 field (i.e., 149.74\deg${<}\alpha{<}$150.65\deg,
and 1.77\deg${<}\delta{<}$2.65\deg).  Within such a region, we defined 20
10\arcmin$\times$12\arcmin\ boxes, and derived, in each box, the redshift
distribution from $z$\,=\,1.5 to 2.8 with redshift bins of 0.01.  For each
redshift bin, we computed the median number of sources, the mean absolute
deviation from the median, and the 16$^{th}$ and 84$^{th}$ percentile
values, after normalizing the number of sources with spectroscopic redshift
per box to the number of sources found in the G237 region.  The derived
median distribution is shown in Fig.~\ref{fig:specz_distribution}.  Based on
this distribution, we would expect 12.8$\pm$3.7, 3.6$\pm$1.7, and
2.1$\pm$1.5 sources at 2.15${\leq}z{\leq}$2.20, 2.15${\leq}z{\leq}$2.16, and
2.19${\leq}z{\leq}$2.20, respectively.  Compared to the number of sources in
these redshift bins in G237, 29, 15, and 8, respectively, the significances
of the redshift peaks in the three bins are 4.4, 6.8, and 4.0$\sigma$,
respectively.  We can thus conclude that there is a significant overdensity
of sources in all three redshift bins.  Including the two sources from our LUCI
sample, there are in total 31 sources with 2.15${\leq}z_{\rm spec}{<}$2.20 or
$|\Delta\varv|\leq$2360\,\kms, yielding an overdensity contrast $\delta_{\rm
gal}$\footnote{The overdensity contrast is defined as $\delta_{\rm
gal}$\,=\,($\rho_{\rm source}-\rho_{\rm field}$)/$\rho_{\rm field}$, where
$\rho_{\rm source}$ is the galaxy density in the source, and $\rho_{\rm
field}$ that in the field.  The overdensity significance is defined as
($\rho_{\rm source}-\rho_{\rm field}$)/$\sigma_{\rm field}$, where
$\sigma_{\rm field}$ is the galaxy density standard deviation in the
field.}\,=\,1.3, and 3.15, with a significance of 4.9$\sigma$ at
2.15${\leq}z{\leq}$2.20, and of 8$\sigma$ at 2.15${\leq}z{\leq}$2.16, respectively.

The list of 31 sources is reported in Table~\ref{tab:specz_sample}, and
their spectra are shown in Fig.~\ref{fig:spectra_2p16},
and~\ref{fig:spectra_2p19}.  We note that in 11 out of 29 objects the redshift
estimate is highly uncertain (quality flag equal to 1), but after a careful
check where we investigated alternative redshift solutions using the EZ
code~\citep{garilli10}, we concluded that no revision was better than the
available estimate.
The detection of two sources at $z\,\simeq$\,2.16 in our LBT observations
and the presence of a peak at the same redshift in the G237 field suggest
that G237 contains a structure at such a redshift.  To
investigate this hypothesis further we examine the spatial distribution,
both projected on the sky and along the line of sight of
all the spectroscopic sources at $z{\simeq}$\,2.15--2.20 (corresponding to
$\Delta \varv$\,=\,4646\,\kms).

\subsection{H$\alpha$ emitters in G237}\label{sec:HAE}

\begin{figure} 
\centering
\includegraphics[width=\linewidth]{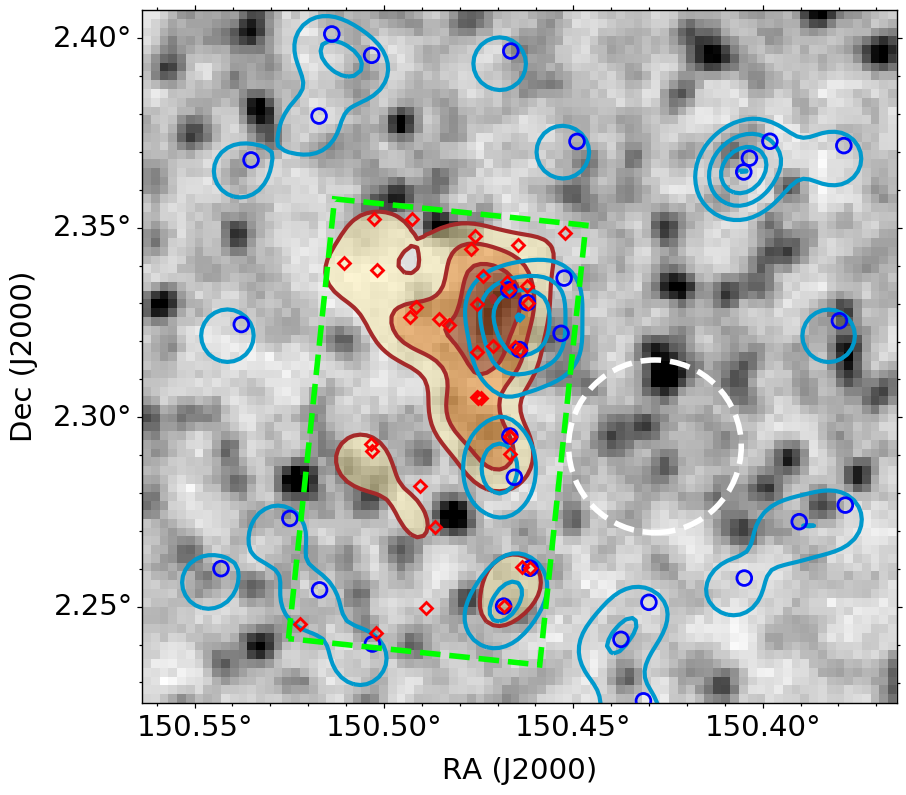}
\caption{{\small Spatial distribution of spectroscopic sources (blue
circles), and HAEs (red diamonds) overplotted on the SPIRE 350$\mu$m image
of the G237 field.  The yellow-brown filled contours show the HAE projected
density, and the turquoise contours the projected density of the
spectroscopic sources.  The density contour levels are shown at 25, 50, 75,
and 100\% of the maximum value.  The dashed green rectangle represents the
MOIRCS field of view where the HAE were selected~\citep{koyama21}.  The
spectroscopic and HAE density peaks are 38\arcsec\ apart.  The white circle
represents a masked region in the L16 catalog due to the presence of a
bright star.}}
\label{fig:HAE_vs_zs}
\end{figure}

The spectroscopic redshift peak found in G237 prompted us to carry out
narrowband imaging observations to identify the \halpha\ emitters (HAEs) in
the field at $z{\simeq}$\,2.16.  MOIRCS/Subaru~\citep{ichikawa06}
observations of G237 using the NB2071 filter ($\lambda_c$\,=\,2.068\,$\mu$m,
$\Delta\lambda$\,=\,0.027\,$\mu$m; corresponding to the \halpha\ line at
$z$\,=\,2.13--2.17) yielded the discovery of 38 HAEs~\citep{koyama21}.  The
spatial distribution of all spectroscopic members, and the HAEs is shown in
Fig.~\ref{fig:HAE_vs_zs}.  Six out of eight spectroscopically confirmed
galaxies with $z{<}$2.17 in the observed MOIRCS 4\arcmin$\times$7\arcmin\
field are also selected as HAEs (see list in Table~\ref{specz_sample}).  The
two spectroscopic members that are not detected as HAE (IDs 58173, and
57730) might have an \Ha\ line with a small equivalent width (EW) (i.e.,
$<$30\AA), or a wrong redshift.

The MOIRCS observations reveal a strong overdensity of HAEs, with a clear
enhancement of massive HAEs at its peak.  The density peaks of both,
the spectroscopic and the HAE samples, are only 38\arcsec\ apart. This
region will be identified as the protocluster core in the next section.

\begin{figure} 
\centering
\includegraphics[width=9cm]{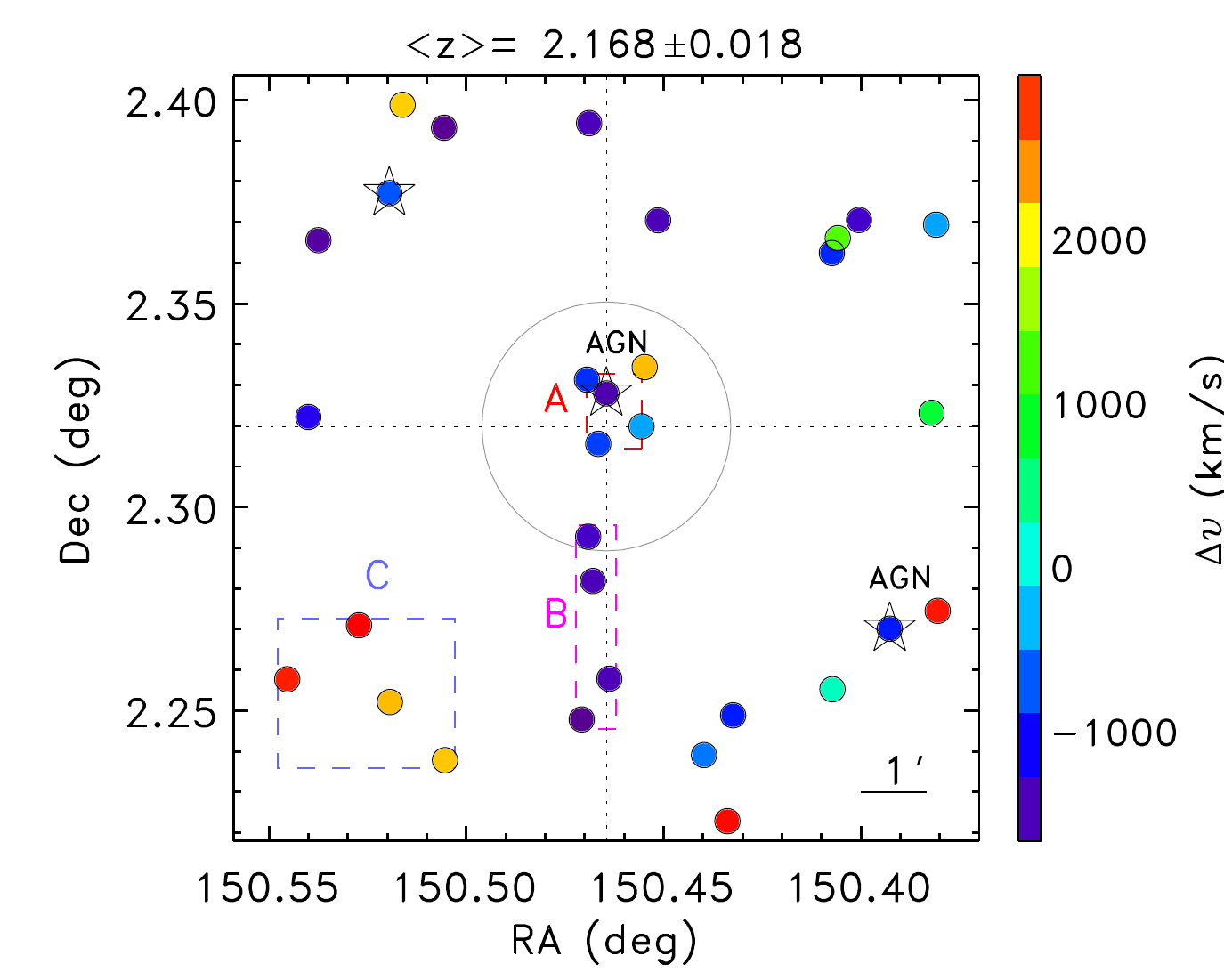}
\caption{{\small Spatial distribution of sources with 
$z$\,=\,2.15--2.20 in the field of \phz712. 
Different colors correspond to the velocity offset from the annotated mean
redshift as indicated by the color bar.  AGN are sources spectroscopically classified AGN, and black stars are X-ray luminous
AGN. A gray circle with radius corresponding to 2\,Mpc comoving at the
annotated mean redshift and centered on the structure center is shown. Dashed rectangles delineate the A, B, and C components
described in~Sect.~\ref{sec:components}, and shown in Fig.~\ref{fig:components}.}}
\label{fig:deltav}
\end{figure}

\subsection{Structure components}\label{sec:components}

\begin{figure} 
\centering
\includegraphics[width=9cm]{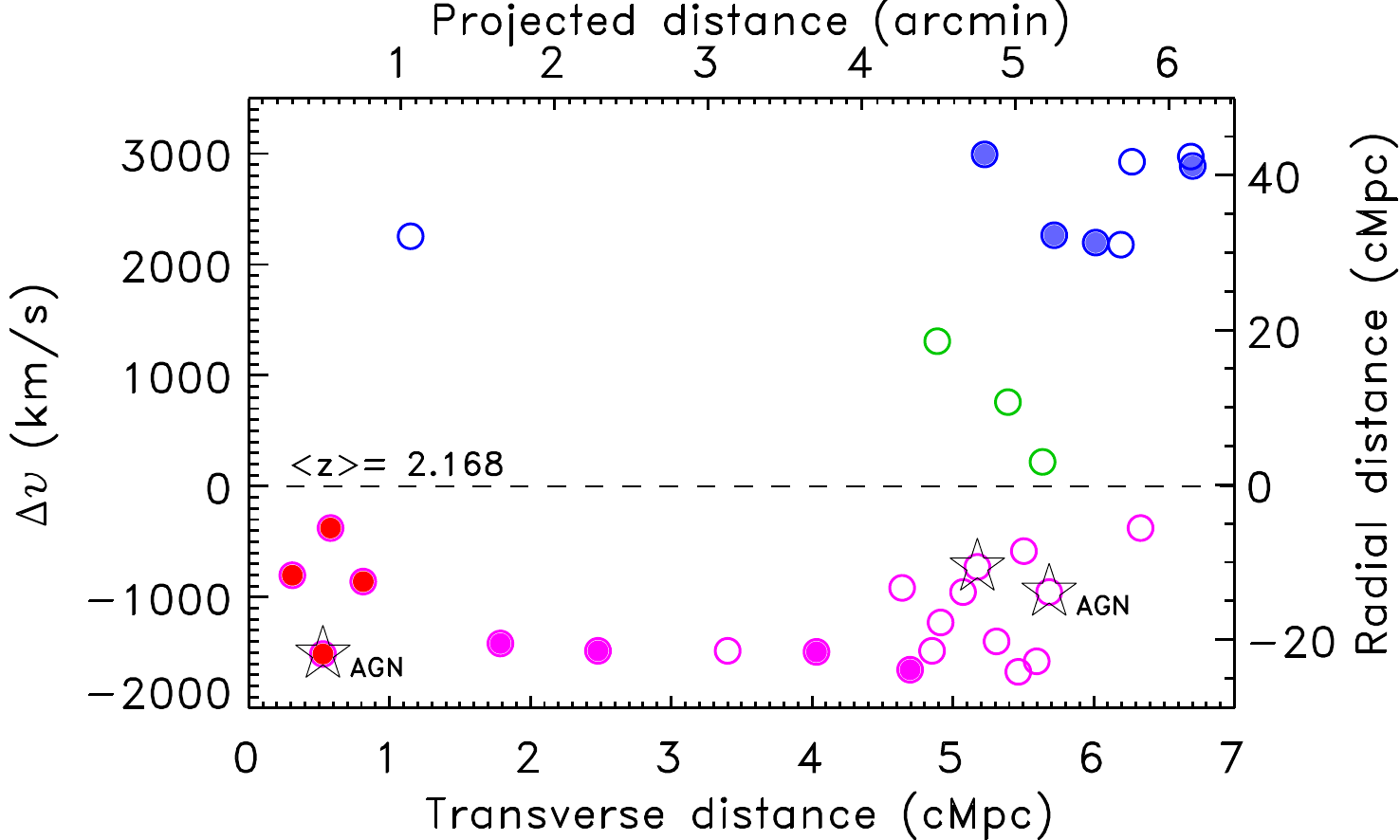}
\caption{{\small Velocity offset or radial distance from the mean structure redshift
(${<}z{>}$=\,2.168; dashed line) as a function of projected distance from the
structure center (defined by the coordinates median). Filled circles represent sources in the
3 components described in Sect.~\ref{sec:components}: A in red, B in magenta,
and D in blue. The remaining members are shown as open circles (magenta if
at 2.15${\leq}z{<}$2.164, blue if at $z{>}$2.19, and green if at
intermediate redshifts, 2.164${<}z{z}$2.19). Spectroscopically identified AGN
are annotated, and X-ray selected AGN are marked with black stars.}}
\label{fig:dv_vs_distance}
\end{figure}

\begin{figure} 
\centering
\includegraphics[width=\linewidth]{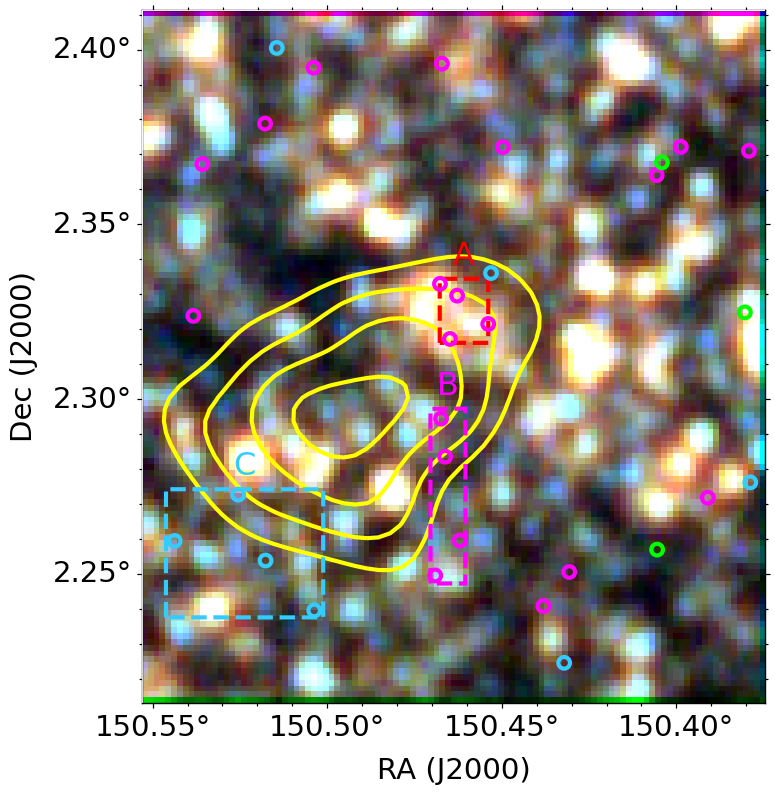}
\caption{{\small Three-color \herschel/SPIRE (R: 500$\mu$m, G: 350$\mu$m, B:
250$\mu$m) 10\arcmin$\times$12\arcmin\ image of the G237 field.  The
yellow contours represent the \planck\ {\tt red-excess} signal at 50, 62.5, 75 and
87.5\% of the peak value. Spectroscopic members are shown as open circles
(magenta: 2.15${\leq}z{<}$2.164, green: 2.164${<}z{<}$2.19, and light blue:
2.19${<}z{<}$2.20).  The regions A, B, and C, enclosing the groups of
sources described in Sect.~\ref{sec:components}, are shown as dashed
rectangles (red and magenta for those at 2.15${\leq}z{<}$2.164, light blue
for those at 2.19${<}z{<}$2.20).}}
\label{fig:components}
\end{figure}

The spatial distribution of all 31 sources at 2.15${\leq}z{<}$2.20 in the
\phz712\ field is shown in Fig.~\ref{fig:deltav}.  We also show the
projected distance along the line of sight in terms of velocity, where
100\,\kms\ at these redshifts correspond to $\sim$1.24\,cMpc.  In
Fig.~\ref{fig:dv_vs_distance}, their distribution in transverse and radial
distance is shown.  The distribution of velocity offsets of all 31 sources
is not gaussian, indicating velocity substructures (see
Fig.~\ref{fig:dv_vs_distance}).  Assuming as reference redshift, the mean
value (i.e., ${<}z{>}$\,=\,2.168$\pm$0.018), the 31 sources can be split in
two substructures with either high ($\varv{\sim}$2000--3000\,\kms) or low
($\varv$ from $-$1700 to 0\,\kms) velocities.  Based on the sources redshift
and projected spatial distribution, we identify three components, as
illustrated in Fig.~\ref{fig:components}.

We first consider the group of sources with negative velocity offsets,
corresponding to 20 sources with redshift 2.15${\leq}z{<}$2.164.  Within
this group, we distinguish a core (A component;
0.8\arcmin$\times$1.1\arcmin\ region; dashed red rectangle in
Fig.~\ref{fig:components}) containing four sources (see red full circles in
Fig.~\ref{fig:dv_vs_distance}).  This region exhibits the highest source
density (4.4 sources arcmin$^{-2}$).  In the same redshift group, there are
four sources at similar redshifts ($\Delta\varv$\,=\,238\,\kms), and located
in an elongated region (0.6\arcmin$\times$3\arcmin; region B) with velocity
offsets increasing with the distance from the core.  These four sources, and
one in the core, ID SL03, are all aligned south of the core, and their
velocity offsets are consistent with them falling and accelerating into the
core.  The remaining 12 sources are more spread out and located in annulus
between 3\,\arcmin, and 6.5\,\arcmin\ from the core.  The group of 20
spectroscopic members with 2.15${\leq}z{<}$2.164 occupies a volume of
9.5\arcmin$\times$9.3\arcmin$\times$1302\,\kms, corresponding to a comoving
volume of $\sim$15.29$\times$14.96$\times$19.1\,cMpc$^3$, or 4376\,cMpc$^3$. 
We note that this volume estimate is not corrected for peculiar velocities
(see Sect.~\ref{sec:halo_mass}).  Among the remaining 11 sources with
positive velocity offsets, eight are at 2.19${<}z{<}$2.20, corresponding to
$\Delta\varv$\,=\,788\,\kms.  Four sources in this group are in a
2.7\arcmin$\times$2.2\arcmin\ region [C component].

There are thus in total 31 sources with 2.15${\leq}z{<}$2.20, corresponding
to $\Delta \varv$\,=\,4646\,\kms over a $\sim$10\arcmin$\times$11\arcmin\
region. The volume occupied by all 31 spectroscopic members is 
$\sim$16$\times$17$\times$67\,cMpc$^3$, or 18532\,cMpc$^3$.

\begin{figure*} 
\centering
\includegraphics[width=0.45\linewidth]{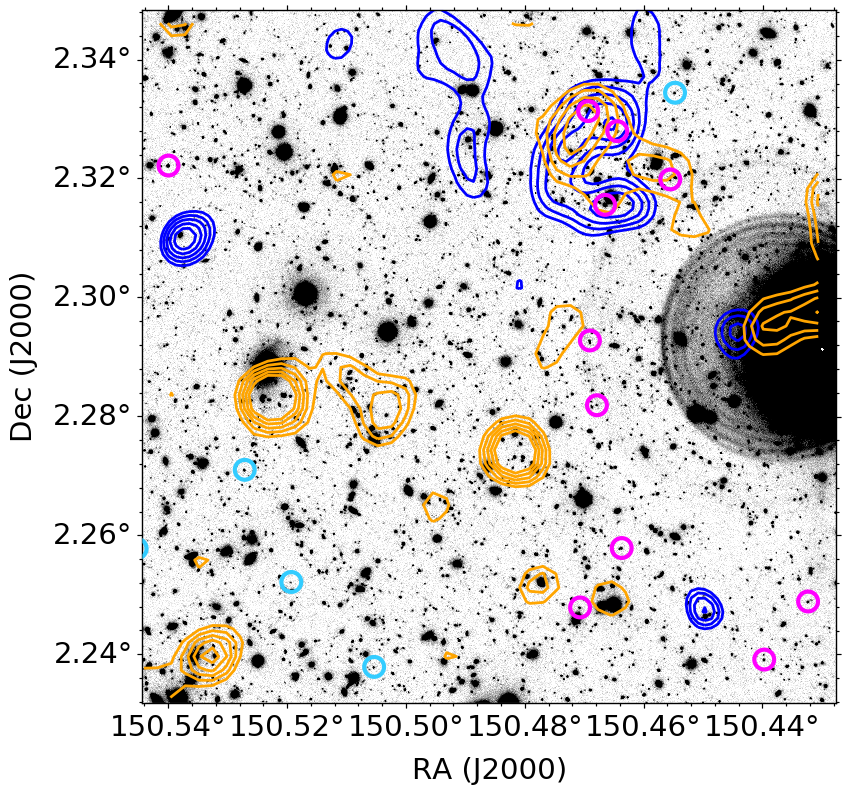}
\includegraphics[width=0.45\linewidth]{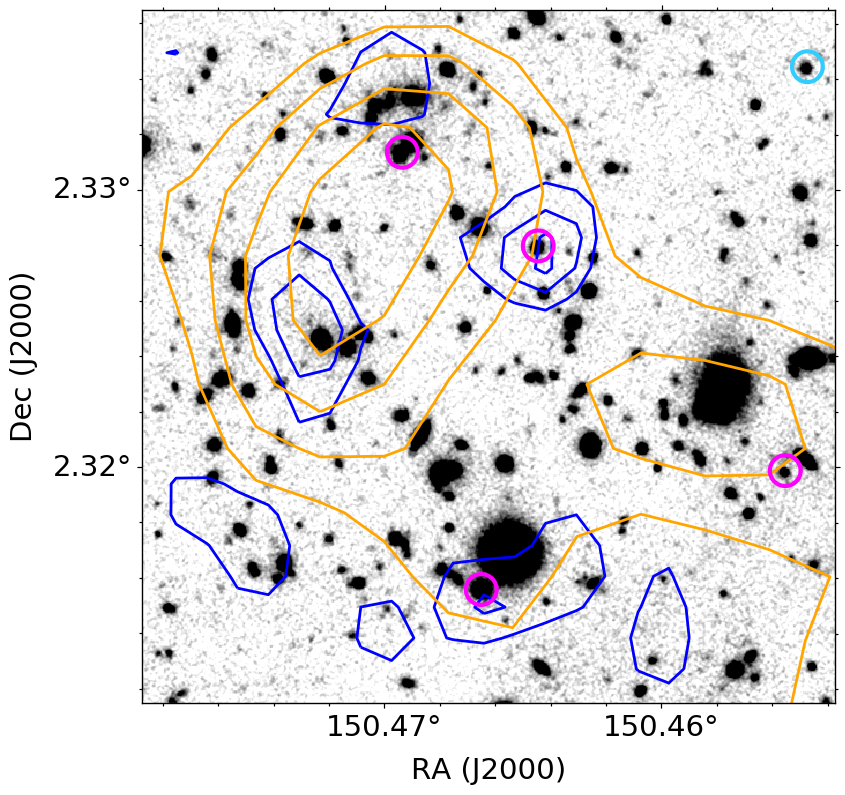}
\caption{{\small UltraVISTA K$_{\mathrm{s}}$-band image of G237.  The
spectroscopic members in G237 are shown as open circles (magenta:
2.15${\leq}z{<}$2.164, green: 2.164${<}z{<}$2.19, and light blue:
2.19${<}z{<}$2.20).  SPIRE 350$\mu$m contours are shown in orange, and XMM
4.5--10\,KeV contours in blue.  The \textit{left panel} corresponds to a
size of 7\arcmin\,$\times$\,7\arcmin, and the \textit{right panel} to a zoom
of 1.5\arcmin\,$\times$\,1.5\arcmin\ where the largest densities of
spectroscopic members and of X-ray and sub-mm sources are situated.}}
\label{fig:K_image_wcont}
\end{figure*}

Based on the spectroscopic sources spatial distribution, we test the
hypothesis that the full sample, and the two groups at
$z{\simeq}$2.15--2.164, and 2.19--2.20 might be structures.  There is no
clear and consensual definition of a high-$z$ structure.  As a reference we
consider the criterion used by~\citet{eisenhardt08} valid for $z>$1 clusters
for which at least five members must be located within a radius of 2\,pMpc
and have spectroscopic redshifts within $\pm$2000\,(1+${<}z{>}$),
corresponding to 4\arcmin, and ${\simeq}\pm$6300\,\kms\ at the redshift of
G237.  The group at 2.15${\leq}z{<}$2.164 has 11 sources within 2\,pMpc, and
that at 2.19${<}z{<}$2.20 has five sources within 2\,pMpc, thus both satisfy
this criterion.  If we consider the full list of 31 spectroscopic sources,
16 of which are located within a 2\,pMpc radius and satisfy the velocity criterion,
we can thus consider the full sample and the two smaller groups as
being structures. In the following, we refer to the group at
redshifts 2.15${\leq}z{<}$2.164 as substructure 1 (ss1), that at redshifts
2.19${\leq}z{<}$2.20 as substructure 2 (ss2), and to all 31 sources as the
full structure.  In Sect.~\ref{sec:halo_mass} we  analyze whether each
of these structures can be considered a protocluster, based on the redshift
at which they will collapse and on the estimated $z$\,=\,0 halo mass.  We
define as center of the full structure, the median coordinates of
all 31 members, $\alpha$\,=\,150.46449\,deg, and
$\delta$\,=\,2.3198330\,deg.

\section{Sub-mm emission in G237}\label{sec:herschel}

\subsection{\planck\ \textit{versus} \herschel}

One of the main questions regarding the PHz sources is the origin of the
sub-mm signal detected by \planck.  SPIRE observations 
provide images at similar wavelengths as \planck, but
with a better sensitivity and a spatial resolution 10 times higher (the SPIRE FWHM at
350\um\ being 25\arcsec). In the following, we compare the \planck\
signal in G237, with the total flux density from all SPIRE sources
detected at 350\um, and 500\um, respectively. The SPIRE observations in the
Cosmos field are from the HerMES survey~\citep{oliver12}. The SPIRE 5$\sigma$
sensitivity limits in the field are 8.0, 6.6, and 9.5\,mJy, at 250,
350, and 500\um, respectively.

We compare \planck\ and SPIRE flux densities in the region where the signal
in the RX map is higher than 50\% of the maximum value (hereafter PHz area). 
This region, shown with yellow contours in Fig.~\ref{fig:components}
overlaid on the SPIRE image of the G237 field, has an area of
29.5\,arcmin$^2$.  To compute the SPIRE flux density we consider all SPIRE
sources within the region drawn from the band-merged (or cross-ID, hereafter
xID) DR4 catalog extracted on Starfinder blind 250\,\micron\
positions~\citep{roseboom10,wang14}.  All selected sources are detected at
more than 3$\sigma$ in at least one of the three SPIRE bands, where $\sigma$
includes both the instrumental and confusion error\footnote{The confusion
noise is estimated to be 5.8, 6.3 and 6.8\,mJy/beam at 250, 350 and 500\um,
respectively~\citep{nguyen10}.}.  There are 46 SPIRE sources in the PHz
area, of which 14 are detected at more than 3$\sigma$ at 350\um, and nine at
500\um.  The total SPIRE flux density at 350 and 500\um\ is the total flux
density obtained from all detected SPIRE sources located in the PHz area. 
The \planck\ flux density at 350\um, and 550\,\um\ is instead obtained by
adding up the signal from all the pixels in the {\tt cleaned} maps within
the PHz area.  The measured total SPIRE and \planck\ flux densities are
reported in Table~\ref{tab:spire_fluxes}.  We note that the \planck\ flux
densities in the PHz area correspond to about 75, and 60\% of the
total flux densities at, respectively, 350, and 550$\mu$m listed in
the PHz catalog~\citep{planck16}, and reported in Sect.~\ref{sec:G237}.  In
comparing SPIRE and $Planck$ flux densities, we need to take into account
the difference in the filter transmission curves.  Based on the average
template derived from the ALESS sample~\citep{dacunha15}, we expect
S$^{SPIRE}_{\rm 350\mu m}$/S$^{Planck}_{\rm 350\mu m}$\,=\,0.84--0.96, and
S$^{SPIRE}_{\rm 500\mu m}$/S$^{Planck}_{\rm 550\mu m}$\,=\,1.01--1.0 at
$z$\,=\,0.5--3.0.

\begin{table*}
\centering
\caption{Number of SPIRE sources and total flux density in the G237 PHz area\label{tab:spire_fluxes}} 
\begin{tabular}{rrrr c cccc} 
\hline\hline
\multicolumn{4}{c}{N SPIRE in PHz area\tablefootmark{a}}& PHz area\tablefootmark{b} &$Planck\,S_{350}$\tablefootmark{c}& SPIRE\,$\sum$S$_{350}$\tablefootmark{d}&$Planck\,S_{550}$\tablefootmark{c}&SPIRE\,$\sum$S$_{500}$\tablefootmark{d} \\
xID & 350\micron\ & 500\micron\ & red &    (arcmin$^2$) &      (mJy)       &       (mJy)            &        (mJy)      &   (mJy) \\
\hline
   46 &        14  &          9 &   13 &       29.5  &      518$\pm$42  &            722$\pm$45  &    349$\pm$34     &            404$\pm$35  \\
\hline
\hline
\end{tabular}\\
\tablefoot{
\tablefoottext{a}{\small The number of xID, 350\micron,  and 500\micron\ sources in the PHz
designate the number of SPIRE sources with at least one 3$\sigma$ detection
in any band, at 350\micron, or at 500\micron, respectively. The red
sources are the xID sources with red sub-mm colors (i.e., $S_\mathrm{350\mu
m}/S_\mathrm{250\mu m}{>}$\,0.7 and $S_\mathrm{500\mu m}/S_\mathrm{350\mu
m}{\geq}$\,0.6).}
\tablefoottext{b}{\small The PHz area is defined as all adjacent
pixels with a value above 50\% of the maximum value in the RX map (see
text).}
\tablefoottext{c}{\small The \planck\ flux densities, $Planck\,S_{\lambda}$, have been computed
by adding the flux density values of all pixels within the PHz area.}
\tablefoottext{d}{\small The SPIRE flux densities have been computed by adding the flux density of all
xID sources within the same region.}
}
\end{table*}

The SPIRE and \planck\ flux densities at 500$\mu$m are consistent within 1$\sigma$,
while the SPIRE 350$\mu$m flux density is $>$3$\sigma$ larger than the \planck\
350\um\ flux density.  This comparison indicates that the SPIRE observations are
deep enough to detect all sources that contribute to the \planck\
flux density~\citep{herranz13,clements14}, and that the difference at 350$\mu$m
might be due to the signal from low-$z$ galaxies that were removed in the
\planck\ {\tt cleaned} images~\citep[see details on the cleaning procedure
in][]{planck16}.

\subsection{Sub-mm source density}\label{spire_density}

\begin{figure}[h!]
\centering
\includegraphics[width=\linewidth]{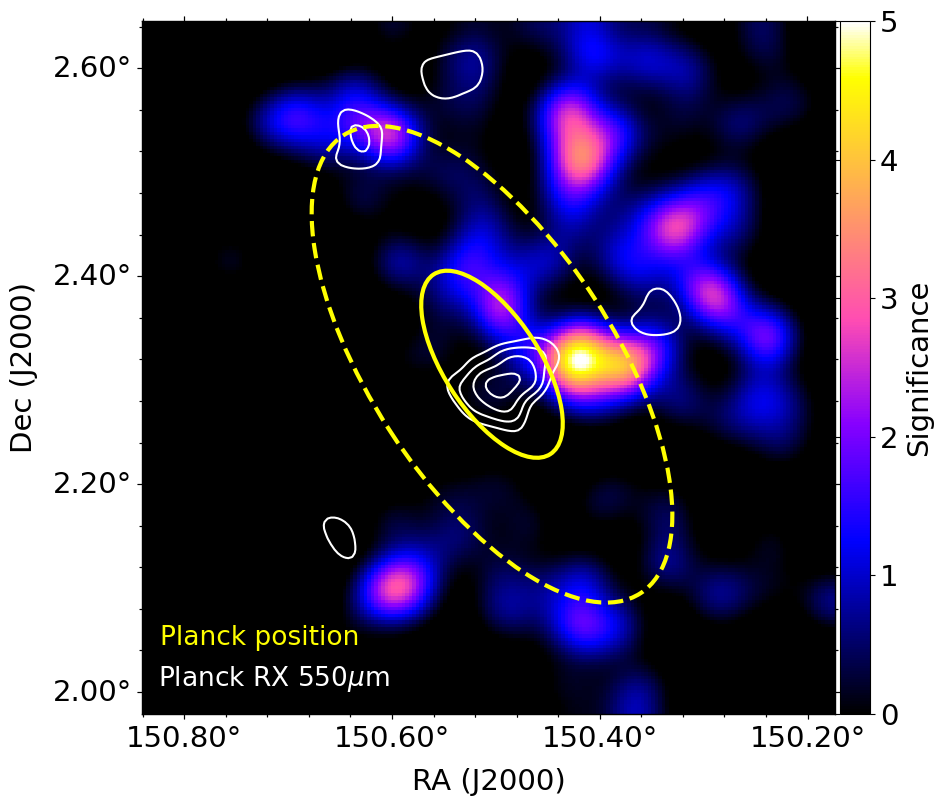}
\caption{{\small 40\,\arcmin$\times$40\,\arcmin\ map centered on G237
showing the significance of the flux-weighted density contrast of the
red SPIRE sources based on the distance distribution to the 10$^{th}$ nearest
neighbors.  White contours represent the $Planck$ signal in the 550$\mu$m
{\tt red-excess} map at 50, 62.5, 75, and 87.5\% of the peak value.  The yellow
ellipse represents the double Gaussian fit (solid: full axis equal to FWHM,
dashed: half axis equal to 3$\sigma^{pos}$) obtained when measuring the position and
extent of G237 in the \planck\ maps. The map and the contours have been smoothed with a
3\,\arcmin\ gaussian filter.}}
\label{fig:submm_overdensity} 
\end{figure}

Here, we investigate whether G237 is overdense of red SPIRE
sources, as found for the majority of PHz sources~\citep{planck15}.  To this
end, we consider all red SPIRE sources over a 1\deg$\times$1\deg\
region detected at $>$1$\sigma$ in all three SPIRE bands, and at $>$3$\sigma$ in
at least one band where $\sigma$ includes the instrumental and confusion
noises.  For each source we compute a flux-weighted local density given by
the distance distribution to the nearest 10$^{th}$ neighbors as:
\begin{equation} 
\delta_i = \frac{W_i}{\pi\,d_{i,10}^2} \sum_{j=1}^{10} \exp\left[-0.5\left(\frac{d_{i,j}}{d_{i,10}}\right)^2\right],
\label{clements_eq} 
\end{equation} 
where $d_{ij}$ is the distance to the $j^{th}$ source, $d_{i,10}$ is
the distance from the $i^{th}$ source to the 10$^{th}$ nearest neighbor, and
$W_i$ is the $i^{th}$ source weighting factor.  This factor is the ratio
between the $i^{th}$ source flux density minus the minimum flux density in
the field, and the difference between the maximum and the minimum flux
densities, where all flux densities are at 350$\mu$m~\citep[adapted from
][]{clements14}.  From the distribution of $\delta_i$ values, we then
compute a mean background density ($\rho_\mathrm{bck}$) as the 3$\sigma$
clipped mean and an associated uncertainty given by the rms
(rms$_\mathrm{bck}$).  Based on the positions of the red SPIRE
sources, and on their $\delta_i$ values, we generate a two-dimensional
density map in a 40\arcmin$\times$40\arcmin\ region centered on G237 after
convolution with a Gaussian kernel with a FWHM of 3\,\arcmin.  From the
density map we derive a density contrast map, and a density significance map
(see definition in footnote 6).  The overdensity significance map of the
red SPIRE sources, shown in Fig.~\ref{fig:submm_overdensity}, has a
peak at $\gtrsim$5$\sigma$ centered at $\alpha$\,=\,150.4186\deg,
$\delta$\,=\,$+$2.3206\deg.  In Table~\ref{tab:her_density}, we list the
maximum density contrast and significance values situated within
3$\sigma^{pos}$ from the \planck\ position (see dashed yellow ellipse in
Fig.~\ref{fig:submm_overdensity}), and the corresponding coordinates.  We
also list the area of contiguous pixels at $>$3$\sigma$ significance around
the maximum in the density significance map.  This overdensity peak is at
$\sim$5\,arcmin from the \planck\ position, which corresponds to
$\sim$1--2$\sigma^{pos}$ considering the source extension, which is given by
an ellipse with major and minor axis with $\sigma^{pos}$ of
5.3$\pm$0.4\,\arcmin, and 2.3$\pm$0.2\,\arcmin, respectively (see yellow
ellipses in Fig.~\ref{fig:submm_overdensity}).

\begin{table}
\begin{center}
\caption{Red SPIRE overdensity parameters\label{tab:her_density}}
\begin{tabular}{c c c c c} 
\hline\hline
\multicolumn{2}{c}{Maximum} & Area ($>$3$\sigma$) & $\alpha_{\rm max}$ & $\delta_{\rm max}$ \\
$\delta$    &  Signif.  & (arcmin$^2$)        &     (deg)      &     (deg)       \\
\hline
   10.4  &  5.1   &  12.8  &  150.4186   &  2.3206 \\
\hline
\hline
\end{tabular}\\
\end{center}
\end{table}

We note that this method differs from that used in~\citet{planck15} to
characterize the red SPIRE source overdensity in the PHz sample.  First,
they measured the overdensity only of the red SPIRE sources within the PHz
area (or \planck\ IN region; see previous section), and the mean density was
measured in $\sim$20\arcmin$\times$20\arcmin\ images after excluding the PHz
area.  Finally, the uncertainty associated with the mean density field used
to compute the overdensity significance was the standard deviation, rather
than the rms.  In G237, this method would not yield a significant
overdensity because the overdensity is situated just outside the PHz area. 
An offset between the red SPIRE source overdensity peak and the PHz area,
although not common, is observed in several PHz sources.  Such an offset is,
however, not surprising as the \planck\ {\tt red-excess} signal and the
SPIRE red sources do not trace exactly the same information.  We note that
the \planck\ source position is the result of a double Gaussian fit on the
550$\mu$m {\tt cleaned} map and not on the {\tt red-excess} one, and G237 is
among the most extended PHz, with only 18\% of all PHz having a major axis
with FWHM$>$12.4\arcmin, as measured in G237.  Considering the source
extension, the \planck\ position and the overdensity peak are consistent.

In the next section, we analyze the spatial distribution and properties of
the SPIRE sources in the field to understand the nature of such an
overdensity, and its possible association with the spectroscopic structure
at $z\sim$2.16.

\subsection{SPIRE member candidates}\label{sec:spire_srcs}

Here, we consider all the SPIRE sources in G237 (the
10\arcmin$\times$11\arcmin\ region shown in Fig.~\ref{fig:deltav}) with at
least a $\geq$3$\sigma$ detection and examine the available ancillary data
to assess whether any of them might be associated with the
spectroscopic sources. To this end, we take advantage of the available
deep VLA 3\,GHz, and MIPS[24$\mu$m] data~\citep{smolcic17,sanders07}, to
identify the most likely counterpart to each SPIRE source. We can
indeed assume that a source detected in the radio at 3\,GHz, or at 24\um\ is
the most likely counterpart to the SPIRE source as the emission at these
wavelengths and the sub-mm radiation are mainly powered by star formation
and, thus, highly correlated~\citep{dejong85,wunderlich87,sopp91,rieke09}.

\begin{figure}[ht!] \centering
\includegraphics[width=\linewidth]{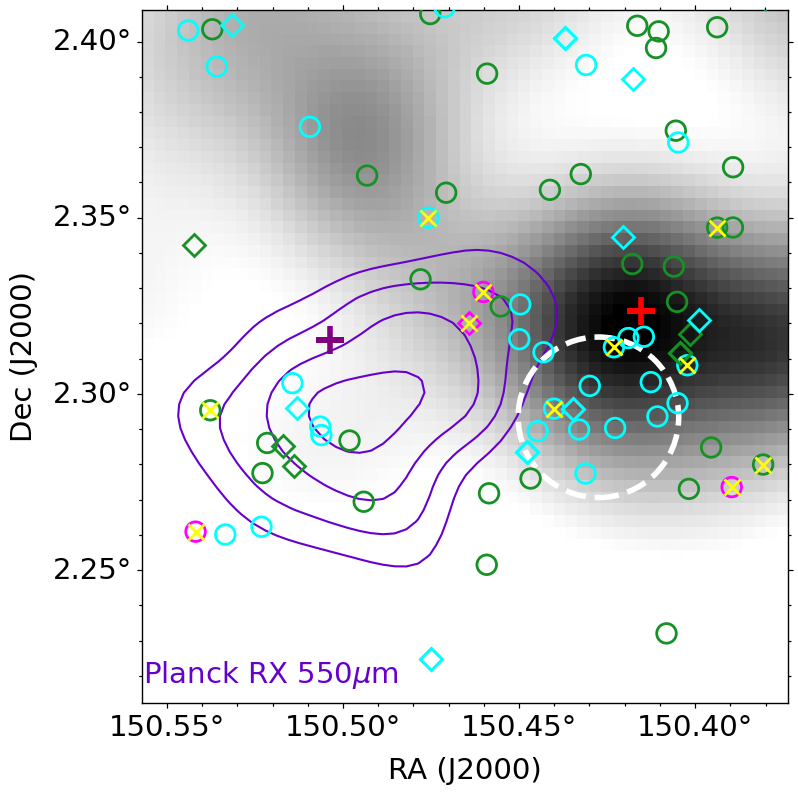}
\caption{{\small Density map of red SPIRE sources, with the
\planck\ {\tt red-excess} contours overlaid in purple (at 50, 62.5, 75,
87.5\% of the highest value).  The dashed white circle represents the 1.4\arcmin\
radius circle masked in the L16 Cosmos catalog due to a bright star.  The
\planck\ position is marked with a thick purple plus sign, while the
position where the density of red SPIRE
sources peaks is marked with a thick red plus sign.  The other symbols
represent 86 SPIRE sources with colors and redshifts consistent with
$z{\sim}$2.16 (circles are sources with a unique VLA and/or MIPS[24\um]
counterpart, diamonds are sources with multiple VLA counterparts): magenta
if 2.15${\leq}z_{\rm spec}{<}$2.20 (4); green if $z_{\rm phot}{>}$1.4 and
red (40), cyan if no redshift is available and red (42). 
We note that about 17 (40\%) of the sources with no redshift are likely at
$z{<}$1.4; see text for more details. A yellow cross is overplotted on
sources with 2.25\deg${<}\delta{<}$2.35\deg\ and with $z_{\rm spec}$ or with
a redshift estimate between 1.8 and 2.5 (11 sources).  Such a redshift
estimate was obtained by requiring the observed radio flux to be reproduced
assuming the radio-FIR correlation.}} 
\label{fig:herschel_members}
\end{figure}

There are 187 SPIRE sources, of which 183 have at least one VLA 3\,GHz or
MIPS[24\um] counterpart within 11\,arcsec, 94 with a unique VLA$+$MIPS
counterpart, eight with 2, and one with 3, and 18 have a unique VLA
counterpart with no MIPS[24\um] match.  Out of the remaining 62 SPIRE
sources with no VLA counterpart, 43 have a unique MIPS counterpart, 15 have
two, and four have three MIPS matches.  Thus, 155/183 (85\%) sources have a
single counterpart at 3\,GHz, 24\um, or in both bands.  The remaining 28 are
matched with 61 VLA/MIPS[24\um] sources.  There are only four SPIRE sources
without a match in either bands, but they are close to the region borders,
and have faint SPIRE flux densities ($\leq$20\,mJy).  We do not consider
these sources in the following as their contribution to the \planck\ flux
density would be low, and their association with the protocluster unlikely. 
Finding VLA or MIPS counterparts allows us to identify the matched source in
the L16 catalog.  In addition, for the MIPS-detected sources, deblended
SPIRE flux densities are available.  To find sources that might be members
of the $z{\sim}$2.16 structure out of all potential SPIRE counterparts, we
select those with 2.15${\leq}z_{\rm spec}{<}$2.20 (4 sources; see
Table~\ref{tab:spire}), those with $z_{\rm phot}{>}$1.4 and with sub-mm
colors consistent with being red (40 sources), and those with no redshift
information and red sub-mm colors (42 sources).  We remove 78 sources with
spectroscopic redshifts outside the redshift range of 2.15${\leq}z_{\rm
spec}{<}$2.20, five because they are not red even if $z_{\rm phot}{>}$1.4,
five because they are not red and no redshift is available, and 42 with
$z_{\rm phot}{<}$1.4.  The spatial distribution of the selected 86 SPIRE
sources is shown in Fig.~\ref{fig:herschel_members}.  These SPIRE member
candidates clump in the region where there is a significant overdensity of
red sources (see Sect.~\ref{spire_density}, and
Table~\ref{tab:her_density}).  In this region there is a bright star that
prevents us from detecting any source in the optical and NIR, and thus none
of those SPIRE sources has a redshift estimate (see cyan symbols in
Fig.~\ref{fig:herschel_members}).  To take into account the lack of redshift
estimate due to the masked region, we apply a statistical correction to
these 42 SPIRE sources without redshift.  Based on the fraction of SPIRE
sources in a large (1.16\deg$\times$1.23\deg) region of the COSMOS field
with \herschel, MIPS, and VLA coverage, we find that that $\sim$40\% of
counterparts of red SPIRE sources have a photometric or spectroscopic
redshift $<$1.4.  We can thus assume that 17 out of those 42 SPIRE sources
might be at redshift $<$1.4.  There are thus 69 SPIRE counterparts in the
G237 field with redshift consistent with being related to the structure.

Assuming a redshift of 2.16, we estimate the total (8--1000\,$\mu$m) IR
luminosities ($L_{\rm IR}$) of the SPIRE member candidates by fitting their
sub-mm spectral energy distribution (SED) with a single-temperature modified
blackbody model.  Fits were performed using the {\tt cmcirsed}
package~\citep{casey12c}, and assuming a dust emissivity-index $\beta$ equal
to 1.8.  From the IR luminosities, SFR estimates are derived assuming the
relationship in \citet{kennicutt98a}, modified for a \citet{chabrier03} IMF
\citep{chabrier03}, SFR[\msun\,yr$^{-1}$]\,=\,9.5\,$\times$\,10$^{-11}$\,L$_{\rm IR}$[\lsun]. 
The total SFR, obtained by summing the contribution from all of the 69
sources, is 14\,140$^{+670}_{-540}$\,\msun\,yr$^{-1}$, consistent with the
\planck\ estimate (i.e., \planck\ SFR\,=\,10\,173\,\msun\,yr$^{-1}$;
see~Sect.~\ref{sec:G237}).  If we limit the sample to the sources with
2.25\deg${<}\delta{<}$2.35\deg, where the \planck\ signal peaks and the
overdensity is located, we obtain 44 sources and a total SFR of
9232$^{+579}_{-468}$\,\msun\,yr$^{-1}$.  We note that to obtain these
estimates, we used as total SFR of the sources with no redshift, the value
obtained by averaging the total SFR obtained in 100 subsamples containing
only 60\% of the sources randomly selected and without repetitions.  Thus,
the \planck\ flux density and derived total SFR can be reproduced from the
SPIRE sources in the field if we select the red ones and assume a redshift
of $\sim$2.2, as implied by the sub-mm colors.  However, it is highly
unlikely that all these sources belong to the same structure.

To check the estimated SFRs, we compare the observed 3\,GHz fluxes with
those derived assuming the radio-FIR relation obtained using equations (5),
and (6) in~\citet{magnelli15} (i.e.,  $q_{\rm
IR}$\,=\,(2.35$\pm$0.8)$\times$(1${+}z$)$^{-0.12\pm0.04}$), and assuming a
radio spectrum with a power-law index of $-$0.8.  We first checked that the
radio fluxes predicted by this relation were consistent with the observed
ones using the subsample of SPIRE sources with a radio counterpart and a
spectroscopic redshift (37 sources).  We then derived the redshift that
yielded a radio flux consistent with the observed one, or with the 5$\sigma$
radio upper limit in case of non detection, for the 86 selected SPIRE
candidate members.  The derived redshifts cover a broad range with median
value of 1.4$\pm$0.7.  If we restrict the considered region to
2.25\deg${<}\delta{<}$2.35\deg (56 sources), and select only the SPIRE
sources with an estimated redshift, as derived from the radio flux and the
radio-FIR relation, between 1.8 and 2.5, we find only 11 sources (see yellow
crosses in Fig.~\ref{fig:herschel_members}).  The total SFR associated with
these 11 sources is $\sim$2410$^{+135}_{-110}$\,\msun\,yr$^{-1}$.  This
analysis implies that a significant fraction ($\gtrsim$62\%) of the selected
red SPIRE sources might be at $z{<}$2.15, $\sim$18\% in the
background, and only a dozen ($\lesssim$20\%) might be associated with the
structure at $z{\simeq}$2.16.  These results are consistent with the
simulations carried out by~\citet{negrello17} showing that a large fraction
of PHz sources might suffer from source confusion.  Nevertheless, the PHz
fields might still contain structures at $z$\,=\,1.5--3, but with lower
total SFRs than derived from the \planck\ signal, as it is the case for
G237.

\section{Properties of the spectroscopic members}\label{sec:analysis}

In this section, we consider all the sources in the two redshift peaks and
at intermediate redshifts as belonging to a unique structure with the components
described in Sect.~\ref{sec:components}.  We analyze their main properties,
such as morphology, stellar mass, SFR, AGN activity, as a function of location
within the structure.  The goal, in this section, is to investigate any relationship between
their properties and the environment.  For this analysis we use the
L16 multiwavelength catalog, and additional catalogs and images from the
IRSA public repository\footnote{The Cosmos spectroscopic redshift catalog is
available on this page: {\it
http://irsa.ipac.caltech.edu/data/Cosmos/overview.html.}}.

\begin{table*}
\centering
\caption{SPIRE detected spectroscopic targets}\label{tab:spire}
\begin{tabular}{lr ccccccc}
\hline\hline
 ID      & SPIRE     & $z_{spec}$ & S$_{\rm 24\mu m}$ & S$_{\rm 250\mu m}$ & S$_{\rm 350\mu m}$ &S$_{\rm 500\mu m}$ & Log(L$_{\rm IR}$/\lsun) & SFR$_{\rm IR}$ \\
         & ID        &            &       ($\mu$Jy)   &       (mJy)        &       (mJy)        &       (mJy)       &                         & (\msun\,yr$^{-1}$)    \\
\hline
  55326  &  13155    &   2.1576   &    172$\pm$19     &    9.6$\pm$2.0     &  $<$19.7           &  $<$15.3          &     12.12$\pm$0.02  &   126$\pm$5 \\
  58057  &  16974    &   2.1517   &    208$\pm$15     &    8.2$\pm$2.0     &  $<$19.7           &  $<$15.5          &     12.05$\pm$0.02  &   106$\pm$5 \\
  SL03   &   9741    &   2.1592   &    263$\pm$25     &   12.3$\pm$2.0     &  17.8$\pm$4.0      &  15.2$\pm$4.3     &     12.24$\pm$0.02  &   165$\pm$9 \\
  60984  &  10299    &   2.1982   &    425$\pm$16     &   16.0$\pm$2.0     &  15.7$\pm$2.5      &  12.1$\pm$3.6     &     12.36$\pm$0.02  &   216$\pm$9 \\
\hline
\hline
\end{tabular}
\end{table*}

\subsection{Sub-mm properties}\label{sec:submm_cntp}

PHz\,G237.0+42.5 was discovered by \planck\ because of its bright and red
sub-mm emission.  Spectroscopic observations in the G237 field reveal an
overdensity of sources at 2.15${\leq}z{<}$2.20.  The connection between the
sub-mm emission and the overdensity at $z{\simeq}$2.16 is likely, but it
could also be due to a projection effect with a foreground overdensity
providing additional flux.  To investigate such a connection we examine the
spectroscopic members that might be sub-mm detected.  Six SPIRE
sources have a spectroscopic member located within the SPIRE beam
($<$11\,\arcsec), IDs SL01, SL03, 55326, 58057, 60718, and 60894. To assess
whether they are the counterparts to the SPIRE sources, we examine the
radio[3\,GHz] and MIPS[24$\mu$m] images and catalogs available in the
field~\citep[L16, ][]{smolcic17}.  Since the emission in the sub-mm, in the
radio and in the mid-IR (MIR) are mainly powered by SF, we can assume that
they are produced by the same galaxy.  The higher spatial resolution in the
radio and 24$\mu$m images helps to pinpoint the optical/NIR counterpart to
the SPIRE source.  A 24$\mu$m counterpart is found in the SPIRE beam of all
6 SPIRE sources with an associated spectroscopic member, in four cases the
spectroscopic member matches the 24$\mu$m source and we can thus confirm the
match between the SPIRE source and the spectroscopic galaxy.

A 3\,GHz counterpart is found in only two of the four SPIRE sources with
spectroscopic member, and they both have a 24$\mu$m counterpart, but
different from the spectroscopic galaxy.  In three cases, a faint radio
source is visible in the images, and it matches the spectroscopic galaxy,
but none is present in the 3\,GHz 5$\sigma$ public catalog.

Based on this analysis, we assign the measured SPIRE flux to the
spectroscopic IDs SL03 (SPIRE: 9741, MIPS: 192876), 58057 (SPIRE: 16974,
MIPS: 677305), 55326 (SPIRE:13155, MIPS: 182700), and 60984 (SPIRE: 10299,
MIPS: 631549).  The SPIRE flux densities and estimated IR (8--1000\,\um)
luminosities of these four sources are listed in Table~\ref{tab:spire}.  For
the remaining spectroscopically confirmed galaxies we consider the measured SPIRE flux
density as an upper limit.

In conclusion, four out of 31 members are detected by SPIRE and have
L(IR)$>$10$^{12}$\,\lsun, and three are in ss1.  Assuming a volume density of
(3$\pm$2)$\times$10$^{-4}$\,Mpc$^{-3}$ for DSFGs with similar luminosities
and 1.3${<}z{<}$3.6~\citep{casey14}, the expected number of DSFGs in our
substructure volume is 1.3$\pm$0.9.  Thus, the number of DSFGs corresponds to
an overdensity contrast $\delta$DSFG\,=\,(3$-$1.3)/1.3\,=\,1.3, with a significance
of (3$-$1.3)/0.9\,=\,1.9$\sigma$.

Overdensities of DSFGs at $z{\sim}$2 are often associated with protocluster
candidates~\citep{blain04}, but confirming their membership through
spectroscopic identification is a challenging process~\citep{casey17}.  In
the G237 field, we carried out targeted spectroscopic observations with
LBT/LUCI of five SPIRE sources, and confirmed membership of only one DSFG
($\sim$20\%).  \citet{lacaille19} carried out spectroscopic observations of
56 DSFGs in two protoclusters at $z{\sim}$2.3, and 2.85 (HS\,1700$+$64, and
HS\,1549$+$19, respectively).  They were able to measure a spectroscopic
redshift for five DSFGs (9\%), and confirm membership of four DSFGs (7\%).

On the other hand, such an association seems to be valid only for the
bright DSFGs~\citep[i.e.,  S$_{\rm 870\mu m}\gtrsim$5\,mJy; ][]{garcia_vergara20}, corresponding to SPIRE flux
densities $\gtrsim$20\,mJy assuming a dust temperature of 30--35\,K at
$z{\sim}$2.  The four spectroscopically identified SPIRE sources in G237 have
SPIRE flux densities $\sim$10--18\,mJy, thus lower than expected to trace massive
structures.  The brightest DSFG, out of the 11 member candidates selected
in Sect.~\ref{sec:spire_srcs}, has a SPIRE flux density of $\sim$50\,mJy.  This
DSFG is bright enough to trace a massive structure, but its membership to
G237 needs to be spectroscopically confirmed.

\subsection{Members colors and classification}\label{sec:colors}

To characterize the SF activity and extinction level of the selected members
we examine their location within two well known diagnostic diagrams, the
rest-frame NUV$-$r $versus$ r$-$K, and U$-$V $versus$ V$-$J diagrams.~\citep{williams09, arnouts13}.  The
two diagrams are widely used in the literature to separate quiescent
galaxies from SFGs up to high redshifts~\cite[$z{\lesssim}$5;
][]{lemaux20,shahidi20}.  A galaxy position in these diagrams depends on its
SFR, extinction, specific SFR (sSFR\,=\,SFR/$\mathcal{M}$), age, and
metallicity~\citep[for further details, see ][]{arnouts13,davidzon16}.

\begin{figure*}[h!]
\centering
\includegraphics[width=9cm]{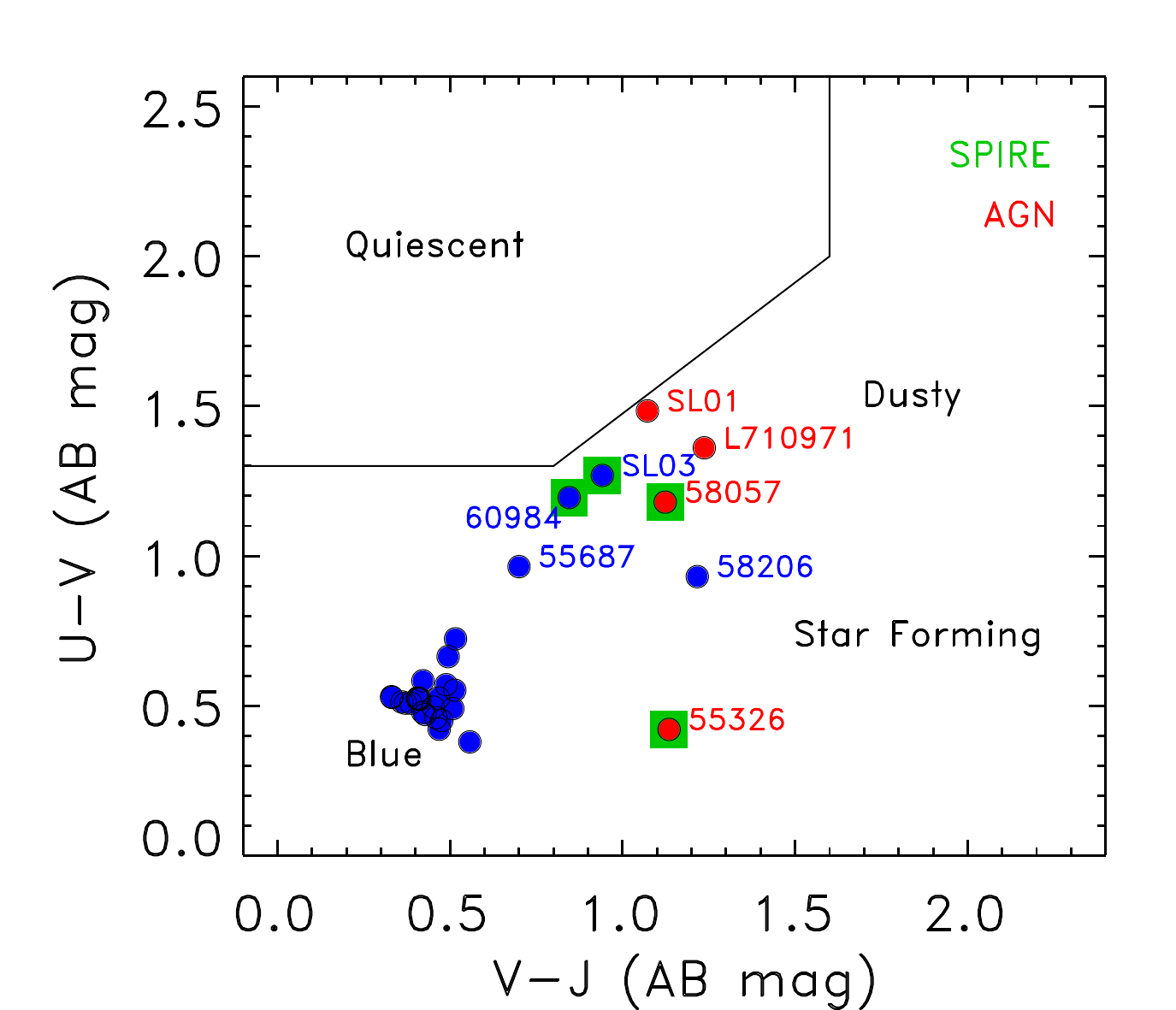}
\includegraphics[width=9cm]{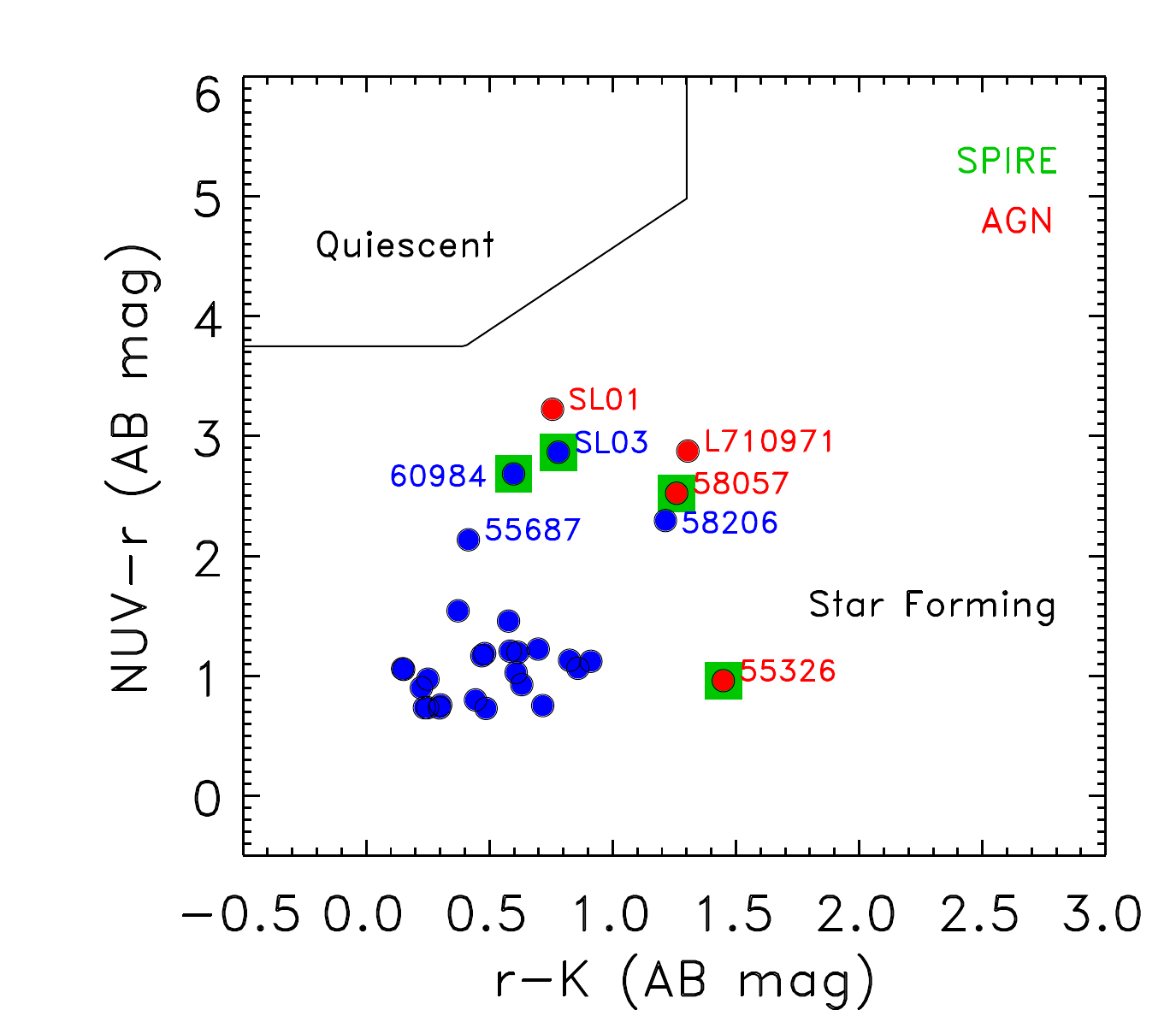}
\caption{{\small Rest-frame U$-$V $vs$ V$-$J colors ({\it left panel}) and
rest-frame NUV$-$r $vs$ r$-$K colors ({\it right panel}) for the 31
spectroscopic members (full blue circles; AGN are shown in red, and SPIRE detected sources with a green square).  The black
line separates quiescent galaxies (top left) from SFGs (bottom right) in
both panels~\citep{williams09,arnouts13,davidzon16}. }}
\label{fig:col_diagrams}
\end{figure*}

We show the positions of the 31 spectroscopic members in these color
diagrams in Fig.~\ref{fig:col_diagrams}.  Rest-frame colors are computed
from the best fit SEDs (see~Sect.~\ref{sec:mstar_sfr}).  All members are
located in the SFG region.  The vast majority (87\%) clump in the region
where blue SFGs are located, consistent with young stellar populations, and
low extinction.  The remaining eight sources (their IDs are annotated in the
figure) are more scattered within the SFG region and exhibit redder colors. 
These eight sources include the four sources detected in the sub-mm, and four AGN 
(red symbols in the figures), of which two are also sub-mm detected
(i.e., 55326 and 58057). The four sub-mm detected sources might be 
redder because of dust extinction.  Regarding the four AGN, their redder colors
might be due to the AGN component or to an older stellar component. 
Galaxies hosting an AGN might be transitioning to the quiescent region as a
consequence of reduced SF activity caused by AGN feedback~\citep[see e.g.,
][]{krishnan17,shimakawa18}.  The four AGN in the sample are, in general,
redder than the SFG members in colors that include a NIR band, such as J or
K, and they are not systematically closer to the quiescent region than the
SFGs (see red full circles in Fig.~\ref{fig:col_diagrams}).  This behavior
suggests that their redder colors are more likely due to the presence of the
AGN component whose contribution increases with wavelength, rather than to a
change in the stellar population.

The similar UV-optical colors of the majority of our spectroscopic members
allow us to carry out an analysis of their spectral properties by co-adding
their spectra.  In the next section, we will analyze the stacked spectrum
and compare it with models to estimate the average stellar age and
metallicity.  The AGN will not be taken into account in the following
analysis.

\subsection{Ultraviolet spectral properties}\label{sec:stack_spectrum}

\subsubsection{Stacked UV spectrum}

\begin{figure*}[h!] 
\centering
\includegraphics[width=\linewidth]{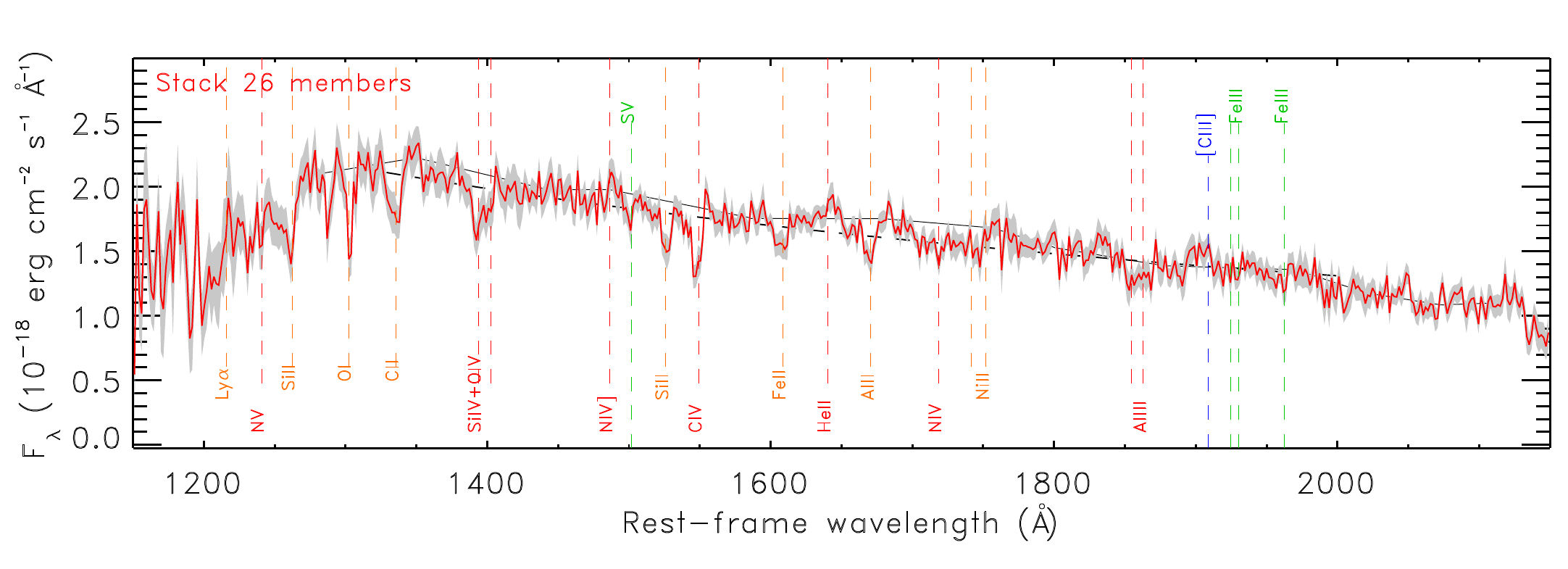}
\caption{{\small Stacked rest-frame VIMOS spectra of 26 member candidates with
$z$\,=\,2.15--2.20 (red solid line). The 1$\sigma$ uncertainty is shown as
gray filled region. The dashed black line is a linear fit obtained to
estimate the UV spectral slope $\beta$, where F$_{\lambda}\propto
\lambda^{\beta}$. The solid black curve represents the pseudo-continuum
obtained by linearly interpolating the error-weighted average flux density in the
"features-free" windows defined in Table~3 in~\citet{rix04}.
The vertical dashed lines represent the location of the main
spectral features as annotated (nebular: blue, ISM: orange, stellar 
photosphere: green, stellar winds: red).}}
\label{fig:spectra_stack}
\end{figure*}

The UV spectrum of SFGs is rich of spectral features carrying
information on several components within a galaxy, like the stellar
population, the ISM, the ionized nebulae~\citep[see
e.g., ][]{fanelli88,kewley19}.  Since the available spectra are noisy and of
moderate resolution, we make a co-added spectrum by stacking all spectra
that cover a similar rest-frame wavelength range and exhibit consistent
properties.  This selection yields 26 SFGs with zCosmos-Deep VIMOS spectra.
The stacked spectrum is obtained by taking the median value after a 3$\sigma$ clipping
procedure and normalizing all spectra using the average flux in two regions
free of strong features, from 1580 to 1830\,\AA, and from 1925 to 2060\,\AA. 
The co-added spectrum is then corrected by Galactic absorption assuming
E(B--V)\,=\,0.145~\citep{Schlafly11}\footnote{https://irsa.ipac.caltech.edu/applications/DUST/},
and the~\citet{cardelli89} extinction law.  The stacked spectrum, shown
in Fig.~\ref{fig:spectra_stack}, covers a rest-frame wavelength range from
1100 to 2200\,\AA, and has a spectral resolution of 1.7\,\AA.

The stacked spectrum shows several spectral features at the expected
wavelengths, providing support to the redshift estimates (10
out of 26 spectra have redshift flag, \textit{zflg}\footnote{Based on the
zCosmos-Bright sample~\citep{lilly07,lilly09}, the estimated redshifts are
expected to be correct in 50\% of the cases for \textit{zflg}\,=\,1, in 85\%
for \textit{zflg}\,=\,2, and in 100\% for \textit{zflg}\,=\,3, and 4.}\,=\,1,
and ten have \textit{zflg}\,=\,2, implying that about six redshift estimates
might be wrong).  In Appendix~\ref{app:spec_comparison}, we show that the
co-added spectra obtained with all 26 non AGN at $z$\,=2.15--2.20,
with the 16 galaxies with \textit{zflg}$\geq$2, and with the 12 galaxies at
$z$\,=2.15--2.16, yield consistent results.

\subsubsection{UV spectral features}

In Figs.~\ref{fig:spectra_stack} and~\ref{fig:spectra_stack_x3} we
highlight the main UV features typically observed in the spectra of SFGs:
absorption lines produced in the ISM; in the photosphere of hot, young, O
and B stars; or in stellar winds, and nebular emission lines produced in
\hii\ regions~\citep[see][]{kinney93,heckman98,gonzalez98,shapley03,leitherer11,zetterlund15,vidal17,feltre20}. 
These features can potentially provide insights on the epoch and source of
ICM enrichment in clusters and protoclusters~\citep{shimakawa15,mantz17},
and on the star formation history of protocluster galaxies~\citep[e.g.,
distinguishing between continuous and bursty star formation
activity; ][]{shimakawa14,casey16,narayanan15}. 

In the following, we analyze the co-added spectrum to estimate the UV
spectral slope, the average extinction, the intrinsic UV luminosity, and the
average stellar metallicity.  The UV-continuum slope $\beta$ is obtained by
fitting the UV spectrum with a power-law (i.e.,
F$_{\lambda}\propto\lambda^{\beta}$).  The slope $\beta$ is sensitive to the
dust reddening, the stellar age, the metal content, and even the escape
fraction~\citep{leitherer99,wilkins16}.  For a SFG with a continuous SF and
age between 10\,Myr and 100\,Myr, and metallicity
\textsc{Z$_{\ast}$}\,=\,0.004--0.02, the intrinsic slope $\beta$, computed
over the 1230--2130\,\AA\ wavelength range, ranges from $-$2.7, to
$-$2.4~\citep[see Fig.~71 in ][]{leitherer99}.  The $\beta$ value derived
from our stack over the 1308--2000\,\AA\ range is $-$1.17$\pm$0.03 (see the
dashed line in Fig.~\ref{fig:spectra_stack},
and~\ref{fig:spectra_stack_x3}).  Assuming that the difference between
observations and model predictions is due to dust extinction and the
\citet{calzetti00} dust prescription, the derived extinction at 1500\AA\
would be A$_{\rm 1500}$\,=\,1.84\,$\times$\,$\Delta\beta$\,=\,2.82, or 2.27
(corresponding to A$_{\rm V}$\,=\,1.1--0.9) for intrinsic
$\beta$\,=\,$-$2.7, or $-$2.4, respectively.  We note that a certain amount
of dust extinction would be required even in case of a SFG with an
instantaneous burst of SF.  Indeed, such a model would produce an UV
continuum with a slope $\beta$ that goes from ${\simeq}-$2.6 for an age of
10\,Myr, to ${\simeq}-$1.3 for an age of 100\,Myr, and metallicity
\textsc{Z$_{\ast}$}\,=\,0.004--0.02.  Assuming an extinction A$_{\rm
1500}$\,=\,2.3--2.8, the extinction-corrected UV (1500\,\AA) luminosity
would be $\lambda$L$_{\rm
1500\AA}$\,=\,(1.08$\pm$0.06)$\times$10$^{44}$\,\ergs.  Based on the
relation between the UV dust extinction and the metallicity given
in~\citet[][i.e., A$_{\rm
1500}$\,=\,$-$2.24$\times$log(\textsc{Z}$_{\ast}$)$-$2.16]{cullen19}, we
estimate an average metallicity \textsc{Z}$_{\ast}$\,=\,0.006--0.011.  A
more accurate estimate of the stellar metallicity is carried out in the next
section.

\subsubsection{Stellar metallicity}\label{sec:metallicity}

The stellar metallicity can be derived from the strength of photospheric
absorption features, expressed in terms of indices.  The relation between
the various UV indices and the stellar metallicity is derived assuming
stellar models with varying star formation histories (continuous or
instantaneous burst), age (typically from 5 to 150\,Myr), IMF, and metallicity (typically
$Z$\,=\,0.05--2\,$Z_{\odot}$)~\citep[e.g., Starburst99 models and BC03;
][]{leitherer99,leitherer11,bruzual03}.  The dependence on stellar age can
be usually neglected as these indices become constant after the first
$\sim$30\,Myr.  Several UV indices have been investigated in the
literature~\citep{rix04,sommariva12,vidalgarcia17,calabro21}.  The
1425\,\AA, and 1719\,\AA\ indices are among the cleanest and least
contaminated stellar metallicity tracers, and insensitive to interstellar
absorption.  They are thus useful to trace stellar population properties,
especially at higher metallicities~\citep{vidalgarcia17}.  Another reliable
metallicity estimator is the 1978\,\AA\ index.  This is associated with a
blend of numerous FeIII features between $\lambda$\,=\,1935\,\AA, and
2020\,\AA, and it is broader and stronger than most UV indices, and
sensitive to the iron abundance.  The 1978\,\AA\ index is clearly visible in
the spectra of starburst galaxies and strong in high metallicity
environment~\citep{heckman98,rix04}.  The 1425\,\AA, 1719\,\AA, and
1978\,\AA\ indices are the least affected by stellar age, dust, IMF, nebular
continuum or interstellar absorption, they are broad and can thus be
detected even at low spectral resolution~\citep[i.e., R$\sim$600;
][]{lilly07,lilly09}, but they are shallow and require
high S/N spectra.

The most significant absorption features, well visible in the stacked UV
spectrum, are the \siiv\ and the \civ\ features, respectively around 1400,
and 1550\,\AA.  They both exhibit a P-Cygni profile typical of high density
stellar winds from giant and supergiant
stars~\citep{shapley03,vidalgarcia17,feltre20}.  Although their strength
correlates with the metallicity of the parent stars, it is difficult to use
these features as metallicity indicators as they are blended with strong
interstellar absorption, and affected by other processes, including the
relative fractions of O and B stars~\citep{leitherer11}.  

\citet{sommariva12} defined a new line index, 1501\AA, associated with the
\sv\ line, an absorption feature that originates in the photosphere of hot
stars, and considered a robust metallicity indicator because almost
independent of age and IMF.  In addition, it is less affected by
uncertainties in the continuum determination because it is defined in a
small wavelength window.  Based on these considerations and on the quality
of our stacked spectrum, we use only the 1425\,\AA, 1501\AA, 1719\,\AA, and
1978\,\AA\ indices and the relationships between these indices and the
metallicity \textsc{Z/Z$_{\odot}$} derived in~\citet{sommariva12}
and~\citet{calabro21}.

\begin{figure*} 
\centering
\includegraphics[width=18cm]{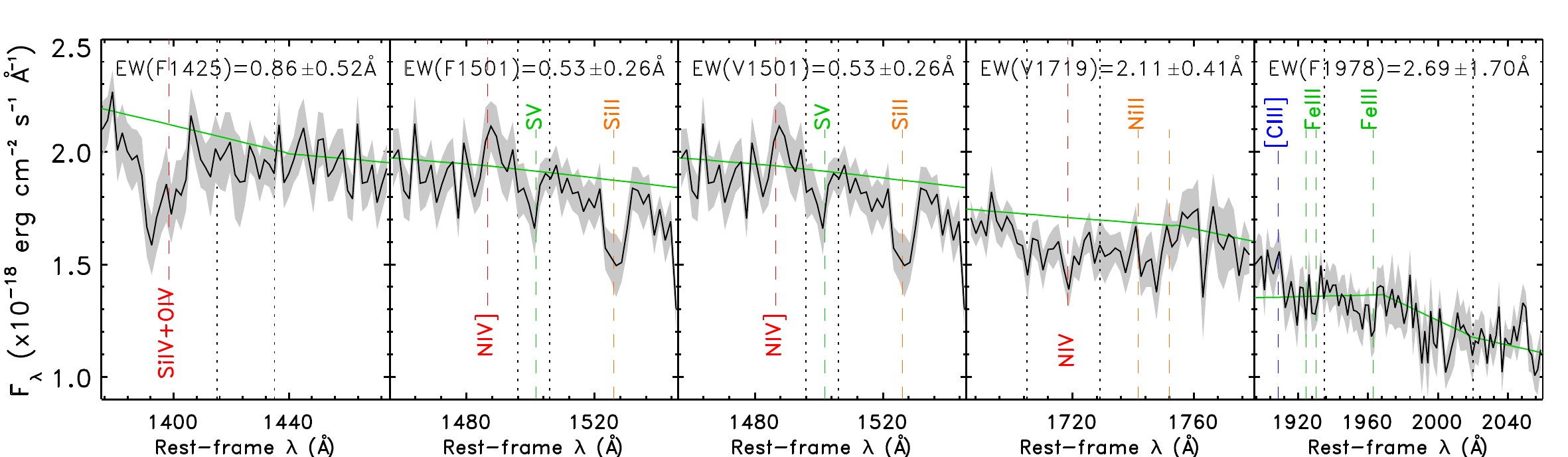}
\caption{{\small Stacked spectrum (black solid line) and 1$\sigma$
uncertainty (gray filled area) and pseudo-continuum (green solid line) use
to compute the indices listed in Table~\ref{tab:indices}.  The wavelength
interval used to compute the EW is delimited by two black vertical dotted
lines.  The name and location of the main spectral features are indicated by
dashed lines as in Fig.~\ref{fig:spectra_stack}.  The final EWs, computed as
the median and 1$\sigma$ uncertainty measured after perturbing the spectrum
1000 times, are annotated on top of each panel.}}
\label{fig:indices}
\end{figure*}

The indices are obtained by measuring the EW within well
defined wavelength ranges~\citep[see Table 2 in ][]{calabro21}. 
Following~\citet{sommariva12,calabro21}, we compute a pseudo-continuum by
linearly fitting the error-weighted mean density fluxes computed in narrow
($\Delta\lambda\simeq$3\,\AA) windows~\citep[see Table~\,3 in ][]{rix04}. 
Because of our low spectral resolution ($R{\sim}$600) we extended the width
of such windows by $\pm$3\,\AA\ ($\Delta\lambda\simeq$11\,\AA).  The derived
continuum is shown as solid black line in Fig.\ref{fig:spectra_stack}, and
as a green line in Fig.~\ref{fig:indices}.  The EW and associated
uncertainty of each index correspond to the median value, and the standard
deviation obtained from 1000 computations, generated by perturbing the
spectrum.  The perturbation constitutes in adding noise to each wavelength
element extracted from a normal distribution with a standard deviation
defined by the error on the spectrum at that wavelength element.  For each
perturbed spectrum, an EW is measured computing a new pseudo-continuum.  In
Fig.~\ref{fig:indices}, we show the spectrum around each index and the
computed pseudo-continuum used for the EW measurement of the unperturbed
spectrum.  The estimated metallicities are given in Table~\ref{tab:indices}.

\begin{table}
\begin{center}
{\renewcommand{\arraystretch}{1.4}
\caption{UV spectral indices}\label{tab:indices}
\begin{tabular}{lcccc}
\hline\hline
 Index &  EW (\AA) & Log(\textsc{Z}$_{\ast}$/\textsc{Z}$_{\odot}$)  & \textsc{Z}$_{\ast}$/\textsc{Z}$_{\odot}$ & Ref.\tablefootmark{a} \\
\hline
 F1425    &  0.86$\pm$0.52    &   $-$0.64$^{+0.58}_{-0.82}$  &  0.23$^{+0.64}_{-0.19}$ & (1) \\
 F1501    &  0.53$\pm$0.27    &   $-$0.56$^{+0.59}_{-0.86}$  &  0.27$^{+0.78}_{-0.24}$ & (1) \\
 V1501    &  0.53$\pm$0.27    &   $-$0.79$^{+0.32}_{-0.46}$  &  0.16$^{+0.18}_{-0.11}$ & (2) \\
 V1719    &  2.11$\pm$0.41    &   $-$0.33$^{+0.18}_{-0.20}$  &  0.47$^{+0.24}_{-0.17}$ & (2) \\
 F1978    &  2.69$\pm$1.70    &   $-$0.57$^{+0.69}_{-1.23}$  &  0.27$^{+1.05}_{-0.25}$ & (1) \\
\hline
 \multicolumn{2}{l}{Error-weighted mean} &   $-$0.44$^{+0.16}_{-0.16}$  & 0.36$^{+0.16}_{-0.11}$ &\\
\hline
\hline
\end{tabular}\\
\tablefoot{
\tablefoottext{a}{\small The relationship between the metallicity and a
specific index was taken from either (1)~\citet{sommariva12}, or (2)~\citet{calabro21}.}
}}
\end{center}
\end{table}

\begin{figure} 
\centering
\includegraphics[width=8cm]{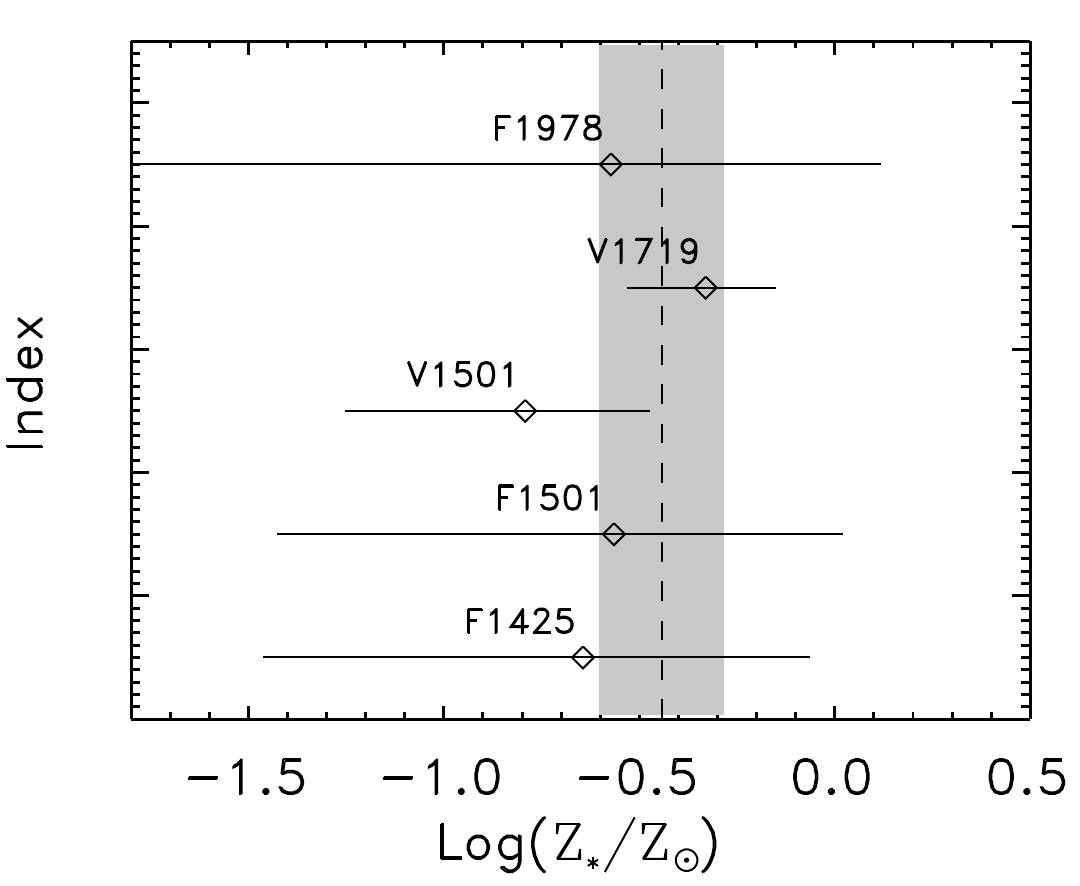}
\caption{{\small Stellar metallicity values derived from the co-added UV
spectrum of 26 SFG members using the UV indices described in
Sect.~\ref{sec:metallicity}, and listed in Table~\ref{tab:indices} (black
diamonds). The vertical dashed line represents the error-weighted mean, and
the gray area the associated 1$\sigma$ uncertainty. }}
\label{fig:histo_metallicity}
\end{figure}

Despite the weakness of the selected features in deriving
UV indices and the uncertainty in determining the continuum, all estimates
are consistent within 1$\sigma$ (see Fig.~\ref{fig:histo_metallicity}), and yield, on average, an error-weighted
average metallicity
\textsc{Z}$_{\ast}$/\textsc{Z}$_{\odot}$\,=\,0.36$^{+0.16}_{-0.11}$, corresponding to
\textsc{Z}$_{\ast}$\,=\,0.005$^{+0.002}_{-0.002}$~\citep[assuming
\textsc{Z}$_{\odot}$\,=\,0.0142; ][]{asplund09}.

\subsubsection{Comparison with SFG UV spectra}

To investigate whether the observed UV spectral properties are peculiar to
our protocluster members or consistent with those observed in field galaxies,
we compare our stacked spectrum with those obtained in other SFGs from the
literature.  This comparison is useful to identify similarities with
specific types of galaxies that can potentially help us to interpret our
spectrum in a model-independent way, and might also pinpoint possible artifacts
in our stack due to the low number of stacked sources and to the spectral
noise. 

\begin{figure*} 
\centering
\includegraphics[width=18cm]{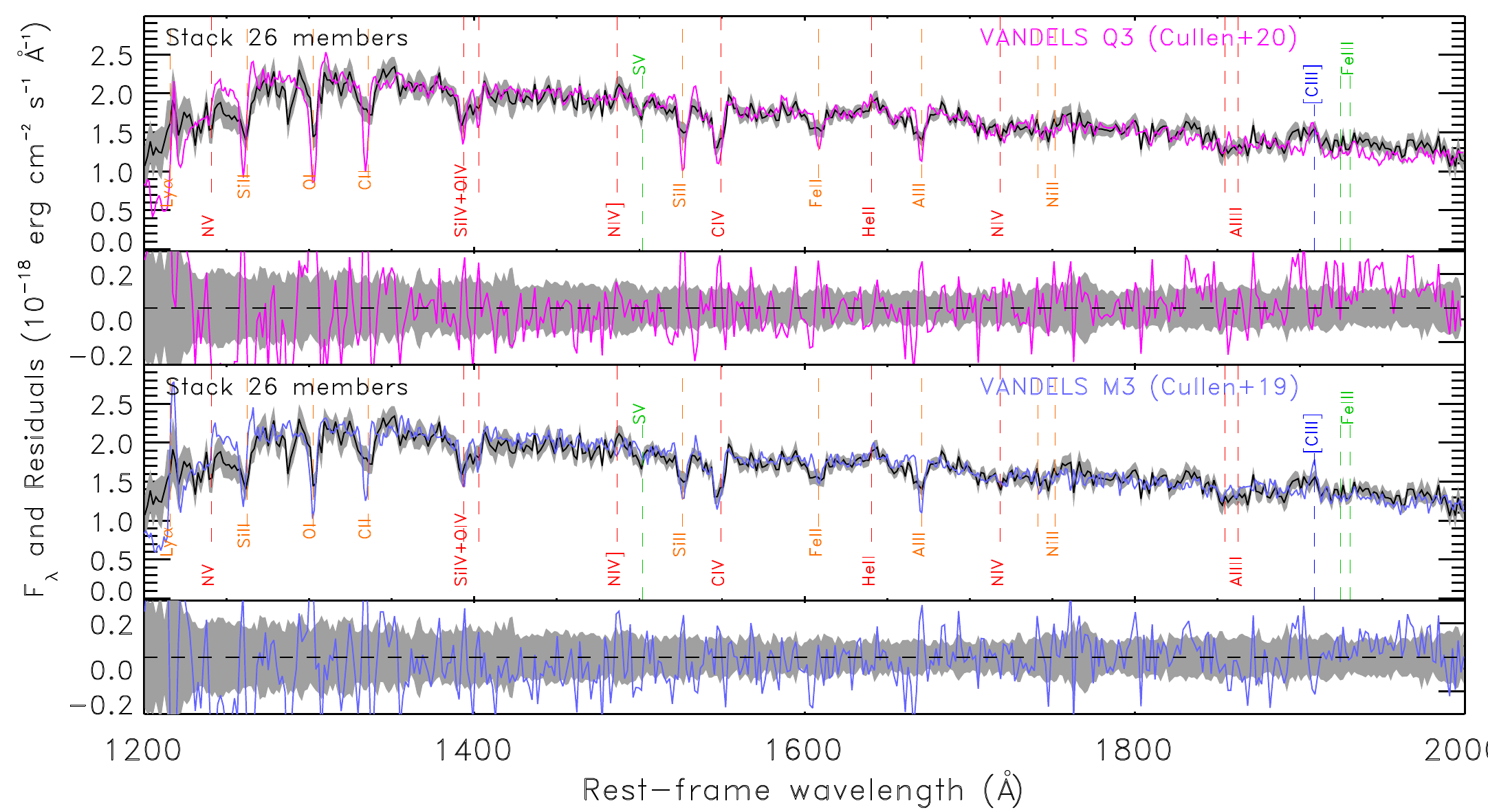}
\caption{{\small Stacked spectrum (black solid line) and 1$\sigma$
uncertainty (gray filled area) compared with composite spectra obtained from
stacking the VANDELS spectra of two LAE subsamples: Q3 (\textit{top
panel}: magenta line); and m3 (\textit{bottom panel}: blue
line)~\citep{cullen19, cullen20}.  Below each panel residuals are plotted and
the 1$\sigma$ uncertainty (gray shaded area) associated with our stack.  The
main spectral features are highlighted with vertical dashed lines and
annotated.}}
\label{fig:stack_vs_obs}
\end{figure*}

High signal-to-noise UV spectra for SFGs at $z\gtrsim$2 are available from
stacking Lyman Break Galaxies~\citep[LBG; ][]{shapley03}, and \lya\
emitters~\citep[LAE; ][]{cullen19, cullen20}.  We collected 12 composite
spectra for LAEs at $z{\simeq}$3.5 divided in bins of stellar mass, or \lya\
EW, and one for LBGs at $z{\simeq}$3.  Half of these spectra
have continuum slopes inconsistent with ours, among the remaining half only
two composite spectra exhibit both continuum slope and spectral features
consistent with our stacked spectrum.  The two most similar composite
spectra are obtained from the VANDELS-m3 LAE sample~\citep[i.e., 153
galaxies with log($\mathcal{M}$/\msun)\,=\,9.20--9.50, and
2.30${\leq}z{\leq}$5.0; ][]{cullen19}, and the VANDELS-Q3 sample~\citep[LAEs
with log($\mathcal{M}$/\msun)\,=\,9.54$\pm$0.44, and 3.0${\leq}z{\leq}$5.0;
][]{cullen20}.  The m3 and Q3 composite spectra are compared with our stack
in Fig.~\ref{fig:stack_vs_obs}.  The two samples of LAEs with UV spectra
similar to our stack, m3 and Q3, are best fitted with stellar models with
metallicity \textsc{Z}$_{\ast}$\,=\,0.0017$\pm$0.0001, and
0.0021$\pm$0.0002, respectively~\citep[assuming
\textsc{Z}$_{\sun}$\,=\,0.0142; ][]{asplund09}.  These are, respectively,
about 0.5, and 0.4\,dex lower than the average metallicity of our
protocluster members.  This discrepancy is consistent with the difference in
stellar mass, and redshift between our sample and the m3 and Q3 samples. 
Indeed, the mean stellar mass of the 26 SFGs used in the stacked spectrum is
Log($\mathcal{M}$/\msun)\,=\,10.0$\pm$0.44 (see~Sect.~\ref{sec:mstar_sfr}),
about 0.5\,dex larger than the LBG median stellar mass.  There is well know
relation between stellar mass and metallicity in SFGs that evolves with
redshift~\citep[see e.g., ][]{calabro21} that reconciles the value 
obtained for our protocluster members with those derived for these LAE
samples.

On the other hand LAEs are, typically, more metal poor than SFGs at the same
stellar mass and redshift~\citep{finkelstein11}, suggesting that our members
might be more metal poor than typical SFGs at the same redshift.  Similar
results have been obtained in other protoclusters at
$z{\sim}$2~\citep{chartab21,sattari21}, and interpreted as a sign of
pristine cold gas accretion, but this is not observed in all
protoclusters~\citep[see e.g., ][]{shimakawa15}.

\subsection{Members stellar masses and SFRs}\label{sec:mstar_sfr}

To determine the activity level and evolutionary phase of the selected
protocluster members, we estimate their stellar mass and SFR and
compare them with the scaling relations that describe normal SFGs, starburst
galaxies, and quiescent galaxies. 

To estimate stellar masses and SFRs of the spectroscopic members, we fit
their SEDs with galaxy models.  The SEDs are obtained from the L16 Cosmos
multiwavelength catalog.  We use 3\arcsec\ diameter aperture-corrected
magnitudes, and apply the relative photometric and systematic offsets
following L16 equation (9).  Magnitudes are also corrected for foreground
galactic extinction using the given reddening $E(B-V)$ values and extinction
factors following L16 equation (10).  We use data in 20 broad-band filters,
including CFHT/MegaPrime $u$, CFHT/WIRCam $H$, and $K_{\rm_s}$,
SUBARU/SuprimeCamera $B$, $V$, $r$, $i$, $z^{++}$, VISTA/VIRCAM $Y$, $J$,
$H$, $K_{\rm s}$ from the ultra-VISTA 3$^{rd}$ data
release~\citep{mccracken12}, \spitzer/IRAC [3.6, 4.5, 5.8, 8.0$\mu$m] and
MIPS [24, 70, and 160$\mu$m], and \herschel/SPIRE [250, 350, and 500$\mu$m]
bands.  A flux estimate in the SPIRE band from either HerMES or L16 is
available for ten of the 31 spectroscopic members.  Those listed in L16 are
all detections at $>$3$\sigma$ and have been obtained using as prior the
MIPS[24$\mu$m] position.  These are available for four sources at 250$\mu$m,
and for two at 350, and 500$\mu$m (see Table~\ref{tab:spire}). In the
remaining cases, we use the HerMES data and treat them as upper limits even
if the flux measurement is $>$3$\sigma$ because the selected source might not be
the only counterpart to the SPIRE flux.
Upper limits to the SPIRE fluxes are equivalent to 5$\sigma$, where $\sigma$
includes instrumental and confusion noise of the Cosmos HerMES
observations~\citep[see Sect.~\ref{sec:phz_sample};][]{oliver12}.

We use the Code Investigating GALaxy Emission~\citep[CIGALE;
][]{noll09,ciesla15,cigale19}, with redshift equal to the spectroscopic one,
to fit the observed SEDs.  The model assumptions are: an exponentially
declining star formation history with a random burst of star formation,
the~\citet{bruzual03} stellar synthesis population models,
a~\citet{chabrier03} IMF, solar metallicity, continuum and line nebular
emission, the~\citet{calzetti00} dust attenuation prescription,
the~\citet{draine07} dust emission models (module \textsf{dl2014}), and an
AGN component as modeled by~\citet{fritz06}.  \citet{ciesla15} show that not
taking an AGN component into account when fitting broad-band SEDs of AGN
host galaxies might result in a bias in the estimated stellar mass and SFR. 
Since some of the selected sources might host an AGN, even if not detected
in the spectrum or in the X-ray, we perform fits both with and without an
AGN component for all sources.  Based on the resulting $\chi^2$, the
addition of an AGN component yields a better fit only in two cases, IDs
55326, and 58057.  The SEDs of the other two AGN of the sample, IDs SL01
and L710971, do not produce better fits with an additional AGN component.

In Fig.~\ref{fig:cig_seds}, we show the best fit SEDs and in
Table~\ref{tab:cig_params} the stellar masses and SFRs obtained from the
best fits.
\begin{figure}[htbp!]
\centering
\includegraphics[width=8.5cm]{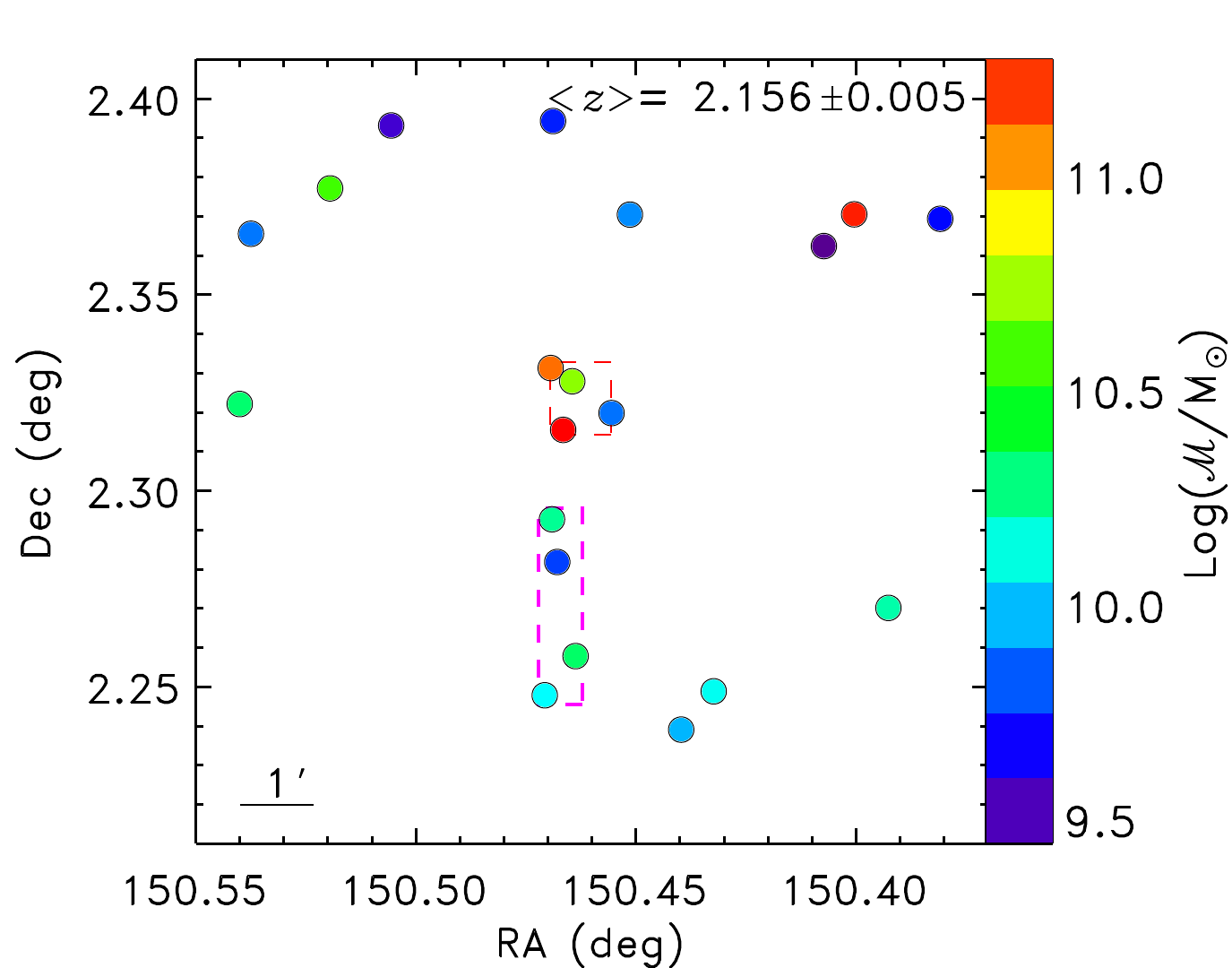}
\includegraphics[width=8.5cm]{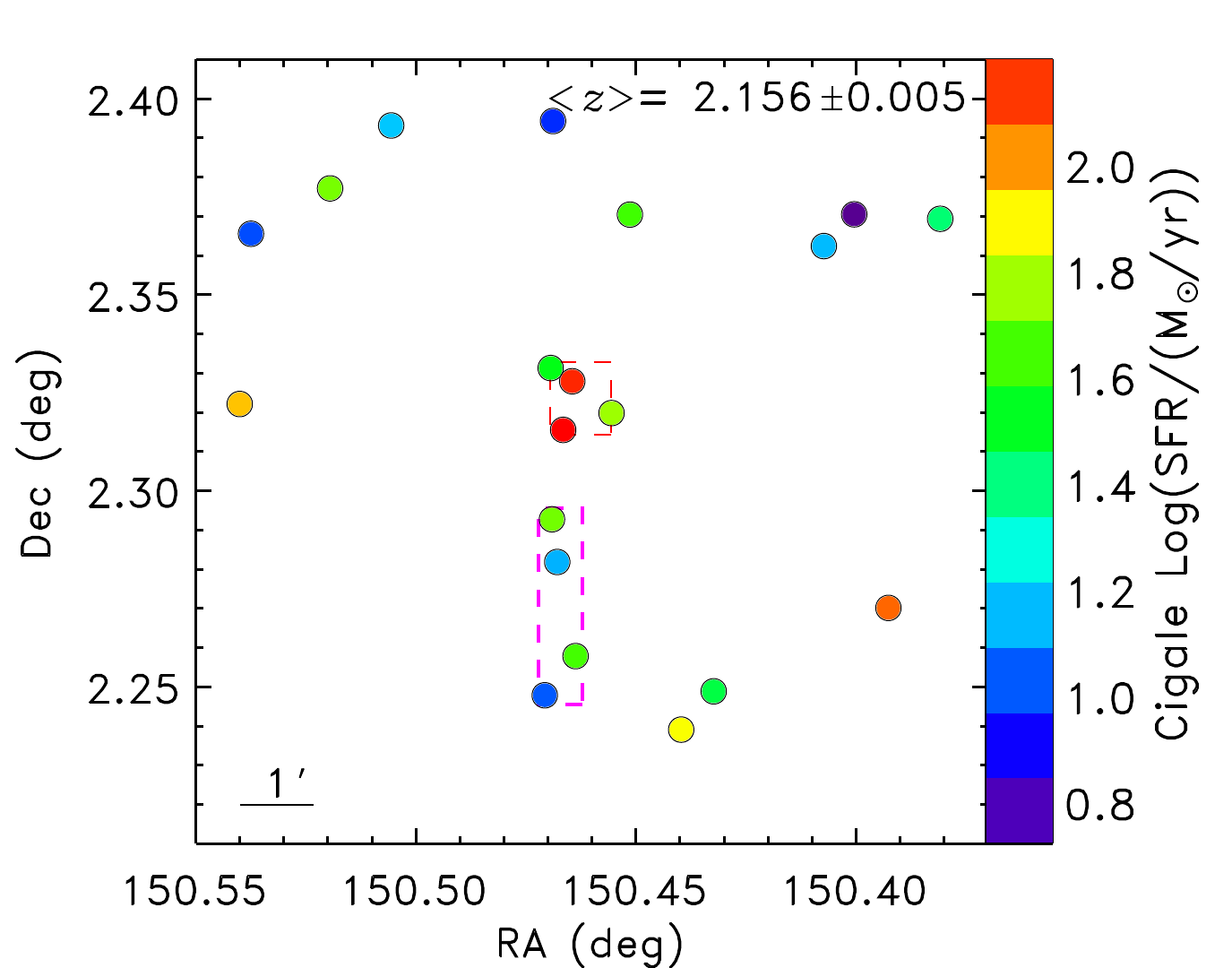}
\caption{{\small Spatial distribution of all spectroscopic sources with
$z$\,=\,2.15--2.16 with colors corresponding to the estimated stellar mass
(\textit{top panel}), SFR (\textit{bottom panel}).
Dashed magenta rectangles delineate the structure components A, B
described in~Sect.~\ref{sec:components}, and shown in Fig.~\ref{fig:components}.}}
\label{fig:cigale_mass_sb}
\end{figure}

\begin{figure} 
\centering 
\includegraphics[width=\linewidth]{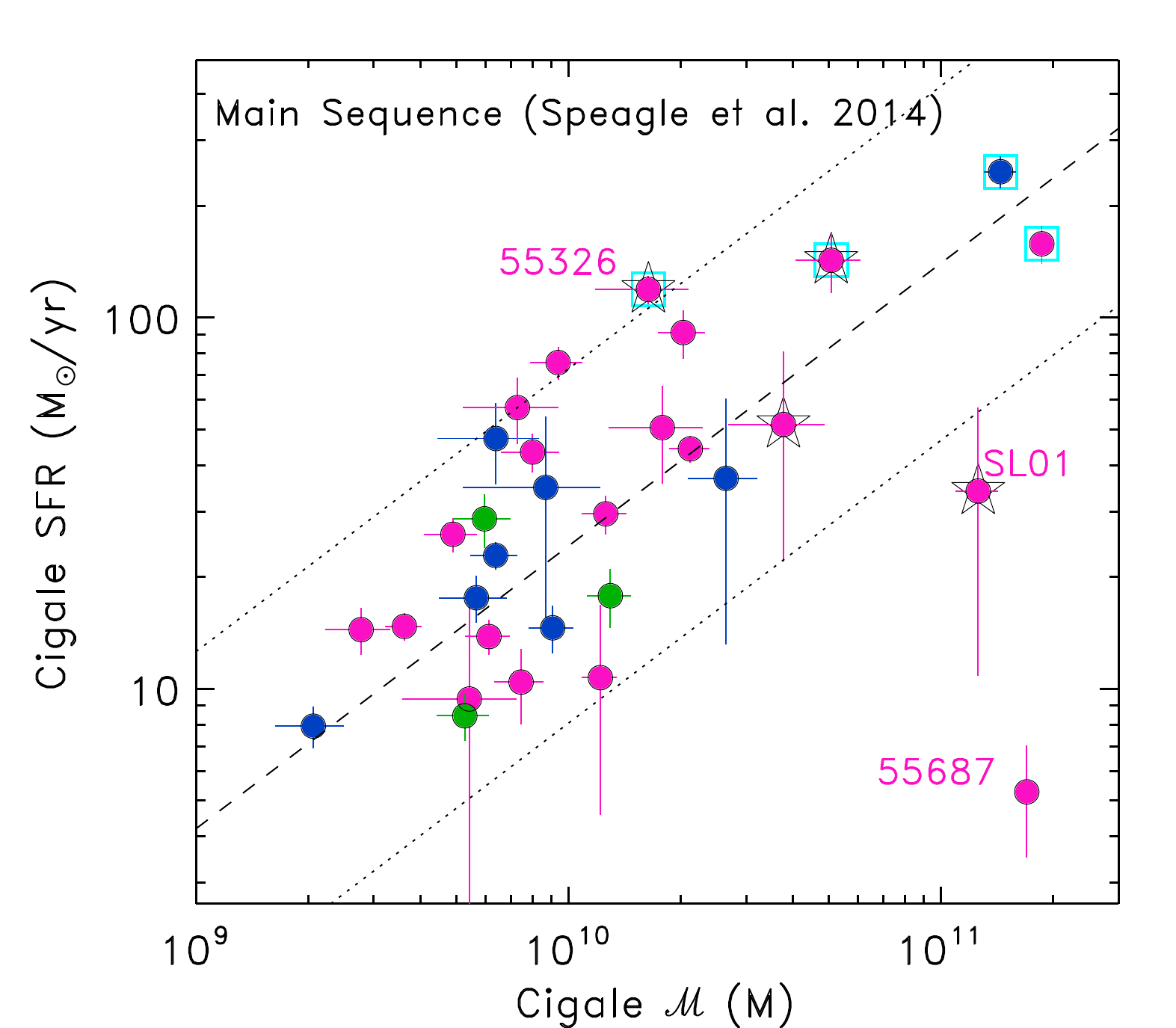} 
\caption{SFR as a function of stellar mass for the 31 spectroscopic members
(full circles, magenta: 2.15${\leq}z{\leq}$2.164, green:
2.164${<}z{<}$2.19, and blue: 2.19${\leq}z{\leq}$2.20).
AGN are indicated with a black star, and sources detected by SPIRE with a
cyan square. The black dashed line is the MS relation as $z$\,=\,2.17 as
formulated by~\citet{speagle14} (corrected for a \citet{chabrier03} IMF), and the black dotted lines represent a
factor of 3 scatter around the MS.}
\label{fig:cigale_ms} 
\end{figure} 

\begin{table}
\centering
\caption{CIGALE best-fit parameters of the spectroscopic members in the G237
field}\label{tab:cigale}
\begin{tabular}{l rrc}
\hline\hline
\multicolumn{1}{c}{ID}     & \multicolumn{1}{c}{$\mathcal{M}$}      & \multicolumn{1}{c}{SFR}        & Log(SFR/SFR$_{\mathrm MS}$)\tablefootmark{a} \\
          & \multicolumn{1}{c}{(10$^{10}$\,\msun)}  & \multicolumn{1}{c}{(\msun\,yr$^{-1}$)} &            \\
\hline
\multicolumn{4}{c}{ss1: $z$\,=\,2.15--2.164}\\
\hline
55326\tablefootmark{b}  &  1.64$\pm$0.46  &   119$\pm$10   & \;\;0.46     \\
  55687    & 17.01$\pm$0.85  &   5.3$\pm$ 1.8 & $-$1.66     \\
  56014    &  0.28$\pm$0.06  &    14$\pm$2    & \;\;0.14     \\
  56915    &  1.26$\pm$0.17  &    30$\pm$4    & $-$0.05     \\
  57557    &  0.80$\pm$0.14  &    43$\pm$5    & \;\;0.26     \\
  58020    &  2.12$\pm$0.26  &    44$\pm$4    & $-$0.05     \\
58057\tablefootmark{b}  &  5.08$\pm$1.00  &   143$\pm$26   & \;\;0.17     \\
  58173    &  0.61$\pm$0.08  &    14$\pm$2    & $-$0.14     \\
  58200    &  0.54$\pm$0.18  &   9.4$\pm$8.3  & $-$0.27     \\
  58206    &  1.79$\pm$0.51  &    51$\pm$15   & \;\;0.06     \\
  58266    &  1.22$\pm$0.13  &  10.7$\pm$6.1  & $-$0.48     \\
  59504    &  0.36$\pm$0.04  &  14.7$\pm$1.3  & \;\;0.06     \\
  60718    &  0.75$\pm$0.11  &  10.4$\pm$2.4  & $-$0.33     \\
  60811    &  2.03$\pm$0.29  &    91$\pm$14   & \;\;0.28     \\
L710971    &  3.78$\pm$1.10  &    52$\pm$30   & $-$0.18     \\
  SL01     & 12.60$\pm$1.67  &    34$\pm$23   & $-$0.76     \\
  SL03     & 18.67$\pm$1.09  &   158$\pm$19   & $-$0.22     \\
  54698    &  0.49$\pm$0.08  &    26$\pm$3    & \;\;0.20     \\
  57138    &  0.94$\pm$0.15  &    76$\pm$8    & \;\;0.45     \\
  57730    &  0.73$\pm$0.21  &    57$\pm$12   & \;\;0.41     \\
\hline                                                             
\multicolumn{4}{c}{2.164${<}z{<}$2.19}\\                                
\hline                                                             
  54765    &  0.53$\pm$0.08  &     9$\pm$1    & $-$0.31     \\
  55951    &  1.30$\pm$0.17  &    18$\pm$3    & $-$0.29     \\
  56005    &  0.60$\pm$0.10  &    29$\pm$5    & \;\;0.18     \\
\hline                                                             
\multicolumn{4}{c}{ss2: $z$\,=\,2.19--2.20}\\                           
\hline                                                             
  54677    &  2.65$\pm$0.56  &    37$\pm$24   & $-$0.21     \\
  56950    &  0.87$\pm$0.35  &    35$\pm$19   & \;\;0.13     \\
  57695    &  0.64$\pm$0.19  &    47$\pm$12   & \;\;0.37     \\
  59496    &  0.57$\pm$0.12  &    18$\pm$3    & $-$0.02     \\
  59906    &  0.64$\pm$0.09  &    23$\pm$2    & \;\;0.05     \\
  60026    &  0.21$\pm$0.04  &     8$\pm$1    & $-$0.03     \\
  60355    &  0.91$\pm$0.12  &    15$\pm$2    & $-$0.26     \\
  60984    & 14.47$\pm$1.45  &   247$\pm$24   & \;\;0.05     \\
\hline                                                        
\multicolumn{4}{c}{full: 2.15${<}z{<}$2.20}\\                           
\hline                                                        
 Total     &  96$\pm$3       &   1485$\pm$71  & \;\;0.22     \\ 
\hline
\hline
\end{tabular}\\
\tablefoot{
\tablefoottext{a}{\small Logarithm of the ratio between the estimated SFR and the expected SFR
assuming the main sequence relation~\citep{speagle14} at the source redshift
and stellar mass corrected for a \citet{chabrier03} IMF}.
\tablefoottext{b}{\small The AGN fraction, or contribution to the bolometric luminosity is
5$\pm$2\% in 55326, and 6$\pm$4\% in 58057.}
}
\label{tab:cig_params}
\end{table}

The estimated stellar masses range from
$\sim$0.2--20$\times$10$^{10}$\,\msun.  Compared to the characteristic mass
$\mathcal{M}^{\ast}$ obtained by fitting the Schechter mass function of SFGs
at 2.0${<}z{<}$2.5~\citep[i.e.,
$\mathcal{M}^{\ast}$\,=\,4$^{+1.6}_{-1.0}{\times}$10$^{10}$\,\msun;
][]{davidzon17}, most (81\%) of the spectroscopic members are less massive,
and only 13\% (4 galaxies) are more massive than $\mathcal{M}^{\ast}$, with
stellar masses $\simeq$1--2$\,{\times}\,10^{11}$\,\msun.  The four galaxies
with $\mathcal{M}{>}$10$^{11}$\,\msun\ include one AGN, and two SPIRE
detected sources.  The relatively low stellar masses for the other galaxies
suggests that these have not yet reached their maximal mass, and might be at
an early stage of evolution, as indicated by their UV spectral properties.

We examine whether there is a relation between the stellar mass and the
location of a galaxy within the structure as previously found for the HAEs
in the structure by~\citet{koyama21}.  We find that the mean stellar mass in
the core (i.e., $<$Log($\mathcal{M}^{\rm core}$/\msun)$>$\,=\,10.7$\pm$0.4)
is 0.5\,dex larger than the average of all ss1 members (i.e.,
$<$Log($\mathcal{M}^{\rm ss1}$/\msun)$>$\,=\,10.2$\pm$0.4), but this
difference is not highly significant considering the wide range of stellar
masses.  We plot the spatial distribution of the members of ss1 in
Fig.~\ref{fig:cigale_mass_sb}.  Two of the most massive galaxies (IDs: SL01,
and SL03) are in the core, while the four least massive ones are in the most
external regions.  We note that the three galaxies in the core are also HAEs. 
The CIGALE mass estimates are consistent with those estimated
in~\citet{koyama21} for IDs SL01, and SL03, and 60\% lower for ID 58057. 
Such a difference is probably due to the AGN component included in our fit,
and not considered in~\citet{koyama21}.  This analysis agrees with
the results reported by~\citet{koyama21}, the most massive members are
located in the structure core, although the enhancement of stellar
mass in the core is less pronounced than found with the HAEs.\\

In terms of SFRs, they range from 5 to $\sim$250\,\msun\,yr$^{-1}$, with mean
$<$SFR$^{\rm ss1}{>}$\,=\,33$\pm$25(\msun\,yr$^{-1}$) for all ss1 members, and 
$<$SFR$^{\rm core}{>}$\,=\,73$\pm$52\,\msun\,yr$^{-1}$ in the core. There is thus also
a higher average SFR in the core than in the whole ss1 structure. This is also illustrated in the middle
panel of Fig.~\ref{fig:cigale_mass_sb}, where we show the spatial distribution of
the sources in ss1 as a function of their SFR. 
The figure shows the core having more members with high SFRs than the outer regions, 
but as in the case of stellar mass the difference is not highly significant.
The estimated total SFR for the six HAEs in our spectroscopic sample is 544\,\msun\,yr$^{-1}$ from the
dust-corrected \halpha\ luminosity, and 440$\pm$43\,\msun\,yr$^{-1}$ from
our SED fitting.  The six objects in common include two AGN, but, interestingly,
the SFRs estimated for the two AGN are consistent, while the \halpha-derived
SFRs are systematically larger than our estimates for the non-AGN HAEs that
are not detected by SPIRE in the sub-mm.  This comparison suggests that our
CIGALE-based SFR estimates might be underestimated by 20\% on average.
The sources with the highest SFRs (i.e., SFR$>$100\,\msun\,yr$^{-1}$) are the
four detected by SPIRE, IDs 55326, 60984, SL03, and 58057. 

To establish whether the estimated SFRs are consistent with normal SFGs, or
starbursting galaxies, we compute the offset from the main sequence (MS)
relation at $z$\,=\,2.17~\citep{speagle14,elbaz11,rodighiero11}.  Such an
offset is the ratio between the estimated SFR and the expected SFR based on
the object stellar mass and assuming the MS relation at the source redshift. 
The SFR as a function of stellar mass for all 31 spectroscopic members is
shown in Fig.~\ref{fig:cigale_ms}.  The dotted lines represent the scatter
around the MS, which is about a factor of 3.  We find that 94\% of the
spectroscopic members are on the MS, with only two sources below, in the
region where quenching galaxies are expected.  None of the selected members
is in the starburst region.  No clear correlation is observed between the
MS-offset and a source location within the structure.  One AGN, ID SL01, is
below the MS and it is also the closest source to the quiescent wedge in the
UVJ and NUV$r$K diagrams.  This source might be going through a quenching
phase.  We note that since this source is very faint in the ACS/F814W-band
image (see~Sect.~\ref{sec:morph}), it is not possible to investigate whether
a morphological transformation is occurring, leaving a bulge-dominated
galaxy.

\subsection{AGN in the structure}\label{sec:agn}

Determining the density and power of AGN in high-redshift structures is of
great interest as feedback from AGN is often invoked as one of the main
mechanisms to halt star-formation in their host galaxies, and to ionize and
enrich the intergalactic medium.  To identify AGN activity in the G237
spectroscopic members, we consider the spectra and the X-ray data available
in the Cosmos field~\citep[the \chandra\ and XMM-Newton X-ray catalogs;
][]{cappelluti09,civano16}.  We could also consider the AGN fraction
estimated from the CIGALE SED-fitting, but the scarcity of MIR detections,
makes this parameter too unreliable to securely identify AGN.  Nonetheless,
CIGALE identifies AGN in only two sources, both already known AGN based on
other diagnostic criteria.
In ss1, three sources (IDs 55326, 58057, and SL01) exhibit broad emission lines
(i.e., \halpha, \lya, \civ) in their spectra, implying the presence of a
broad line (type 1 or unobscured) AGN.  Two of these three AGN are also X-ray
detected.  There is a third source that is X-ray detected, ID L710971,
but with no strong \civ\ emission, implying that it is an obscured (type 2)
AGN.  We thus find, in total, four AGN out of 20 members (20$\pm$10\%), of
which three (15$\pm$9\%) are X-ray sources.  The total (0.5--10\,keV) (not
corrected for absorption) X-ray luminosities of the three X-ray detected sources
are all $>$4$\times$10$^{43}$\,erg\,cm$^{-2}$\,s$^{-1}$, implying that they
are powered by an AGN~\citep{alexander05}.  The fraction of X-ray detected
AGN (15\%) is consistent with those measured in other two protoclusters at
$z{<}$2.25~\citep[17$^{+8.6}_{-6.0}$\% in Cl\,00218.3$-$0510 at $z$\,=\,1.6,
and 17$^{+16.2}_{-9.1}$\% in 2QZ\,1004$+$00 at $z$\,=\,2.23;
][]{krishnan17,lehmer13}, and higher than in protoclusters at higher
redshift~\citep[4$^{+8}_{-2}$\% in HS1700$+$643 at $z$\,=\,2.31, 2$\pm$1.3\%
in USS1558$-$003 at $z$\,=\,2.53, and 5.1$\pm$3.3\% in SSA22 at
$z$\,=\,3.09; ][]{digby10,macuga19,lehmer09}.
When considering the 11 sources at 2.164${<}z{<}$2.20, no X-ray detected
source nor broad line AGN is found.  Thus, including these additional
sources, the AGN fraction in the entire structure would be 13$\pm$6\%, still
consistent with an excess of AGN activity with respect to the field.
Two of the four SPIRE-detected members, IDs, 58057 and 55326, are X-ray luminous
sources (L$_{\rm X}{>}$7$\times$10$^{43}$\,\ergs), corresponding to
50$\pm$35\,\%.  This fraction is much higher than the fraction of X-ray
detected SMGs in the field~\citep[i.e., 11--29\% SMGs are X-ray detected in
the field with any X-ray luminosity; ][]{johnson13,georgantopoulos11,laird10}. 
The simultaneous sub-mm and X-ray emission in these objects, although based
on a few sources, is intriguing and will be discussed later.

\begin{figure} 
\centering
\includegraphics[width=\linewidth]{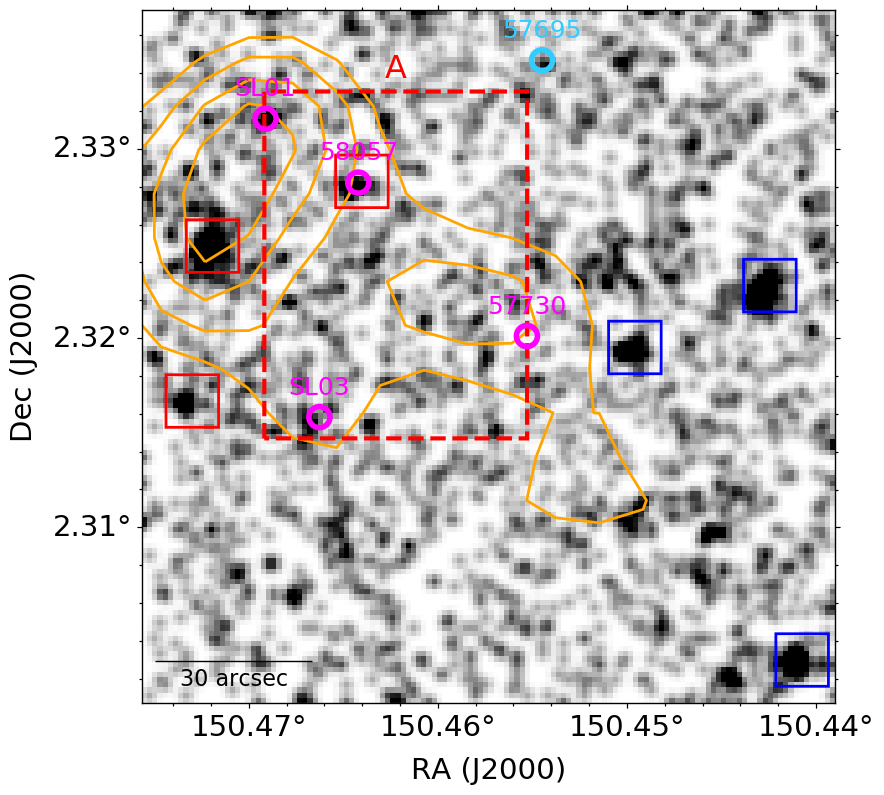}
\caption{{\small \chandra\ 0.5--7\,KeV rebinned 2.2\arcmin$\times$2.2\arcmin\
image of the protocluster core.  Spectroscopic members are
shown with 3\arcsec\ radius circles (magenta if 2.15${\leq}z{<}$2.164, light blue if
2.19${<}z{<}$2.20, and green if 2.16${<}z{<}$2.19), and X-ray detected sources
with 10\arcsec$\times$10\arcsec\ boxes (red if detected by both
\chandra\ and
XMM-Newton, and blue if detected only by \chandra). There are several X-ray
sources missing in the released \chandra\ Cosmos catalog, including IDs
SL01, and probably SL03, and 57695. SPIRE 350$\mu$m contours are shown in
orange. The red rectangle shows the boundaries of the protocluster core
(component A in Sect.~\ref{sec:components}).}}
\label{fig:cxo_img}
\end{figure}

In the 10\arcmin$\times$12\arcmin\ region centered on G237, there are 74 X-ray
sources, or 0.62$\pm$0.07\,arcmin$^{-2}$.  Out of these 74 X-ray sources, 45
have a spectroscopic redshifts and three (4\%) of them are structure members. 
For the remaining 29 X-ray sources, a photometric redshift is available for
24 sources and none of them falls in the structure redshift range
2.15--2.20.  We can thus say that there are at least three X-ray detected
members in the Cosmos released X-ray catalogs~\citep{cappelluti09,civano16},
and eight at the most, if we include the five sources for which no redshift
estimate is available.  

The three confirmed members have X-ray (0.5--10\,keV) luminosities greater than
4.4$\times$10$^{43}$\,erg\,cm$^{-2}$\,s$^{-1}$, and they all belong to ss1. 
Assuming a volume density of (0.2--1)$\times$10$^{-4}$\,Mpc$^{-3}$ for X-ray
sources with similarly high luminosities at $z{\sim}$2--3
~\citep{silverman08,fotopoulou16,ranalli16}, the expected number of X-ray
sources in the volume of ss1 (4376\,Mpc$^3$) is $\leq$0.44 sources.  We thus
find a number of luminous X-ray sources in the structure that is $\sim$7
times larger than expected, which corresponds to an X-ray source overdensity
of $\delta_{\rm X-ray}$\,=\,(3-0.44)/0.44\,=\,5.8.

In a 2.2\arcmin$\times$2.2\arcmin\ region centered on the protocluster core
there are six X-ray sources.  In Fig.~\ref{fig:cxo_img}, we show the
\chandra\ 0.5--7\,keV image of such a region and superimpose all the
detected X-ray sources from the Cosmos released \chandra\ and XMM-Newton
X-ray catalogs~\citep{cappelluti09,civano16}, and the structure
spectroscopic members.  Out of the six X-ray sources, only one (14\%), ID
58057, belongs to the structure, the others five are at $z{<}$2 (based on
two spectroscopic, and three photometric redshifts).  The \chandra\ image
shown in Fig.~\ref{fig:cxo_img} shows, however, the presence of other X-ray
sources that are not present in the public \chandra\ catalog, including some
spectroscopic members, ID SL01, and possibly IDs SL03, and 57695.  No other
spectroscopic sources with visible X-ray emission in the \chandra\ image,
and missing in the released X-ray catalog, have been found in the rest of
the field.  SL01, and SL03 belong to ss1, and 57695 to ss2.  ID SL01 was
already taken into account in the AGN fraction estimate as it is
spectroscopically classified as an AGN, if we also include ID SL03, the AGN
fraction in ss1 increases to 25$\pm$11\%.  There is thus strong evidence of
an excess of X-ray sources in the protocluster core.  Including the two
spectroscopically classified AGN also located in the core, we can state that
this region shows an enhanced fraction of AGN.  The mechanisms that favor
the presence of AGN might be also responsible for the enhanced level of SF
activity.

\subsection{Members morphology}\label{sec:morph}

A morphological analysis of the galaxy members within assembling structures
can reveal whether, when and where galaxies experience morphological
transformation and which processes are the main drivers (e.g.,  mergers,
close encounters, gas stripping, etc..).  A morphological classification in
early-type (ET), disk, and irregular galaxy is available from the Cosmos
morphological catalogs for 26
members\footnote{https://irsa.ipac.caltech.edu/data/Cosmos/overview.html}. 
This morphological classification is based on the HST/ACS F814W-band images
(corresponding to $\sim$\,2500\,\AA\ in the rest-frame), and on the
asymmetry, concentration, clumpiness parameters, and the Gini index as
described in~\citet{cassata07,tasca09}. For five sources a
morphological classification is not available.  Two of them exhibit an
elongated structure in the HST/ACS F814W-band image, and can thus be classified
as disks.  The remaining three sources are extremely faint and extended to
be morphologically classified through a visual inspection. The
morphological classification of all 31 members is listed in
Table~\ref{tab:specz_sample}.  Postage stamps (6\arcsec$\times$6\arcsec) in
the HST/ACS F814W-band and in the UltraVISTA/YJH bands of all member candidates
are shown in Figs.~\ref{fig:cutouts1},~\ref{fig:cutouts2}, and
~\ref{fig:cutouts3}.  

It is also interesting to investigate the presence of a nearby
galaxy that might be indicative of an interactive system, like a merger.  We
searched for potential companion galaxies using the L16 catalog and by
visually inspecting the postage stamps.  We consider a pair of galaxies to
be likely interacting if the projected distance is $\leq$15\,pkpc
(equivalent to $\leq$1.8\arcsec\ at the structure redshift).  We note that
larger physical separations are observed in interacting pairs, but would be more
consistent with a pre-merger phase~\citep[see e.g., ][]{ventou17}.  We find
in total six sources with a potential companion, three from the L16 catalog (IDs
58206, 59906, and 54698), and three additional ones by the images visual
inspection (IDs 58020, 58266, and 60811; see
Fig.~\ref{fig:cutouts1}--\ref{fig:cutouts3}, and
Table~\ref{tab:specz_sample}).  The photometric redshifts, available only
for three companions, are too uncertain to confirm or rule out any
interaction within each pair.

All members are either disks or irregular galaxies, as expected for a
SFG. There is only one exception, ID 55326, classified as
ET, but since it hosts an AGN, it is possible that the light
profile is dominated by the AGN point-like emission.  We do not see any
difference in the morphological classification in the different
substructures (e.g., in the core, or in the outer regions).  About 40\% of
members are classified as irregular galaxies, and 19\% have a close
companion, suggesting that several members might be or have been
experiencing an interaction. Three out of the four sources in the elongated
structure (B component in Sect.~\ref{sec:components}) have a companion.
Although the pair fraction in this component is much higher than in the rest
of the structure, it would be difficult to draw any conclusion being based
on a handful of sources and lacking spectroscopic confirmation for all
pairs.

\begin{table*}
\centering
\caption{\phz712\ spectroscopic members}\label{specz_sample}
\setlength{\tabcolsep}{5.pt}
\begin{tabular}{l cccc rlcll}
\hline\hline
   ID     &$\alpha_{J2000}$ &  $\delta_{J2000}$ &  NUMBER & $z_{\mathrm spec}$ & \textit{zflg}\tablefootmark{a}  & Type\tablefootmark{b} & mag(Ks)         & $z_{\mathrm phot}$ & Comp./Note\tablefootmark{c} \\
          &(deg)            &  (deg)            &         &               &                &          &  (AB)           &              &  \\
\hline
\multicolumn{10}{c}{Substructure 1: $z$\,=\,2.15--2.164}\\
\hline
  55326   &    150.39265    &      2.270105     &  639096 &    2.1576     & 14  &   ET    &   22.189$\pm$0.022 &   1.8205  & CXO-ID: 769 \\   
  55687   &    150.40045    &      2.370529     &  707504 &    2.1529     &  1  &   Disk  &   20.455$\pm$0.006 &   1.4264  & \\ 
  56014   &    150.40732    &      2.362427     &  701268 &    2.1580     &  1  &   Irr   &   23.221$\pm$0.049 &   2.3446  & \\   
  56915   &    150.43237    &      2.248860     &  624965 &    2.1576     &  2  &   Disk  &   22.483$\pm$0.020 &   2.1418  & \\   
  57557   &    150.45144    &      2.370448     &  706805 &    2.1520     &  1  &   Disk  &   22.779$\pm$0.023 &   1.6211  & \\   
  58020   &    150.46384    &      2.257829     &  631032 &    2.1519     &  3  &   Irr\tablefootmark{d}   &   21.902$\pm$0.011 &   2.1821  & B/HAE \\   
  58057   &    150.46451    &      2.327942     &  677305 &    2.1517     & 13  &   Disk  &   21.612$\pm$0.008 &   2.0626  & A/CXO-ID: 1265, HAE \\   
  58173   &    150.46791    &      2.281895     &  646606 &    2.1520     &  2  &   Irr   &   22.984$\pm$0.029 &   1.654   & B/non HAE \\   
  58200   &    150.46892    &      2.394305     &  721845 &    2.1520     &  2  &   \nodata   &   23.410$\pm$0.042 &   1.7865  & \\   
  58206   &    150.46911    &      2.292743     &  653383 &    2.1527     &  3  &   Disk\tablefootmark{d,e}  &   22.675$\pm$0.015 &   1.99    & B/HAE \\   
  58266   &    150.47072    &      2.247844     &  625320 &    2.1502     &  2  &   Irr\tablefootmark{d}   &   22.489$\pm$0.037 &   2.1169  & B/HAE \\   
  59504   &    150.50562    &      2.393203     &  721608 &    2.1500     &  2  &   Disk  &   22.715$\pm$0.027 &   2.2462  & \\   
  60718   &    150.53749    &      2.365551     &  703833 &    2.1510     &  2  &   Disk\tablefootmark{e}  &   22.638$\pm$0.030 &   1.5001  & \\ 
  60811   &    150.54004    &      2.322157     &  673721 &    2.1547     &  1  &   Disk\tablefootmark{d}  &   21.729$\pm$0.012 &   1.4376  & \\   
  L710971 &    150.51958    &      2.377024     &  710971 &    2.1600     &  1  &   \nodata   &   22.645$\pm$0.025 &   2.6792  & CXO-ID: 1746 \\
  SL01    &    150.46945    &      2.331292     &  679779 &    2.1586     & 14  &   \nodata   &   21.379$\pm$0.008 &   2.1975  & A/HAE \\   
  SL03    &    150.46652    &      2.315597     &  669706 &    2.1592     &  3  &   Disk  &   20.584$\pm$0.005 &   2.2488  & A/HAE \\   
  57138   &    150.43973    &     2.239109      &  618296 &    2.1615     &  4  &   Disk  &   22.198$\pm$0.013 &    2.1485  & \\  
  57730   &    150.45557    &     2.319829      &  672385 &    2.1637     &  1  &   Disk  &   22.566$\pm$0.020 &    1.4109  &   A/non HAE \\
  54698   &    150.38090    &     2.369391      &  706004 &    2.1637     &  2  &   Irr\tablefootmark{d}   &   22.667$\pm$0.053 &    0.8765  & \\  
\hline  
\multicolumn{10}{c}{2.164${<}z{<}$2.19} \\                
\hline                    
  54765   &    150.38211    &     2.323130      &  673531 &    2.1757     &  1  &   Disk  &   23.731$\pm$0.123 &    1.9198  & \\  
  55951   &    150.40584    &     2.366063      &  703555 &    2.1815     &  1  &   Disk  &   22.441$\pm$0.020 &    2.1324  & \\  
  56005   &    150.40720    &     2.255304      &  629074 &    2.1700     &  2  &   Disk  &   22.899$\pm$0.029 &    2.1825  & \\  
\hline  
\multicolumn{10}{c}{Substructure 2: $z$\,=\,2.19--2.20} \\       
\hline                
  54677   &    150.38052    &     2.274505      &  641998 &    2.1986     &  4  &   Irr   &   21.728$\pm$0.022 &    2.2415  & \\  
  56950   &    150.43380    &     2.222876      &  607540 &    2.1991     &  1  &   Irr   &   22.641$\pm$0.021 &    1.4602  & \\  
  57695   &    150.45481    &     2.334413      &  682054 &    2.1915     &  1  &   Irr   &   22.506$\pm$0.020 &    2.1538  &  non HAE \\  
  59496   &    150.50540    &     2.237819      &  617293 &    2.1909     &  4  &   Disk  &   22.770$\pm$0.021 &    2.0405  &  C/non HAE \\  
  59906   &    150.51611    &     2.398838      &  725166 &    2.1907     &  2  &   Irr\tablefootmark{d}   &   22.555$\pm$0.024 &    2.1756  & \\  
  60026   &    150.51934    &     2.252120      &  626986 &    2.1916     &  2  &   Disk  &   23.665$\pm$0.046 &    2.216   & C/non HAE \\  
  60355   &    150.52715    &     2.270970      &  639939 &    2.1993     &  4  &   Irr   &   22.503$\pm$0.017 &    2.2373  & C \\  
  60984   &    150.54539    &     2.257720      &  631549 &    2.1982     &  1  &   Irr   &   20.738$\pm$0.004 &    1.7994  & C \\  
\hline
\hline
\end{tabular}\\
\tablefoot{
\tablefoottext{a}{\small \textit{zflg} is a flag that indicates the reliability of the redshift
estimate.  Based on the zCosmos-Deep sample~\citep{lilly07}, the estimated
redshifts are expected to be correct in 50\% of the cases for \textit{zflg}\,=\,1,
in 85\% for \textit{zflg}\,=\,2, and in 100\% for \textit{zflg}\,=\,3, or
4. In case of AGN spectral features in the spectrum, \textit{zflg} is incremented by 10.}
\tablefoottext{b}{\small Morphological type based on the ACS I-band image
(Disk, ET: early-type, Irr: irregular) from the Cosmos morphological
catalogs.  A morphological type is not available when a source is too
faint.}
\tablefoottext{c}{\small The letter A, B, or C indicates the component described in
Sect.~\ref{sec:components}. \chandra\ IDs are listed in case of X-ray
detection.  Sources within the Subaru/MOIRCS FOV are listed as HAE if
selected as such, or as non-HAE otherwise~\citep[see Sect.~\ref{sec:HAE}, and ][]{koyama21}.}
\tablefoottext{d}{\small Presence of a nearby galaxy at $<$1.8\arcsec,$\sim$15\,kpc in the ACS I-band image.}
\tablefoottext{e}{\small Morphological classification derived from visual inspection of the ACS I-band images
because not available in the Cosmos morphological catalogs.}
}
\label{tab:specz_sample}
\end{table*}

\section{Discussion}\label{sec:discussion}

\subsection{Is G237 a protocluster?}\label{sec:halo_mass}

A high-$z$ overdensity is considered a protocluster if it will collapse and
relax by $z$\,=\,0, and its predicted halo mass at $z$\,=\,0, M$_{\rm
h}$($z$=0), is $>$10$^{14}$\,\msun~\citep{chiang13}.  In addition to being
used to confirm the protocluster nature of a high-$z$ overdensity, M$_{\rm
h}$($z$=0) is also useful to establish a connection between a protocluster
and the analogous of its descendant, to explore the diversity across
protoclusters, and the effect of the halo mass on the protocluster assembly
and on the evolution of its members.
There are various methods in the literature to estimate the $z$\,=\,0 halo
mass~\citep[see e.g., ][]{steidel98,cucciati14,long20}.  Here, we adopt the
widely used analytical method first applied by~\citet{steidel98}.  This
method provides an estimate of the total mass of the descendant cluster at
$z$\,=\,0 from the (comoving) volume $V$ occupied by the
overdensity, the mean density of the Universe $\bar{\rho}$ at the
protocluster redshift, and the mass overdensity of the protocluster
$\delta_m$, through the following equation:
\begin{equation}
{\rm M_h}\,=\,\bar{\rho}\,{\rm V}\,(1 + \delta_m), 
\label{eq:mhalo}
\end{equation}
\noindent  where $V$ is the true volume, $V{\equiv}V_{\rm
apparent}/C$, $\delta_m$ is the mass overdensity that is related to the
galaxy overdensity $\delta_{\mathrm{gal}}$ through
1${+}b\delta_m$\,=\,$C$(1$+\delta_{\mathrm{gal}}$), where $b$ is the linear
galaxy bias parameter, and
$\bar{\rho}$\,=\,3.97$\times$10$^{10}$\,\msun\,Mpc$^{-3}$ for the
cosmological parameters used in this work.  The galaxy overdensities,
$\delta_{\mathrm{gal}}$, derived as described in Sect.~\ref{sec:cosmos} for
the spectroscopic members in the full structure, and in the two
substructures, and the corresponding comoving volumes are reported in
Table~\ref{tab:overdensity_prop}. The parameter $C$ takes into account the
redshift distortions caused by peculiar velocities.  Assuming that the
structure is just breaking away from the Hubble flow, and
following~\citet{steidel98}, $C$ can be approximated by
$C$\,=\,1$+f-f$(1$+\delta_m$)$^{1/3}$, where $f$ is the rate of growth of
perturbations at the redshift of the protocluster~\citep{lahav91}.  In the
cosmology adopted in this paper, $f$\,=\,0.96 at the mean redshift of the
overdensity, $z$\,=\,2.168~\citep{lahav91}.  We note that this method assumes
that galaxies selected in a certain way predominantly occupy halos of the
same mass, and this is expressed by the bias parameter.  Here, we chose the
bias parameter measured for galaxies at $z{\simeq}$2 with stellar masses
$\mathcal{M}{>}$10$^9$\,\msun~\citep[i.e.,  $b$\,=\,1.74; ][]{chiang13}. 
The $C$, and $\delta_{\rm m}$ values derived assuming these
parameters for the different substructures, and the derived halo masses,
M$_{\rm h}$($z$=0), are also listed in Table~\ref{tab:overdensity_prop}. 
The uncertainty on M$_{\rm h}$($z$=0) is derived from the uncertainty on
$\delta_{\mathrm{gal}}$.  This is derived assuming the galaxy field
density $\pm$ 1$\sigma$, and by taking into account the redshift flag
assigned to the spectroscopic redshifts in counting the galaxy
members\footnote{The number of spectroscopic members is weighted by the
redshift flag as it follows, $n$ galaxies with \textit{zflg}\,=\,1 count as
0.5${\times}n$, $n$ galaxies with \textit{zflg}\,=\,2 as 0.85${\times}n$, and those
with \textit{zflg}$\geq$3 remain unchanged.}.

The same estimate can be derived considering the overdensity of sub-mm
galaxies ($\delta_{\mathrm{SMG}}$\,=\,1.3) in ss1 and their bias
parameter~\citep[i.e., $b$\,=\,1.3, ][]{miller15}.  In this case, we obtain
$C$\,=\,0.81, and $\delta_m$\,=\,0.67, and M$_{\rm
h}$($z$=0)\,=\,3.6$^{+7.2}_{-1.3}{\times}$10$^{14}$\msun.  This is
consistent with the value derived from the spectroscopic members
overdensity.  

To establish whether these structure will collapse and virialize by
$z$\,=\,0, we investigate the evolution of the structures overdensity in the
linear regime of a spherical collapse model and compare the estimated linear
matter enhancement at $z$\,=\,0, $\delta_{\mathrm L}$($z$=0) with the
critical collapse threshold of $\delta_{\mathrm
c}$\,=\,1.686~\citep{percival05}.  The $\delta_{\mathrm L}$($z$) values are
derived from the $\delta_{\mathrm m}$ estimates using equation (18)
in~\citet{mowhite96}.  The $\delta_{\mathrm L}$($z$=0) values are obtained
by multiplying $\delta_{\mathrm L}$($z$) by the linear growth
factor\footnote{The linear growth factor for the chosen cosmology was
obtained from the ICRAR's Cosmology Calculator
(https://cosmocalc.icrar.org/).} scaled by the cosmic scale
factor~\citep[see eq.  (7) in ][]{cucciati18}.  The derived $\delta_{\mathrm
L}$($z$=0) values are reported in Table~\ref{tab:overdensity_prop}.  The
derived $\delta_{\mathrm L}$($z$=0) exceeds the collapse threshold of
$\delta_{\mathrm c}$\,=\,1.686 for the two substructures, and it does not
for the full structure.  Therefore, both substructures are likely to
virialize independently and form a larger, but likely not virialized
structure.  The $z$\,=\,0 mass of the two substructures are comparable to
the mass of a Virgo-type cluster~\citep[i.e.,
$\sim$(3--10)$\times$10$^{14}$\,\msun; ][]{fouque01}.

\begin{table*}
\centering
\caption{Overdensity properties\label{tab:overdensity_prop}} 
\setlength{\tabcolsep}{3.0pt}
{\renewcommand{\arraystretch}{1.4}
\begin{tabular}{l c r r c c c r ccccr} 
\hline\hline
Structure & $z$ range   & \multicolumn{1}{c}{N}    & \multicolumn{1}{c}{$<$N$>$} & $\delta_{\mathrm gal}$ & $\sigma_{\delta_{\mathrm gal}}$ & Extent & \multicolumn{1}{c}{V$_{\mathrm app}$}    & $C$        & $\delta_{\mathrm m}$ & $\delta_{\mathrm L}$($z_{\rm obs}$) & $\delta_{\mathrm L}$($z$=0) &M$_{\mathrm h}$($z$=0) \\
       &      & \multicolumn{1}{c}{G237} & \multicolumn{1}{c}{Field}   &                    &                            &     & \multicolumn{1}{c}{(cMpc$^3$)} &  &  &    &    &                            (10$^{14}$\,\msun) \\
\hline
 full   & 2.15--2.20  &  29  &    10.5$\pm$3.4   &  1.8  &  5.4  &  9.9\arcmin$\times$10.6\arcmin\ &  18532  & 0.80 & 0.72 & 0.46 & 1.15 & 15.7$^{+6.0}_{-3.2}$    \\
 ss1 & 2.15--2.164 &  18  &     3.3$\pm$2.7   &  4.5  &  5.4  &  9.5\arcmin$\times$9.3\arcmin\  &   4376  & 0.66 & 1.48 & 0.70 & 1.75 &  6.6$^{+26.0}_{-3.5}$   \\
 ss2 & 2.19--2.20  &   8  &     1.6$\pm$1.3   &  3.9  &  5.0  &  9.9\arcmin$\times$10.6\arcmin\ &   3724  & 0.68 & 1.35 & 0.67 & 1.69 &  5.1$^{+19.7}_{-1.9}$   \\
\hline
\hline
\end{tabular}\\
}
\tablefoot{\small N is the number of sources with spectroscopic redshift in the Cosmos field from the catalogs listed in Section~\ref{sec:cosmos}.
We note that the sources from our LBT/LUCI program are not taken into account to estimate the galaxy overdensity. $<$N$>$ is the average number of sources with spectroscopic redshift in the Cosmos field after excluding the G237 region. $\delta_{\mathrm gal}$, and $\sigma_{\delta_{\mathrm gal}}$ are, respectively, the density contrast and significance of the galaxies in G237 in the given redshift ranges. Extent is the area projected on the sky where the galaxies in the defined redshift ranges are distributed, and V$_{\mathrm app}$ the volume obtained by multiplying the extent by the distance along the line of sight given by the redshift range. $C$ is the parameter that takes into account the redshift distortions~\citep[see text and ][]{steidel98}. $\delta_{\mathrm m}$ is the matter overdensity obtained as described in Section~\ref{sec:halo_mass}, and used in eq.~\ref{eq:mhalo} to derive the $z{=}$0 halo mass 
M$_{\mathrm h}{(}z{=}$0) reported in the last column. $\delta_{\mathrm L}{(}z_{\rm obs}{)}$, and $\delta_{\mathrm L}{(}z{=0)}$ are the linear matter enhancements at the observed redshift and at $z$\,=\,0, respectively. An overdensity is expected to collapse when $\delta_{\mathrm L}$ reaches the collapse threshold $\delta_{\mathrm c}$\,=\,1.686~\citep{percival05}. Thus only the two substructures, and not the full structure, are expected to collapse by $z{=}$0.}
\end{table*}

\subsection{G237 contribution to the cosmic SFR density}\label{sec:sfrd}

The total SFRs of the structures in G237, obtained by adding the SFR of all
spectroscopic members, are 1485$\pm$71\,\msun\,yr$^{-1}$ if we consider the
full structure, 1002$\pm$58\,\msun\,yr$^{-1}$ for ss1, and
429$\pm$41\,\msun\,yr$^{-1}$ for ss2.  If we consider the HAEs presented
in~\citet{koyama21} and their SFRs derived from the dust-corrected \halpha\
luminosity, we have an additional SFR of 715\,\msun\,yr$^{-1}$.  This
estimate does not include the SFR from the six sources in common between the
two samples.  Including the SFR from the HAEs, the total SFRs become
2200$\pm$71\,\msun\,yr$^{-1}$, and 1717$\pm$58\,\msun\,yr$^{-1}$,
for the full structure, and for ss1, respectively.  The estimated SFRs
include the four spectroscopically confirmed members detected in the sub-mm
by SPIRE.  These are relatively faint sub-mm sources, with S$_{\rm 350\mu
m}{<}$18\,mJy, and SFR$\sim$110--220\,\msun\,yr$^{-1}$.  Since not all SFG
members in the structure have been identified, even after considering the
HAEs, the estimated total SFRs are only a lower limit.  We note that the HAE
selection is limited to $z{<}$2.17, and to a 7\arcmin$\times$4\arcmin\
region.  In addition, some heavily extincted SFGs might not be detectable as
HAEs, and a large fraction of the field, where the density of SPIRE source
peaks, is affected by a bright star at optical and NIR wavelengths.  We
estimate how much SFR we might be missing because of dust extinction and of
the masked area, by considering the 11 SPIRE member candidates, minus the 4
already included in the spectroscopic sample (see
Sect.~\ref{sec:spire_srcs}).  The total SFR, including these additional
seven (11 candidates, minus four confirmed), increases to
$\sim$4000\,\msun\,yr$^{-1}$.  We also derive the SFR in the protocluster
core by adding the SFRs of the four spectroscopic members, and two additional
HAEs located in the A component.  All the derived SFRs, and inferred SFR
densities (SFRDs) are listed in Table~\ref{tab:sfr_components}.
\begin{table}
\begin{center}
\caption{G237 structures SFR}\label{tab:sfr_components}
{\renewcommand{\arraystretch}{1.4}
\setlength{\tabcolsep}{2.pt}
\begin{tabular}{p{4cm}ccc}
\hline\hline
 Members    &  N   &   SFR  & SFRD \\
            &      & (\msun\,yr$^{-1}$) & (\msun\,yr$^{-1}$ cMpc$^{-3}$) \\
\hline
Full: 2.15${\leq}z_{\rm spec}{<}$2.20       &  31  &  1485$\pm$71 & 0.080$\pm$0.004 \\
ss1: 2.15${\leq}z_{\rm spec}{\leq}$2.164    &  20  &  1002$\pm$58 & 0.229$\pm$0.013 \\
ss2: 2.19${\leq}z_{\rm spec}{\leq}$2.20     &   8  &  429$\pm$41  & 0.115$\pm$0.011 \\
HAE\tablefootmark{a} (w/o $z_{\rm spec}$ members)&  32  &   715   & 0.163           \\
Full$+$HAE                                  &  63  &  2200$\pm$71 & 0.119$\pm$0.004 \\
ss1$+$HAE                                   &  52  &  1717$\pm$58 & 0.392$\pm$0.013 \\
SPIRE: 1.8${<}z_{\rm phot}{<}$2.5           &  11  &  2408$^{+134}_{-106}$ & 0.130$^{+0.007}_{-0.006}$ \\
Full$+$HAE$+$SPIRE~\tablefootmark{b}        &  70  & 3995$^{+152}_{-128}$ & 0.216$^{+0.008}_{-0.007}$\\
Core\tablefootmark{c}                       &   6  & 436$\pm$41   & 9.6$\pm$0.9 \\
\hline
\hline
\end{tabular}\\
}
\tablefoot{
\tablefoottext{a}{The HAE sample contains 38 sources, of which six are also
present in ss1 and in the full spectroscopic samples, their SFRs are thus
not considered in this subsample.  We note that the HAE SFRs are derived
from the extinction-corrected \Ha\ luminosity~\citep{koyama21}.  }
\tablefoottext{b}{Four SPIRE sources are already included in the full
spectroscopic sample and are thus not considered in the SPIRE subsample for
the total SFR estimate.}
\tablefoottext{c}{We consider the core the A component with four
spectroscopic members, of which three are HAEs, and two additional HAEs. 
The volume of the core is 45\,cMpc$^3$, corresponding to
0.8\arcmin$\times$1.1\arcmin\,=\,1.45$\times$1.89\,cMpc$^2$, and a
transverse distance of 16.4\,cMpc, corresponding to $\Delta
z$\,=\,2.1637$-$2.1517.}
}
\end{center}
\end{table}

Compared to the average cosmic SFR density (SFRD) at
z$\sim$2.16~\citep[$<$SFRD($z$\,=\,2.168)$>$\,=\,0.128\,\msun\,yr$^{-1}$\,Mpc$^{-3}$;
][]{madau14}, the SFRD of ss1 and of the core are, respectively,
3.1$\pm$0.1 and 75$\pm$7 times higher, while the full structure SFRD is
consistent or $\sim$1.6 times higher if the SPIRE member candidates are
included.  Since the full structure includes a much bigger volume than the
two substructures, it is not surprising that its SFRD is closer 
to the average cosmic value.

It is interesting to note that the enhanced SFRD in the core with respect to
the rest of the structure is mainly due to the larger galaxy density
(0.13\,spectroscopic members and HAEs\,cMpc$^{-3}$ in the core $versus$
0.01\,cMpc$^{-3}$ in ss1), and, only in part due to a larger average SFR per
galaxy in the core than in the full structure,  
$<$SFR$>$\,=\,73$\pm$52\,\msun\,yr$^{-1}$ in the core $versus$
35$\pm$27\,\msun\,yr$^{-1}$ in the full sample.  The difference in average
SFR is, however, less important than the number density, and not highly
significant as the SFRs span a wide range, and suffer from large
uncertainties.

Since ss1, and its core are the most significant structures in the G237
field, we compare their SFRD estimates with those of other protoclusters
from the literature at similar
redshifts~\citep{clements14,dannerbauer14,kato16}, and with theoretical
models~\citep{chiang17}.  \citet{chiang17} predict that the contribution
from protoclusters, and protocluster cores to the cosmic SFRD at
$z{\sim}$2.2 is 12--17\%, and 1.2--2.9\%, respectively.  These estimates are
obtained using two semi-analytical simulations~\citep[SAM;
][]{henriques15,guo13}, and are based on all galaxies with
$\mathcal{M}{>}$10$^{8.5}$\,\msun\,yr$^{-1}$.  Based on these simulations,
the predicted SFRDs for an average protocluster, and its core at
$z{\simeq}$2, are $\sim$0.6\,\msun\,yr$^{-1}$\,Mpc$^{-3}$, and
61--91\,\msun\,yr$^{-1}$\,Mpc$^{-3}$, respectively~\citep{chiang17}.  These
estimates assume a spherical volume with average radius of 8\,cMpc, and
0.8\,cMpc, for the protocluster and its core, corresponding to
$\sim$8\arcmin, and 0.8\arcmin, respectively at our redshift.  Our estimate
of SFRD in the core is almost one order of magnitude smaller than the
average value predicted by the models, but the volume we consider is $>$20
times bigger.  The SFRD relative to ss1 (i.e.,
0.39$\pm$0.01\,\msun\,yr$^{-1}$\,Mpc$^{-3}$) is also lower than predicted,
but only by a factor of 1.6, and the considered volume is twice than in the
model.  The smaller SFRDs with respect to the model predictions can be in
part explained by having considered bigger volumes, and possibly by not
having included all star-forming members.  Overall, we can conclude that the
SFRD estimates derived for the ss1, and for the core can be reconciled with
the predictions from theoretical models~\citep{chiang17}.

Compared to the SFRD estimates of other protoclusters at similar redshift
from the literature~\citep[i.e., 500-5000\,\msun\,yr$^{-1}$\,Mpc$^{-3}$;
][]{clements14,dannerbauer14,kato16}, the SFRDs in ss1 or in the core are
much lower.  There are several differences between these works and ours that
should be taken into account before making any comparison.  In these works,
the considered volume is often assumed to be a sphere of diameter equivalent
to the members projected distance, and physical dimensions are considered
rather than comoving, which yields smaller volumes, and thus larger SFRDs by
a factor of $>$10--15 at $z{\sim}$2--2.5~\citep[see e.g., ][]{dannerbauer17}. 
In some cases, the integrated SFR is obtained considering all sub-mm
detected galaxies, even when membership is not spectroscopically
confirmed~\citep[see e.g., ][]{clements14,kato16}.  Finally, their SFR
estimates are systematically higher than ours by a factor of 1.7--1.8 because
they assume a ~\citet{salpeter55} IMF and the L(IR)--SFR relations reported
in~\citet{bell03}.  Because of all these differences we will not carry out a
comparison with the SFRDs reported in the literature for other
protoclusters and warn the reader against using those estimates to identify
extreme sources and rule out models.

\subsection{Quenching signs and the role of the
AGN}\label{sec:agn_quenching}

Protocluster first descendants, clusters at $z{\sim}$1.5--2, are already
dominated by a population of massive quiescent galaxies in excess compared
to the field~\citep{andreon14,andreon16,strazzullo19}.  We can thus expect
these galaxies to start quenching during the protocluster phase.  Our
protocluster, G237 at $z$\,=\,2.16, can give us a first glance at this
elusive quenching population, and at the processes responsible for halting
their star formation. Theoretical studies often invoke AGN feedback as one of the 
main quenching mechanism, especially in the most
massive galaxies~\citep[e.g.,][]{rosito20}.  AGN feedback can occur
mechanically, via a radio jet (radio mode), or radiatively (quasar mode),
through an AGN-driven wind.  The radio mode corresponds to AGN accreting
well below the Eddington rate (Eddington ratio $\lambda_{Edd}{<}$0.01) and
with a radio jet.  The quasar mode instead corresponds to high luminosity
AGN that accrete at near their Eddington ratio and are capable of driving
strong winds~\citep{harrison17}.

Out of the four members hosting AGN activity, three have relatively modest stellar
masses, $\sim$(2--5)$\times$10$^{10}$\,\msun, and large SFRs,
$\sim$50--140\,\msun\,yr$^{-1}$.  The fourth AGN, ID SL01, is instead just
below the MS and, with $\mathcal{M}{\simeq}$1.3$\times$10$^{11}$\,\msun\
and SFR$\sim$34\,\msun\,yr$^{-1}$, turns out to be more massive and less active than the
other AGN members.  This source is in the structure core and
might be at the end of its growing phase and starting to quench.  The
rest-frame optical spectrum of this source shows a slightly broad \halpha\
line, suggesting that the AGN is unobscured.  Its X-ray emission is weak,
visible in the \chandra\ image, but not detected with a high enough
signal-to-noise to be included in the Cosmos \chandra\ released catalog. 
These properties are consistent with the AGN evolutionary scenario where,
following AGN feedback, the line of sight is cleared out from the obscuring
dust, and, both the SF activity and AGN luminosity, decrease due to lack of
fuel~\citep{sanders88}.  ID SL01 is not detected in the deep radio images
available in the field, we can thus rule out the presence of a powerful
radio jet.  To investigate instead whether this source is radiatively
powerful enough to launch a wind that can halt the current star formation,
we estimate its Eddington ratio. This value is derived from the ratio of
the SED-derived bolometric luminosity, and the Eddington luminosity that
depends on the black hole (BH) mass.  Based on the optical luminosity at
5100\,\AA\ in the rest-frame (i.e.,
Log($\lambda$L$_{\lambda}$(5100\AA)/(erg\,s$^{-1}$))\,=\,44.6) and on the width of
the \halpha\ line (i.e., FWHM(\halpha$^{broad}$)\,=\,4004\,\kms),
we estimate a BH mass in SL01 of
$\simeq$2.0$\times$10$^8$\,\msun~\footnote{M$_{\rm
BH}$/\msun\,=\,4.817$\times$[$\lambda$L$_{\lambda}$(5100\AA)/10$^{44}$\ergs]$^{0.7}{\times}$FHWM(\halpha)$^2$, where FHWM(\halpha) is in \kms~\citep{woo02}.}.

The estimated bolometric luminosity, derived from the CIGALE
best fit model, and assuming that is solely due to the AGN, is
2.9$\times$10$^{45}$\,\ergs.  Since the AGN does not dominate the bolometric
emission in SL01, as its SED clearly shows (see Fig.~\ref{fig:cig_seds}),
the SED-derived bolometric luminosity should be considered as an
upper limit to the AGN bolometric luminosity.  The derived upper
limit to its Eddington ratio, $\lambda_{\rm Edd}$\,=\,L$_{\rm bol}$/L$_{\rm
Edd}{<}$\,0.11, is high enough to cause some feedback onto the host galaxy
through the AGN radiation.

We can investigate whether the other AGN members in the structure
could cause AGN feedback.  None of them is detected in the radio,
thus we can rule out the radio mode feedback.  To estimate their Eddington
ratio, we use the bolometric correction as this quantity is correlated with
the Eddington ratio~\citep{lusso12}.  The bolometric correction $k_{\rm
bol}$, can be estimated from the ratio between the AGN bolometric luminosity
and the X-ray luminosity, using $k_{\rm bol}$\,=\,L$_{\rm bol}$/L$_{\rm X}$. 
We derive L$_{\rm bol}$ from the best fit SED, it is thus an upper limit to
the AGN bolometric luminosity.  L$_{\rm X}$ is derived from the 0.5--10\,kev
measured flux, and it is a lower limit since it is not corrected for
absorption.  The estimated upper limits to the $k_{\rm bol}$ values are thus
$<$57, 95, and 26, for IDs 55326, 58057, and L710971, respectively. 
Assuming the relation between $\lambda_{\rm Edd}$, and $k_{\rm bol}$ for
unobscured AGN reported by~\citet{lusso12}, the ensuing $\lambda_{\rm Edd}$
would be $\lesssim$0.3, 0.6, and 0.1, for IDs 55326, 58057, and L710971,
respectively.  The estimated Eddington ratios imply that these AGN can
produce a significant radiative feedback.

The high AGN fraction in G237 suggests a connection between such an activity
and the protocluster environment.  Overall, all AGN members are
powerful enough to exert some radiative feedback, and this might be
quenching the star formation activity in ID SL01, a massive galaxy hosting
an AGN with a lower SFR than expected for a MS galaxy with the
same mass (see Fig.~\ref{fig:cigale_ms}).

\subsection{Structure properties that make a \planck\ high-$z$ source}\label{sec:pc_planck}

The overall SFR of G237 is not extreme, its members are ordinary
SFGs, and the protocluster overall properties are similar to those observed
in other protoclusters.  It is thus legitimate to ask why G237 is
significantly detected by \planck, while most protoclusters from the
literature are not.  To investigate whether there is a specific property
that explains its \planck\ significant detection, we compare G237 with
other protoclusters situated in the Cosmos field at $z{\sim}$2--2.5.  There
are other protoclusters found at much higher redshifts~\citep[e.g.,
][]{capak11,pavesi18,lemaux18,mitsuhashi21}, but we do not analyze them as
the \planck\ high-$z$ source selection is not tailored to find protoclusters
at $z{\gtrsim}$4.

To our knowledge there are no other known structures in the Cosmos field at
exactly the same redshift as G237.  G237 is located to the east of the
central 1\,\sqdeg\ of the Cosmos field, where many deep and extensive sub-mm
and spectroscopic surveys have been carried out.  A bright star, situated in
the field and covering a circular region of 2.4--3.6\arcmin\ in diameter
(depending on the band), prevents the detection of sources in the optical
and NIR that are thus missing in the Cosmos L16 multiband catalog.  These
considerations might in part explain why this structure was not previously
discovered through spectroscopic or high precision photometric redshifts. 
It is also possible that the nature of G237 and of the PHz sources, in
general, is different from those discovered through other techniques.

\begin{figure*} 
\centering
\includegraphics[width=0.75\linewidth,angle=0]{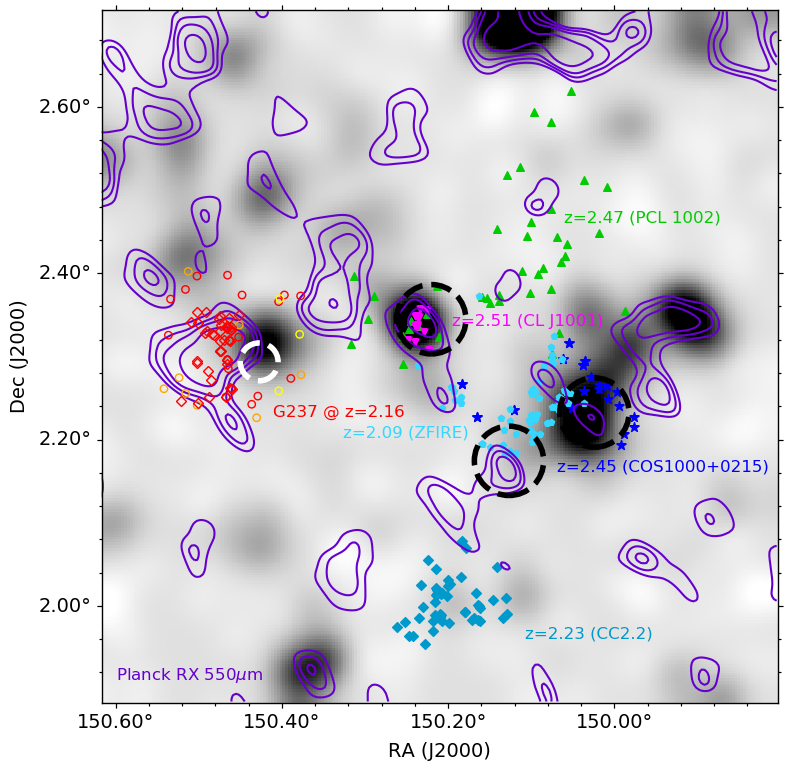}
\caption{{\small Layout of the structures found in a
50\arcmin$\times$50\arcmin\ region in the Cosmos field overlaid on a density
map of red SPIRE sources. \planck\ {\tt red-excess} contours
(at 30, 40, 50, 68, 95 and 99.7\% of the highest value) are shown as purple lines. 
Different symbols are members of the following structures:
COS1000$+$0215 protocluster at $z$\,=\,2.45~\citep[blue stars;][]{diener15,chiang15}, 
CL\,J1001 at $z$\,=\,2.51~\citep[magenta up-side down triangles; ][]{wang16,wang18},
PCL\,1002 at $z$\,=\,2.47~\citep[green triangles;][]{casey15},
CC2.2 at $z$\,=\,2.23~\citep[turquoise blue diamonds;][]{darvish20}, and ZFIRE at $z$\,=\,2.09~\citep[light blue circles; ][]{yuan14,hung16}.
The spectroscopic members in G237 are shown as open circles (red:
2.15${\leq}z{<}$2.164,
yellow: 2.164${<}z{<}$2.19, and orange: 2.19${<}z{<}$2.20), and the
HAEs~\citep{koyama21} as red
diamonds.  The dashed white circle represents the 1.4\arcmin\ radius circle
masked in the L16 Cosmos catalog due to a bright star. Large black dashed
circles (2.5\arcmin\ radius) indicate locations, close to the selected
protoclusters, where a signal $\gtrsim$2$\sigma$ is detected in the \planck\
RX map, but not significant enough to be selected as a PHz source. }}
\label{fig:cosmos_structures_vs_planck} 
\end{figure*}

There are two known structures in Cosmos that are also bright at sub-mm
wavelengths, for example PCL1002 at $z$\,=\,2.47~\citep[with 41 spectroscopically
confirmed members, including seven SMGs, one QSO and 33 LBGs, over
20\,\arcmin$\times$20\,\arcmin; ][]{casey15} and CL\,J1001 at
$z$\,=\,2.51~\citep[with 24 spectroscopically confirmed members, including 1
SMG, and six HAEs, over $\sim$3\,\arcmin$\times$3\,\arcmin;
][]{wang16,wang18}.  These structures are considered part of a
proto-supercluster at $z$\,=\,2.45~\citep{cucciati18} that includes 96
spectroscopically confirmed members, initially discovered as an overdensity
of LAEs over a region of 1$\times$1\,arcmin$^2$~\citep[COS1000$+$0215;
][]{diener15,chiang15,civano12,kriek15,trump09,casey15}.  There are two other
significant structures in the Cosmos field at $z{\simeq}$\,2, CC2.2 at
$z$\,=\,2.23, discovered as overdensity of HAEs~\citep{darvish20}, and ZFIRE
at $z$\,=\,2.085, discovered in the Z-FOURGE photometric redshift
survey~\citep{spitler12,yuan14,hung16}.  The main properties of these
protoclusters are listed in Table~\ref{tab:cos_sample}, and the locations of
their confirmed members are shown in Fig.~\ref{fig:cosmos_structures_vs_planck}.

All these structures have estimated total SFRs larger than what we measure
in G237, and are massive structures (M$_{\rm
h}$($z$=0)$\simeq$3--80$\times$10$^{14}$\,\msun).  However, none of them is
detected as a \planck\ high-$z$ source.  In
Fig.~\ref{fig:cosmos_structures_vs_planck}, we show the density map of
red SPIRE sources in a 50\arcmin$\times$50\arcmin\ region where all
above structures are located, and the contours of the \planck\ RX map at
550\,\um.  The figure shows that the \planck\ RX map detects the
overdensities of red SPIRE sources.  The red SPIRE sources
overdensity in G237 is slightly off the \planck\ RX peak.  We note that the
bright star in the G237 field is close to the position of the SPIRE
overdensity, thus no galaxy can be investigated in that region in the
optical/NIR.  There are three \planck\ sources detected at $\sim$2--3$\sigma$
level in the \planck\ maps at the position of CL\,J1001, where some members
of the PCL\,1002 structure are also present, and close to COS1000$+$0215,
and ZFIRE (see black dashed circles in
Fig.~\ref{fig:cosmos_structures_vs_planck}).  The association with these
structures is probably due to the presence of bright SPIRE sources that
belong to these two structures.  The sub-mm signal in the vicinity of
COS1000$+$0215 might indicate that the structure is more extended and
include DSFGs that are not detectable as LAE.  Emission line galaxies (e.g., 
LAEs, and HAEs) are rarely detected at sub-mm wavelengths, and there is some
evidence that they might occupy different regions than DSFGs in
protoclusters~\citep[see e.g., ][]{umehata15,umehata17,lee17}. 
Interestingly, there are nine DSFGs in the ZFIRE protocluster, but this
structure is not detected by \planck.  A possible explanation is that the
DSFG members are not sufficiently clustered.  Indeed they have been found by
a dedicated search of DSFGs with spectroscopic redshift in the region around
ZFIRE~\citep{hung16}.  On the contrary, in G237 two out of four DSFGs are in the
core, and are also detected as HAEs.

This analysis indicates that the total SFR, the halo mass, and the extension
do not determine whether a protocluster will be detected by \planck\ and
selected as a PHz.  The main key factor is possibly the presence of
a large concentration of DSFGs at $z{\sim}$2, but this can be also
due to projection effects or to amplification of line-of-sight structures. 
We would need to analyze more PHz sources to confirm such a claim.

This comparison also highlights the importance of understanding the
connection between protoclusters identified through different selection
techniques, and the different types of member galaxies (massive galaxies,
LAEs, HAEs, DSFGs, and AGN) and how these vary as a function of halo mass,
and redshift.  We can expect that galaxies of a certain type are mostly
associated with a specific evolutionary stage, for example sub-mm galaxies tend to
be more massive, and thus in a more advanced phase than LAEs.  Protoclusters
found through different galaxy types might then be at different maturity
levels.  These questions can only be addressed by studying a large number of
protoclusters covering a significant range of redshifts, masses, and
selections, but this would be a highly demanding endeavor due to the need of
multiwavelength deep data over wide areas.

The lack of starbursting members in G237 is hard to reconcile with the
\planck\ selection, which was intended to find such systems. 
Interestingly, based on the protocluster cosmic filling
factor~\citep[i.e., 0.02 at $z{\sim}$2; ][]{chiang17}, protoclusters are
expected to occupy a volume of $\sim$1500\,cMpc$^3$ in a random sightline in
a $\sim$10\arcmin\ field over 2.0${<}z{<}$2.2.  The two substructures in
G237 occupy, combined, a much larger volume, that is $\sim$8100\,cMpc$^3$. 
Thus, having found these two protoclusters in a \planck\ source field is not
fortuitous, and the 5$\sigma$ overdensity of red SPIRE sources (see
Sect.~\ref{spire_density}) is likely associated with the structures. 
Starbursting protoclusters or with a large number of DSFGs at $z{\lesssim}$3
are expected to be in the \planck\ source list, and the best way to find
them might be choosing those with the most significant sub-mm overdensities. 
G237 is among the least overdense \planck\ sources, and this might explain
the low level of starbursting activity, but a study of another \planck\
source~\citep[i.e., PLCK\,G073.4$-$57.5; ][Hill, R., priv. 
comm.]{kneissl19}, characterized by a large overdensity also reveals a
dominance of normal SFG members, rather than starbursts, and a large number
of DSFG members (five detected by SPIRE, and three by SCUBA2).

\subsection{Physical processes that power the star formation activity in G237}\label{sec:fueling}

\begin{figure} 
\centering
\includegraphics[width=\linewidth]{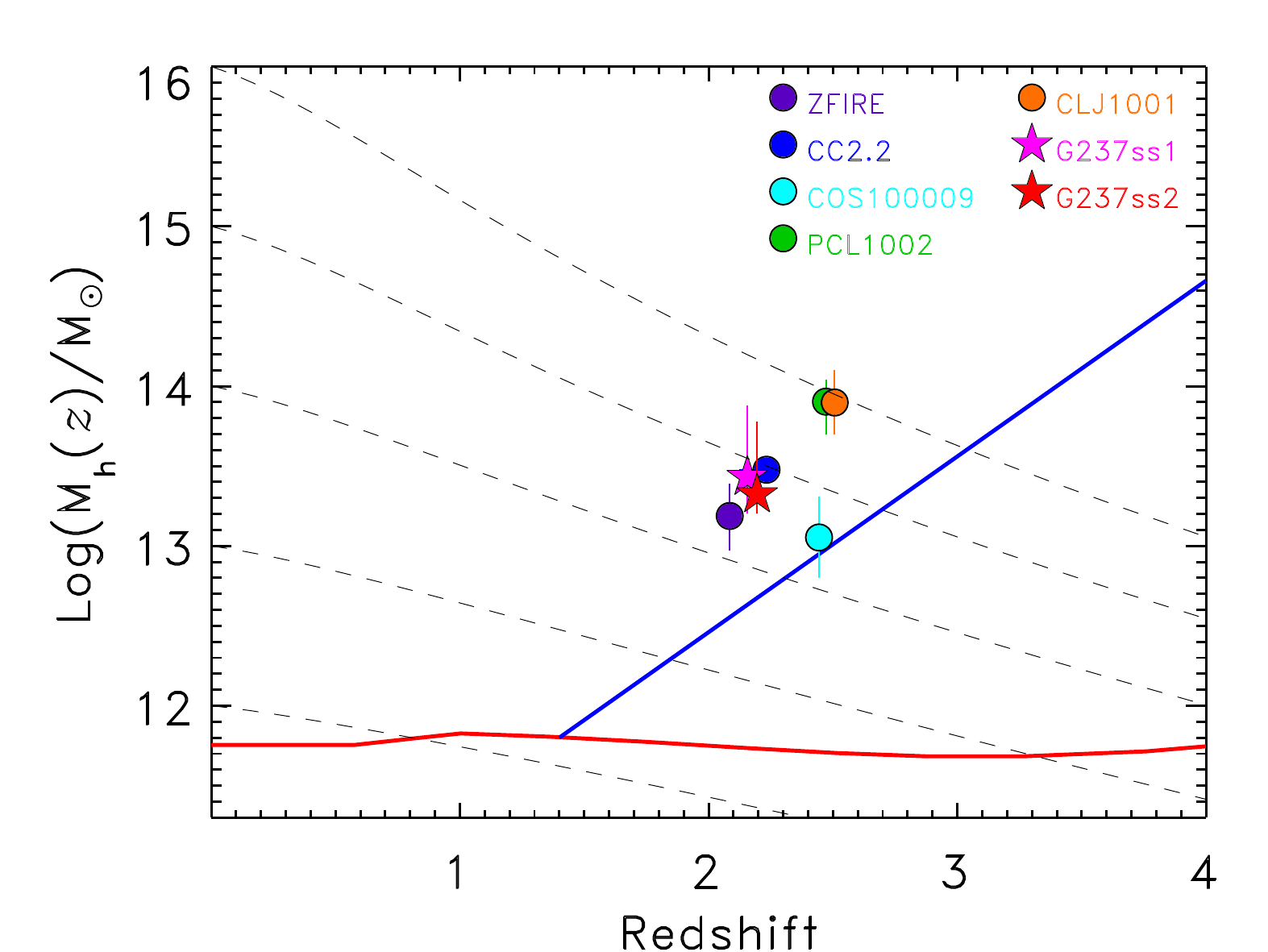}
\caption{{\small Protocluster dark matter halo mass as a function of
redshift for the protoclusters in the Cosmos field at $z{\simeq}$2 listed in
Table~\ref{tab:cos_sample} as annotated, and for the G237 substructures
(ss1: magenta full star, ss2: red full star).  The dashed black lines represent the average halo mass
growth based on the fitting functions formulated by~\citet{behroozi13}. 
Thick lines indicate different gas cooling regimes as predicted
by~\citet{dekel06}: below a critical halo mass of M$_{\rm{h}}^{crit}{\simeq}$10$^{12}$\msun\ 
(red curve) inflowing gas is predominantly
cold allowing galaxies to grow (in addition to
merging); above M$_{\rm{h}}^{crit}$ (red curve), and below a critical
redshift, $z^{crit}$ (to the left of the blue line), cold gas
flows are shock-heated, thus shutting off most of the gas supply to the
galaxies; above M$_{\rm{h}}^{crit}$ (red curve), and above $z^{crit}$ (to the right of the blue
line) cold gas is
able to penetrate the hot gas environment through cold flow streams favoring
galaxy growth and star formation. }}
\label{fig:dekel_diagram}
\end{figure}

\begin{table*}
\centering
\caption{Protoclusters in the Cosmos field at 2.0${<}z{<}$2.5}\label{tab:cos_sample}
{\renewcommand{\arraystretch}{1.4}
\begin{tabular}{llcc cccrr}
\hline\hline
 Name               & Redshift &  N spec.    &  M$_{\rm h}$($z$=0)     & $\Sigma$SFR & Extent & N. DSFG\tablefootmark{a}   & Reference \\
                    & [$z_{min}-z_{max}$] &  members &  (10$^{14}$\,\msun) & (\msun\,yr$^{-1}$) & (arcmin) &  \& AGN    &               \\
\hline
ZFIRE               &   2.085 [2.076--2.104] &     57      & 2.5$^{+2.5}_{-1.2}$ &    4002     & 7.4$\times$10  &  9 \& 4  &    (1,2) \\
CC2.2               &   2.233 [2.122--2.247] &     47      & 9.2                 &    2228     & 9.2$\times$9.2 &  0 \& 3  &      (3) \\
COS1000$+$0215      &   2.444 [2.428--2.456] &     20      & 3.2$^{+4.8}_{-1.9}$ &    \nodata  & 5.0$\times$3.1 &  0 \& 4  &    (4,5) \\
PCL\,1002           &   2.472 [2.463--2.487] &     41      & 77$^{+53}_{-41}$\tablefootmark{b}    &    4447     & 3.0$\times$1.5 &  7 \& 5  &      (6) \\
CL\,J1001           &   2.507 [2.494--2.515] &     18      & 83$^{+96}_{-44}$\tablefootmark{b}    &    2790     & 2.7$\times$3.0   &  1 \& 1  &    (7,8) \\
\hline
G237 ss1            &   2.155 [2.150--2.164] &     20      &  6.6$^{+26}_{-3.5}$ & 1002$\pm$58 & 9.5$\times$9.3  &  3 \& 4  &  (This work) \\
G237 ss2            &   2.195 [2.190--2.199] &      8      &  5.1$^{+20}_{-1.9}$     &  429$\pm$41 & 9.9$\times$10.6 &  1 \& 0  &  (This work) \\
\hline
\hline
\end{tabular}\\
}
\tablefoot{
\tablefoottext{a}{\small Here, we consider a DSFG a galaxy detected in the
sub-mm by SPIRE, SCUBA-2, LABOCA, or AzTEC, even if it might comprise
multiple galaxies as in the case of CL\,J1001 where ALMA finds five sources
associated with a single SPIRE source~\citep{wang16}.}
\tablefoottext{b}{\small For PCL\,1002, and CL\,J1001, the $z$\,=\,0 halo
masses have been derived from the estimated halo mass at the observed
redshift~\citep{casey15,wang16}, these are
(8$\pm$3)\,$\times$10$^{13}$\,\msun\ for PCL\,1002, and
(7.9$^{+4.7}_{-2.9}$)\,$\times$10$^{13}$\,\msun\ for CL\,J1001, and assuming
the mass growth as formulated by~\citet{behroozi13}, to be consistent with
the other structures analyzed here.  These are larger than those derived by
~\citet{casey15}, and~\citet{wang16} assuming a mass-dependent exponential
growth model (i.e., (2$\pm$1)$\times$10$^{15}$\,\msun).} 
\tablebib{\small (1)~\citet{yuan14}, (2)~\citet{hung16},
(3)~\citet{darvish20}, (4)~\citet{diener15}, (5)~\citet{chiang15}
(6)~\citet{casey15}, (7)~\citet{wang16}, (8)~~\citet{wang18}.}
}
\end{table*}

The analysis of the member properties with respect to their location in G237
shows a weak dependence on the environment (see
Section~\ref{sec:mstar_sfr}).  Although most spectroscopic members are blue
SFGs, those in the core are on average more massive and more active.  Also,
the fraction of DSFGs and AGN is much higher in the core (i.e.,  $f_{\mathrm
AGN}^{core}{=}f_{\mathrm DSFG}^{core}{=}$\,50\% $versus$ $f_{\mathrm
AGN}^{full}{=}f_{\mathrm DSFG}^{full}{=}$\,13\%).  The ubiquity of both SF
(traced by the sub-mm emission), and AGN (traced by the X-ray emission)
activity suggests a common mechanism at their origin.  Both SF and AGN
require gas to fuel them, and their enhancement in the core suggests that
the fueling mechanism might be favored in dense regions.  Gas can be
funneled into the center of galaxies through gas-rich mergers or through
gravitational disk instabilities caused by inflowing gas~\citep[e.g.,
][]{keres05,keres09,dekel06,dekel09}.

\begin{table*}
\centering
\caption{Protoclusters from the literature at 1.9${<}z{<}$2.6}\label{tab:lit_sample}
\begin{tabular}{l ccc rr}
\hline\hline
 Name               & Redshift &      N      &  Selection & $\Sigma$SFR & Reference \\
                    &          & members\tablefootmark{a} & method\tablefootmark{b}& (\msun\,yr$^{-1}$)  &           \\
\hline
CARLA\,J1018$+$0530 &   1.952  &      8      &   HzRG     &     512     &       (1) \\
CARLA\,J0800$+$4029 &   1.985  &      9      &   HzRG     &     754     &       (1) \\
Cl\,J1449$+$0856    &   1.990  &     15      &   phot     &    1236     &     (2,3) \\
CARLA\,J2039$-$2514 &   1.999  &      9      &   HzRG     &     945     &       (1) \\
ZFIRE               &   2.085  &      9      &   phot     &    4002     &       (4) \\
MRC\,1138$-$262     &   2.157  &     36      &   HzRG     &    5713     &   (5,6,7) \\
\phz712\ ss1        &   2.155  &     20      &  sub-mm    &    1002     & (This work) \\
HELAISS02           &   2.170  &      4      &  sub-mm    &    1492     &       (8) \\
2QZCluster          &   2.230  &      7      &  sub-mm    &    2004     &       (9) \\
CC2.2               &   2.233  &     40      &   HAE      &    2228     &      (10) \\
HS\,1700$+$64       &   2.310  &     13      &  sub-mm    &    5290     & (9,11,12) \\
HATLAS\,J084933     &   2.412  &      4      &  sub-mm    &    6026     &      (13) \\
PCL\,1002           &   2.472  &     41      &  sub-mm    &    4447     &      (14) \\
4C\,23.56           &   2.488  &     27      &   HzRG     &    3887     &    (6,15) \\
CL\,J1001           &   2.507  &     14      &   phot     &    2790     &      (16) \\
USS\,1558$-$003     &   2.528  &     35      &   HzRG     &    2375     &       (6) \\
HXMM20              &   2.609  &      5      &  sub-mm    &    1877     &       (8) \\
\hline
\hline
\end{tabular}\\
\tablefoot{
\tablefoottext{a}{\small Number of spectroscopically confirmed members for which a SFR estimate is available.}\\
\tablefoottext{b}{\small The listed protoclusters have been found through the following methods,
HzRG: search around a HzRG, phot: identification of an
overdensity of photometrically selected high-$z$ sources, sub-mm:
overdensity of sub-mm sources, HAE: overdensity of \halpha\ emitters in
narrowband images.}\\
\tablebib{\small (1)~\citet{noirot18}, (2)~\citet{smith19},
(3)~\citet{valentino15}, (4)~\citet{hung16}, (5)~\citet{dannerbauer17}, (6)~\citet{tadaki19},
(7)~\citet{hatch09}, (8)~\citet{gomez19}, (9)~\citet{kato16}, (10)~\citet{darvish20},
(11)~\cite{lacaille19}, (12)~\citet{chapman15}, (13)~\citet{ivison13}, (14)~\citet{casey15},
(15)~\citet{lee17}, (16)~\citet{wang18}.}  \\
}
\end{table*}

\begin{figure*} 
\centering
\includegraphics[width=9cm]{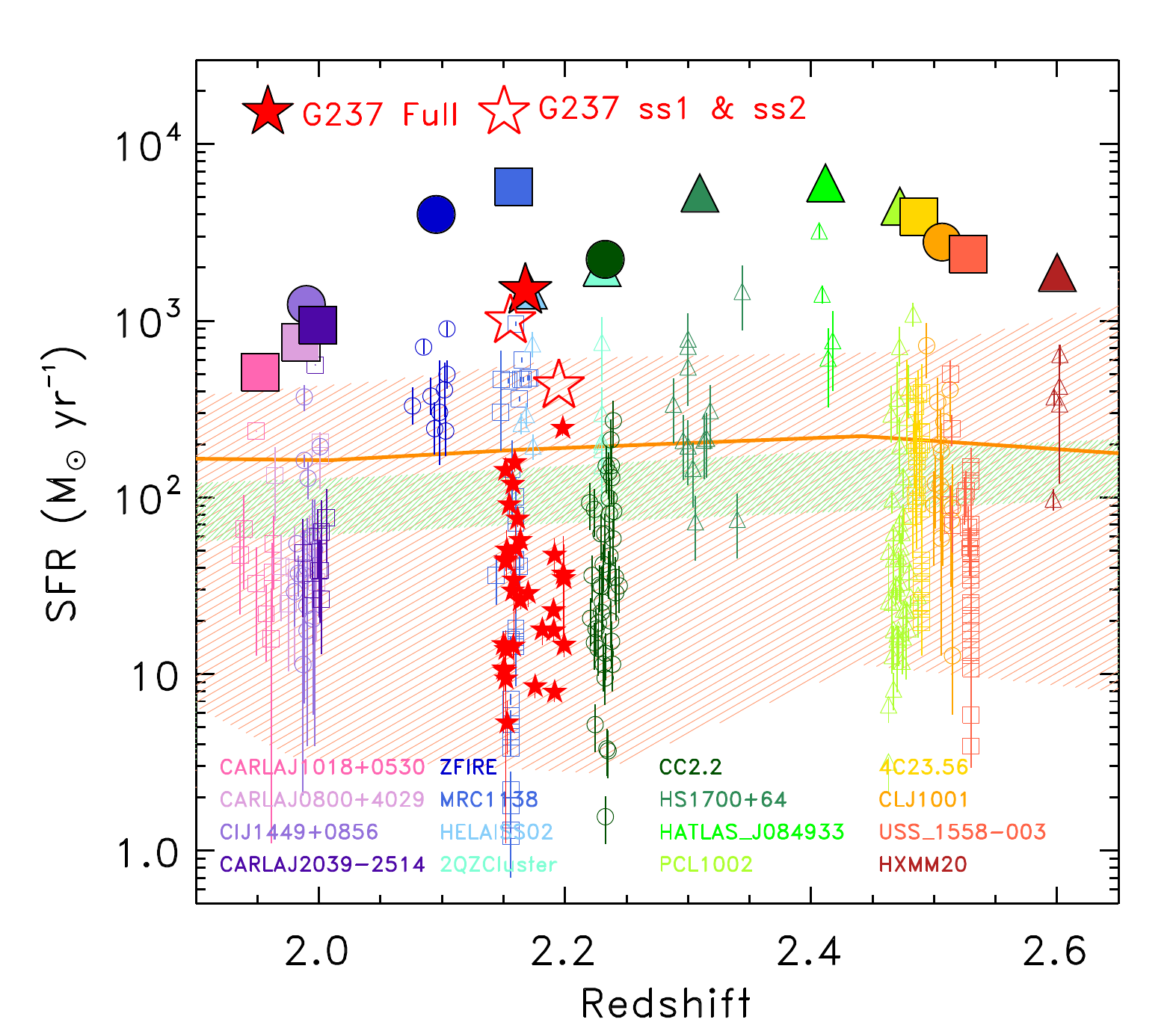}
\includegraphics[width=9cm]{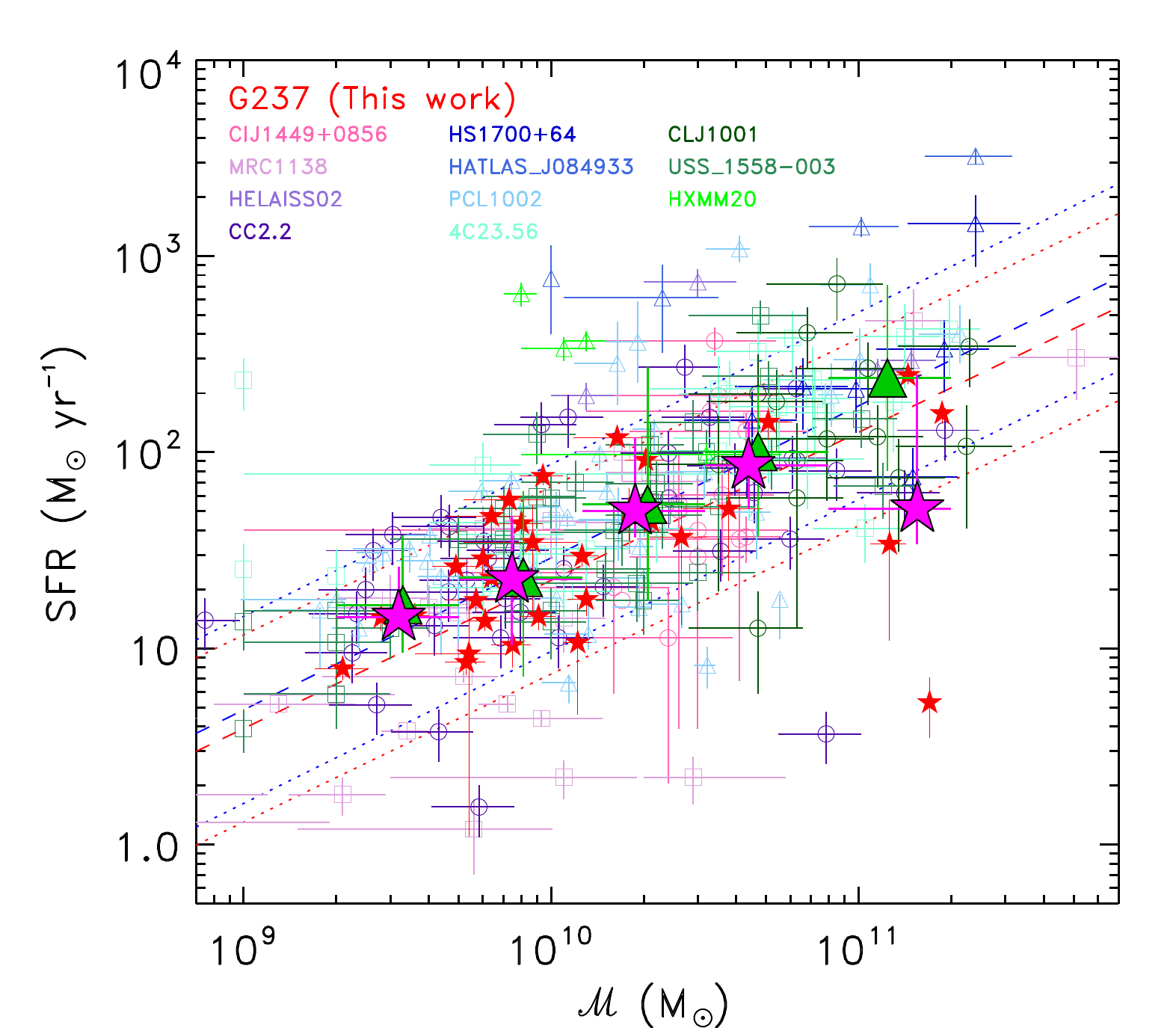}
\caption{{\small SFR as a function of redshifts (\textit{left panel}) and of
stellar mass (\textit{right panel}) of the spectroscopic members in G237
(small filled red stars) and in 16 (11 in the \textit{right panel} because
stellar masses for individual members were not available for all)
protoclusters at 2${\lesssim}z{<}$3 from the literature (open triangles:
sub-mm protoclusters, open circles: color selected (i.e., red optical-NIR
sources or HAEs), or open squares: searches around HzRGs).  The protocluster
names are annotated (for more details see Table~\ref{tab:lit_sample}).  In
the \textit{left panel}, large symbols represent total SFRs.  In the
\textit{left panel}, predictions from the empirical models
of~\citet{behroozi13} (green hatched region), from the 25 most massive
$z$\,=\,0 clusters in the TNG300 simulation (salmon hatched area), and the
median SFR of the five most massive TNG300 simulated clusters (salmon thick
line)~\citep[adapted from ][]{lim21} are shown for comparison.  In the
\textit{right panel}, large full green triangles and large full magenta
stars are mean SFRs per stellar mass bins for the literature samples with
$z{<}$2.5, and for G237, respectively.  In the \textit{right panel}, the
dashed blue and red lines are the MS relation at $z$\,=\,2, and 2.6 as
parameterized by~\citet{speagle14}, respectively, after correcting it for a
\citet{chabrier03} IMF.  The dotted lines represent the scatter around the
MS, equivalent to a factor of 3.}}
\label{fig:lit_sample}
\end{figure*}

In Sect.~\ref{sec:morph}, we find morphologies consistent with signs of
interaction or disturbance in about 40\% of the members, and $\sim$20\% have
also a close companion (see Table~\ref{tab:specz_sample} and
Fig.~\ref{fig:cutouts1}--\ref{fig:cutouts3}).  These properties are
consistent with various merger phases, from pre-merging~\citep[pairs at
$\lesssim$3\arcsec\ at our redshift;][]{ventou17}, to ongoing or post
merging events.  Since
close encounters and interactions are more likely in
overdensities~\citep{jian12}, mergers might be a viable fueling
mechanism.  However, the lack of a clear trend between the morphology and
the spatial distribution questions such an interpretation.  High spatial
resolution imaging in the optical rest-frame, or kinematic measurements of
the gas in these systems would be necessary to determine the role of
mergers.

Alternatively, cold gas streams can also be a viable fueling channel in G237
members.  Cold gas flows are expected to fuel SF in massive halos only at
high redshifts.  Below a critical redshift, $z_{\mathrm{crit}}{\gtrsim}$2,
the streaming cold gas is shock-heated in massive halos
(M$_{\rm{h}}{\gtrsim}$10$^{12}$\,\msun), and does not readily fuel ongoing
SF.  G237 is hosted by a massive halo and it is at a redshift below
$z_{\mathrm{crit}}$.  The same is true for the other protoclusters in the
Cosmos field listed in Table~\ref{tab:cos_sample}.  In
Fig.~\ref{fig:dekel_diagram}, we place the selected protoclusters on a halo
mass $versus$ redshift diagram, and compare them with the different gas
regimes proposed by~\citet{dekel06}.  The halo mass and redshift
marking the transition between the cold streams to the shock-heated regime
are not yet well constrained, theoretically, nor observationally.  Here, we
show the values from~\citet{dekel06}~\citep[but see also][]{lapiner21}.  To
place the selected protoclusters in the diagram, we estimate their halo mass
at the observed redshift.  Those estimates were obtained by taking their
estimated $z$\,=\,0 halo mass and evolving it back in time using the halo
accretion growth based on fitting functions of~\citet{behroozi13} (see
dashed lines in Fig.~\ref{fig:dekel_diagram}), when not available in
the literature~\footnote{For PCL\,1002, and CL\,J1001, we used the estimated
halo mass at the observed redshift reported by~\citet{casey15},
and~\citet{wang16} (i.e., (8$\pm$3)$\times$10$^{14}$\,\msun\ for PCL\,1002,
and (7.9$^{+4.7}_{-2.9}$)\,$\times$10$^{13}$\,\msun\ for CL\,J1001).}.

All selected protoclusters are in the shock-heated regime implying that cold
gas accretion should be inhibited and star-formation will continue as long
as there is cold gas, and might end by starvation.  Assuming a constant SFR
equal to the average value in ss1 (i.e., 33\,\msun\,yr$^{-1}$), and the
relation between SFR and cold molecular gas for normal SFGs as formulated
by~\citet{sargent14}, these galaxies have, on average,
$<$M$_{\mathrm{mol}}{>}\sim$3.4$\times$10$^{10}$\,\msun.  If we assume that
all the molecular gas will be converted into stars, this molecular reservoir
will be depleted in $\sim$1\,Gyr, or at $z{\simeq}$1.6.  This simple
scenario implies no need for cold gas accretion or mergers to sustain the
current SFR in our protocluster members, but does not explain the SFR
enhancement in the core.

\subsection{Member properties in G237 and in other protoclusters}\label{sec:members_comp}

In~Sect.~\ref{sec:mstar_sfr}, we find that most members in G237 are on the MS,
 and no starburst members have been found even among the
SPIRE-detected sources.  Here, we compare G237 member properties with those
of other protoclusters at $z{\sim}$2--3 with spectroscopically
confirmed members and available SFRs, and stellar masses.  These protoclusters
are drawn from three different selection methods: searches around HzRGs,
overdensities of red optical-NIR sources, HAEs, or sub-mm galaxies.  The
list of selected protoclusters for this comparison is reported in
Table~\ref{tab:lit_sample}, along with the number of members with measured SFR and
the integrated SFR. The star formation rates are either derived from FIR-sub-mm data corrected
for a \citet{chabrier03} IMF, from the dust corrected \halpha\ luminosity, or
from the SED fitting.  The SFR as a function of redshift of single members as
well as the integrated SFR of each protocluster are compared with G237 in
Fig.~\ref{fig:lit_sample} (\textit{left panel}). 
Single structure members' SFR span a wide range, $\gtrsim$2 dex.  Sources
with large SFRs ($>$100\,\msun\,yr$^{-1}$) are found in protoclusters drawn from all
selection techniques, although the most intense SFGs
($>$1000\,\msun\,yr$^{-1}$) are
mainly found in sub-mm overdensities, and they are at $z{\sim}$2.4--2.5. 
The spectroscopic members in G237 exhibit a wide range of SFRs, as in other
protoclusters at similar redshift.
We compare the observed total SFRs with those predicted by empirical models
and simulations of protoclusters~\citep{behroozi13,lim21} in Fig.~\ref{fig:lit_sample} and 
find that the observed ones are higher by a factor of $\sim$30 on average.

The SFR as a function of stellar mass for all protocluster members for which
those estimates are available are shown in the \textit{right panel} of
Fig.~\ref{fig:lit_sample}.  The majority of sources lie on the MS.  There
are several sources above (by a factor of 3) the MS, consistent with being
starburst galaxies.  They are mostly sub-mm selected members, and massive
($>$10$^{10}$\,\msun) galaxies.  A few sources are located below the MS,
including two members in G237.  Members below the MS are present at all
stellar masses, they are in MRC\,1138$-$262, in PCL\,1002, and in CC2.2,
fields where deep observations are available, possibly implying that such
members might be present in other structures, but require extensive searches
and deep data to be found.  We also show the average SFR per stellar mass
bin for G237 (full magenta stars), and for the protoclusters at
2.0${<}z{<}$2.5 (full green triangles).  The upper and lower uncertainty
bounds on the average SFRs are estimated by taking the 16th and 84th
percentiles.  The mean SFRs for G237 and for the protocluster from the
literature are both consistent with the MS relation.  In G237, the value at
the largest masses is below the MS, yet consistent with it if we take into
account the associated uncertainty.  This offset might hint to a beginning
of quenching among the most massive galaxies in G237.  It is interesting to
explore whether such a population is also present in other protoclusters
that exhibit properties of more mature structures, like the presence of a
red sequence due to quiescent galaxies, and diffuse X-ray
emission~\citep[i.e., Cl\,J1449$+$0856, and CL\,J1001;
][]{strazzullo16,valentino16,wang16}.  There is indeed one galaxy in each of
these two clusters below the MS, and in the case of Cl\,J1449$+$0856 there
are four additional members, classified as quiescent galaxies, with only an
upper limit to their SFR, that would fall below the MS at high stellar
masses~\citep[${>}10^{10.5}$\,\msun; ][]{strazzullo18}.  These are not shown
in Fig.~\ref{fig:lit_sample} because their SFR estimates are not available. 
There are, however, other protoclusters, including highly star-forming ones,
with several members below the MS, as pointed out earlier.  This highlights
the heterogeneity observed in protocluster members, which might be in part
intrinsic, and in part the result of the variety of techniques, and data
used to identify and characterize them.

In conclusion, the members in G237 exhibit similar SFRs and stellar masses
as those observed in other protoclusters at similar redshifts.  Overall, the
analyzed protoclusters exhibit different fractions, although poorly
constrained, of ordinary SFGs, starbursts, and quenching galaxies.  These
fractions might be related to the protocluster maturity level, but also to
observational biases.  Deep multiwavelength data, able to identify
different types of member galaxies (starburst, SFGs, and quenching) of
protoclusters drawn from different selections are necessary to understand
the origin of the observed diversity.

\section{Summary and conclusions}\label{sec:summary}

We discuss the spectroscopic identification and characteristics of members
of an overdensity at $z{\simeq}$2.16 located in a protocluster candidate
discovered with the \planck\ satellite, PHz\,G237.01$+$42.50 (G237).  An
overdensity of HAEs at the same redshift in this field had been previously
reported by~\citet{koyama21}.  The overdensity contains 31
spectroscopically confirmed galaxies at 2.15${\leq}z_{\rm spec}{<}$2.20
over a 10\arcmin$\times$11\arcmin\ region, consistent with an overdensity
significance of 5.4$\sigma$. Within this overdensity, we identify
two substructures, ss1 with 20 members at 2.15${\leq}z{<}$2.164, and ss2
with eight members at 2.19${\leq}z{<}$2.20.  Both substructures are expected to
collapse by $z$\,=\,0 and become clusters with $z$\,=\,0 halo masses of
5--6$\times$10$^{14}$\,\msun, roughly corresponding to the virial mass of a
Virgo-type cluster.  We thus refer to them as protoclusters.
Most members are normal SFGs with a wide range of stellar masses
($\mathcal{M}$\,=\,2$\times$10$^9$--2$\times$10$^{11}$\,\msun), and are
either disks or irregular galaxies with SFRs consistent with the MS
relation at their redshift.  We find a large AGN fraction of
20$\pm$10\,\% among the identified members in ss1, selected via the
X-ray luminosity or spectral line widths.  All AGN members are powerful
enough to halt the star formation in their hosts through radiative feedback. 
However, we do not see any obvious sign of quenching in the AGN members,
with one exception where the galaxy is situated below the MS.

We find that the protocluster core, besides being denser, includes members
that are, on average, more massive and star-forming and contains a larger
fraction of AGN and DSFGs than the full sample.  Although this result might
indicate a more efficient growth in the densest regions within the
structure, its significance is not high enough to establish an environmental
effect on the growth of these galaxies.

The total SFR estimated by adding the SFR of the spectroscopic members, of
the HAEs found in the structure~\citep{koyama21}, and of the SPIRE member
candidates is $\sim$4000\,\msun\,yr$^{-1}$.  Although this is higher than
predicted by simulations ~\citep{lim21}, it is much smaller than estimated
from the \planck\ data (i.e., 10173\,\msun\,yr$^{-1}$).  By analyzing the
\herschel/SPIRE data in the field, in combination with the available
ancillary data, we explain this discrepancy as an effect of sources
alignment along the line of sight that produces a 5$\sigma$ overdensity of
red SPIRE sources in the field.  This result is in agreement with the
predictions based on the standard scenario for the evolution of large-scale
structure~\citep{negrello17}.

In view of protocluster member selection and identification in future
studies, it is important to point out that the LAE selection technique would
have missed most of the spectroscopic members analyzed here as they do not
exhibit a sufficiently strong \lya\ emission line.  We plan to use the
recently released Cosmic HydrOgen Reionization Unveiled with Subaru (CHORUS)
data~\citep{inoue20} taken with the NB387 narrowband filter to search for
LAEs in this field, and test this hypothesis.  Based on our spectroscopic
sample, the HAE selection technique finds 75\% of all the spectroscopic
members analyzed here at $z{<}$2.17, including the DSFGs.  Thus, it seems to
be an effective technique to find protocluster members with
SFR$\geq$5\,\msun\,yr$^{-1}$.  However, because of the limited width of the
narrowband filters, the narrowband imaging technique only finds members in
a narrow redshift range (e.g., $\Delta z{\simeq}$0.027) and might not
sample the full richness of an overdensity.

Further advances in our understanding of these structures might come from
observations that will become possible with upcoming facilities. 
For example, the \textit{Euclid}~\citep{laureijs11} mission will be
able to detect all spectroscopic members in G237 in its wide optical-NIR
survey, and will reach a 60\% completeness level in detecting structures
like G237 ss1~\citep[M$_{\rm h}(z\,{=}\,2)\simeq{5}\times{10}^{14}$\,\msun;
see Fig.~10 in ][]{euclid_sims}. The \textit{James Webb} Space
Telescope~\citep[JWST; ][]{gardner06}, and ground-based AO imaging
observations in the NIR could provide high spatial resolution observations
in the optical rest-frame to determine whether mergers are the primary
fueling process. Wide field deep NIR spectroscopy with
MOONS~\footnote{https://vltmoons.org/}
on the VLT could provide the identification of a large number of
protocluster members and better constrain their stellar population.  Deep
X-ray observations with Athena~\citep{barret20} might detect the proto-ICM
hot gas permeating these structures, and probe their halo masses and
maturity level. Observations of the molecular gas with IRAM or ALMA 
would provide secure identification of DSFG members and information on the
molecular gas properties (see Polletta et al., in prep.).
A promising future is ahead for the comprehension of the
physical processes occurring in these complex and fascinating objects.

\begin{acknowledgements}
We are thankful to the anonymous referee for a careful reading, useful
comments and suggestions that improved the quality of this
paper.
MP kindly thanks F.~Cullen for providing the VANDELS stacked spectra in
electronic format, and A.  Marchetti for reducing the LBT data.  MP thanks
S.~Molendi, S.~Andreon, C.~Mancini, B.~Garilli, A.~Gargiulo,
L.~P.~Cassar\`a, and P.  Franzetti for useful discussions.  MP acknowledges
the financial support from Labex OCEVU.  BLF gratefully acknowledges support
from the Universit\'{e} de Paris-Saclay.  GV acknowledges financial support
from Premiale 2015 MITiC (PI B.  Garilli).
This work has been carried out thanks to the support of the OCEVU Labex
(ANR-11-LABX-0060) and the A*MIDEX project (ANR-11-IDEX-0001-02) funded by
the "Investissements d'Avenir" French government program managed by the ANR.
%
The work is based on observations obtained with Planck
http://www.esa.int/Planck, an ESA science mission with instruments and
contributions directly funded by ESA Member States, NASA, and Canada.
%
This research has made use of data from HerMES project
(http://hermes.sussex.ac.uk/). HerMES is a Herschel Key Programme utilising
Guaranteed Time from the SPIRE instrument team, ESAC scientists and a mission
scientist.
%
The HerMES data was accessed through the Herschel Database in Marseille
(HeDaM - http://hedam.lam.fr) operated by CeSAM and hosted by the Laboratoire
d'Astrophysique de Marseille.
%
This research has made use of the NASA/IPAC Infrared Science Archive, which
is funded by the National Aeronautics and Space Administration and operated
by the California Institute of Technology.
%
Based on data products from observations made with ESO Telescopes at the La
Silla Paranal Observatory under ESO programme ID 179.A-2005 and on data
products produced by TERAPIX and the Cambridge Astronomy Survey Unit on
behalf of the UltraVISTA consortium.  
%
The LBT is an international collaboration among institutions in the United
States, Italy and Germany.  LBT Corporation partners are: The University of
Arizona o n behalf of the Arizona university system; Istituto Nazionale di
Astrofisica, Italy; LBT Beteiligungsgesellschaft, Germany, representing the
Max-Planck Society, the Astrophysical Institute Potsdam, and Heidelberg
University; The Ohio State University, and The Research Corporation, on
behalf of The University of Notre Dame, University of Minnesota, and
University of Virginia.  We acknowledge the support from the LBT-Italian
Coordination Facility for the execution of observations, data distribution
and reduction.
%
This work is based in part on observations made with the Spitzer Space
Telescope, which was operated by the Jet Propulsion Laboratory, California
Institute of Technology under a contract with NASA.
%
Based on observations collected at the European Organisation for
Astronomical Research in the Southern Hemisphere under ESO programme
175.A-0839.
%
Based in part on data collected at Subaru Telescope, which is operated by the National Astronomical Observatory of Japan.
%
{\em Software:} This research made use of astropy, a community developed core Python package
for astronomy~\citep{astropy}, of APLpy, an open-source plotting package for
Python~\citep{aplpy}, of topcat~\citep{topcat}, and of the HEALPix software~\citep{gorski05}.
%
We acknowledge the use of the ICRAR's, and Ned Wright's Cosmology
Calculators~\citep[https://cosmocalc.icrar.org/; ][]{wright06}.

\end{acknowledgements}

\clearpage

\appendix
\appendixpage

\section{LBT spectra}

The complete list of the LBT spectroscopic targets, with
coordinates, estimated redshifts, and associated quality flags, K-band
magnitudes, and IRAC colors from the L16 catalog, is shown in
Table~\ref{tab:lbt_sample}. In Fig.~\ref{fig:lbt_spectra} we show the
LBT/LUCI spectra of the 15 LUCI selected targets.  The six LBT targets that
had been previously observed by the Cosmos team and the available
zCosmos-Deep~\citep{lilly09} redshifts are listed in
Table~~\ref{tab:lbt_sample}, and relative spectra are shown in
Fig.~\ref{fig:lbt_vimos_spectra}.
\begin{table*}[h!]
\caption{LUCI targets' main properties}\label{tab:lbt_sample}
\centering
\begin{tabular}{r ccc cc ccc}
\hline\hline
 LUCI    &   NUMBER   & $\alpha$ (L16) & $\delta$ (L16)   &$z_{\rm LUCI}$&   $z$   & K$_s$   & S$_{4.5\mu m}$/S$_{3.6\mu m}$& \herschel\  \\
 ID      &     (L16)  & (deg)    &  (deg)      &($z_{\rm VIMOS}$)\tablefootmark{a} & flag\tablefootmark{b} &  (AB)   &      & association \\
\hline                                          
   SL01  &    679779  & 150.469450 &  2.33129  & 2.1586\tablefootmark{c}& 3 & 21.788$\pm$0.004 & 1.083$\pm$0.014  & 7814 \\   
   SL02  &    676097  & 150.472301 &  2.324637 &    2.1240   &        3 &   20.099$\pm$0.001 & 1.398$\pm$0.008  & 2889 \\     
   SL03  &    669706  & 150.466518 &  2.315597 &    2.1592   &        3 &   20.802$\pm$0.002 & 1.192$\pm$0.008  & 9741 \\     
   SL04  &    656918  & 150.472125 &  2.296723 &   \nodata   &  \nodata &   21.341$\pm$0.003 & 1.327$\pm$0.005  & 4103 \\     
   SL05  &    643336  & 150.483696 &  2.276123 &    2.4189   &        2 &   21.188$\pm$0.002 & 1.239$\pm$0.009  & 1572 \\     
   SL06  &    657340  & 150.506844 &  2.297430 &   \nodata   &  \nodata &   21.994$\pm$0.006 & 1.240$\pm$0.011  & \nodata\ \\ 
   SL07  &    665034  & 150.468504 &  2.309183 &   \nodata   &  \nodata &   21.615$\pm$0.004 & 0.741$\pm$0.047  & \nodata\ \\ 
   SL08  &    664921  & 150.480758 &  2.309062 &1.5210 (0.8613) & 4 (1) &   21.839$\pm$0.005 & 1.148$\pm$0.010  & \nodata\ \\ 
   SL09  &    673976  & 150.486251 &  2.320757 &   (0.9246)  &      (3) &   19.816$\pm$0.001 & 0.712$\pm$0.009  & \nodata\ \\ 
   SL10  &    647891  & 150.485378 &  2.280741 &   (0.2227)  &      (4) &   19.711$\pm$0.001 & 0.797$\pm$0.016  & \nodata\ \\ 
   SL11  &    657494  & 150.481342 &  2.297394 &   \nodata   &  \nodata &   21.226$\pm$0.003 & 0.713$\pm$0.025  & \nodata\ \\ 
   SL12  &    658044  & 150.489417 &  2.297995 &   (0.9250)  &      (3) &   20.360$\pm$0.001 & 0.727$\pm$0.012  & \nodata\ \\ 
   SL13  &    659137  & 150.502422 &  2.300324 &   (0.2107)  &  \nodata &   21.689$\pm$0.004 & 0.884$\pm$0.009  & \nodata\ \\ 
   SL14  &    674760  & 150.475890 &  2.323199 &   \nodata   &  \nodata &   21.482$\pm$0.003 & 0.755$\pm$0.036  & \nodata \\  
   SL15  &    647673  & 150.508934 &  2.281465 &   \nodata   &  \nodata &   20.529$\pm$0.001 & 0.761$\pm$0.022  & \nodata\ \\ 
\hline
\hline
\end{tabular}\\
\tablefoot{
\tablefoottext{a}{\small VISMOS spectroscopic redshifts are from the zCosmos spectroscopic
survey~\citep{lilly07,lilly09}.}
\tablefoottext{b}{\small The redshift flag corresponds to the probability of the
redshift estimate of being correct and can be equal to 1 (50--75\% probability),
2 (75--90\% probability), or to 3, and 4~\citep[$>$90\%
probability; ][]{lefevre13}.}
\tablefoottext{c}{\small The LUCI spectrum shows a broad \halpha\ line, implying that this source is an AGN.}
}
\end{table*}

\begin{figure}[ht!]
\centering
\includegraphics[width=\linewidth]{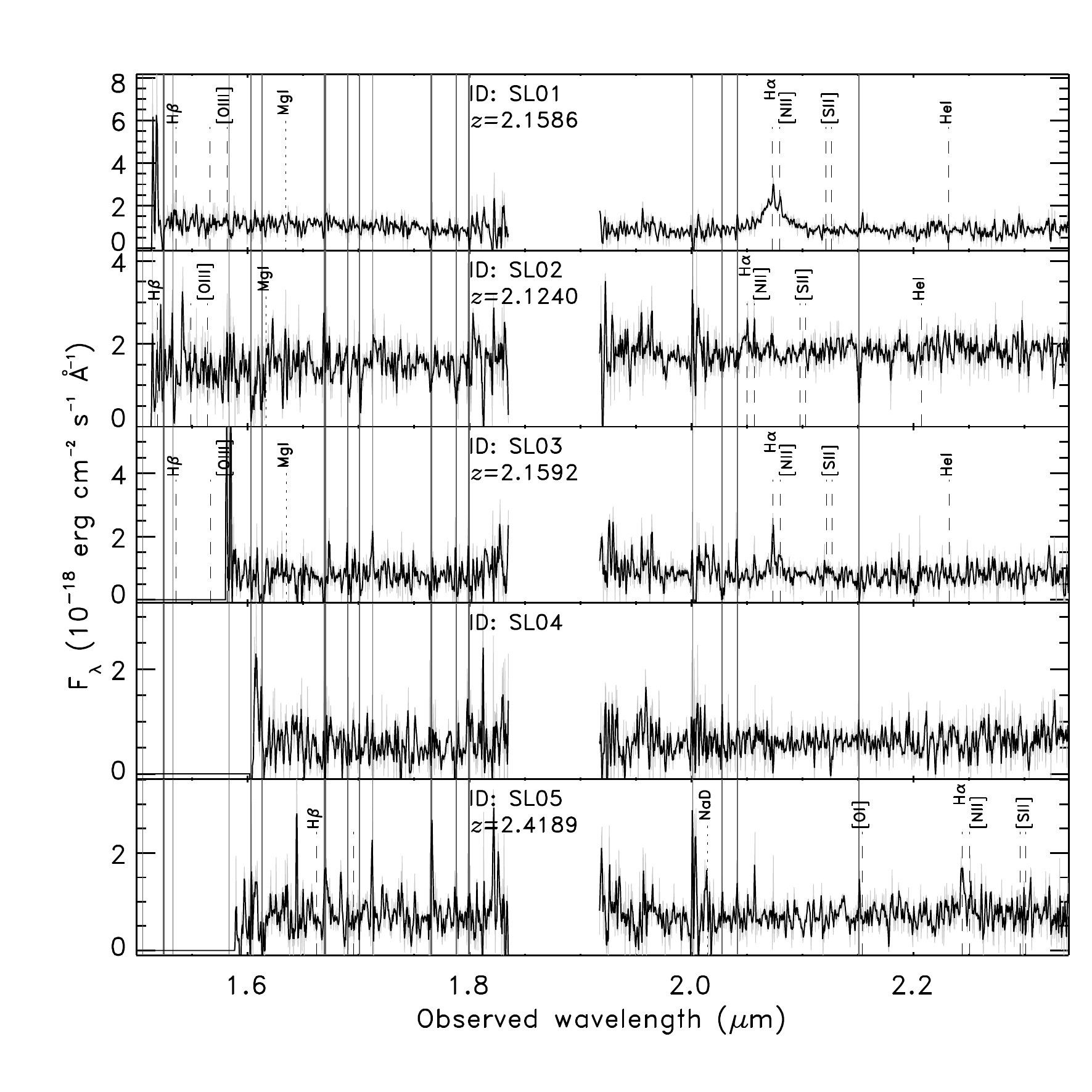}
\caption{{\small LBT/LUCI spectra (gray line, black lines are rebinned
spectra) of the 15 observed targets listed in Table~\ref{tab:lbt_sample}. 
Identifiers are annotated and spectroscopic redshifts are shown when they
could be reliably measured.  In such cases, the main emission and absorption
lines are shown with vertical dashed, and dotted lines, respectively. 
Spectral regions contaminated by bright sky lines are marked with dark gray
vertical bands.  Spectral features expected at the optical redshift provided
by zCosmos-deep spectrum, when available (see Table~\ref{tab:lbt_sample}),
are shown as red vertical dashed lines and annotated.}}
\label{fig:lbt_spectra}
\end{figure}
\begin{figure}\ContinuedFloat
\centering
\includegraphics[width=\linewidth]{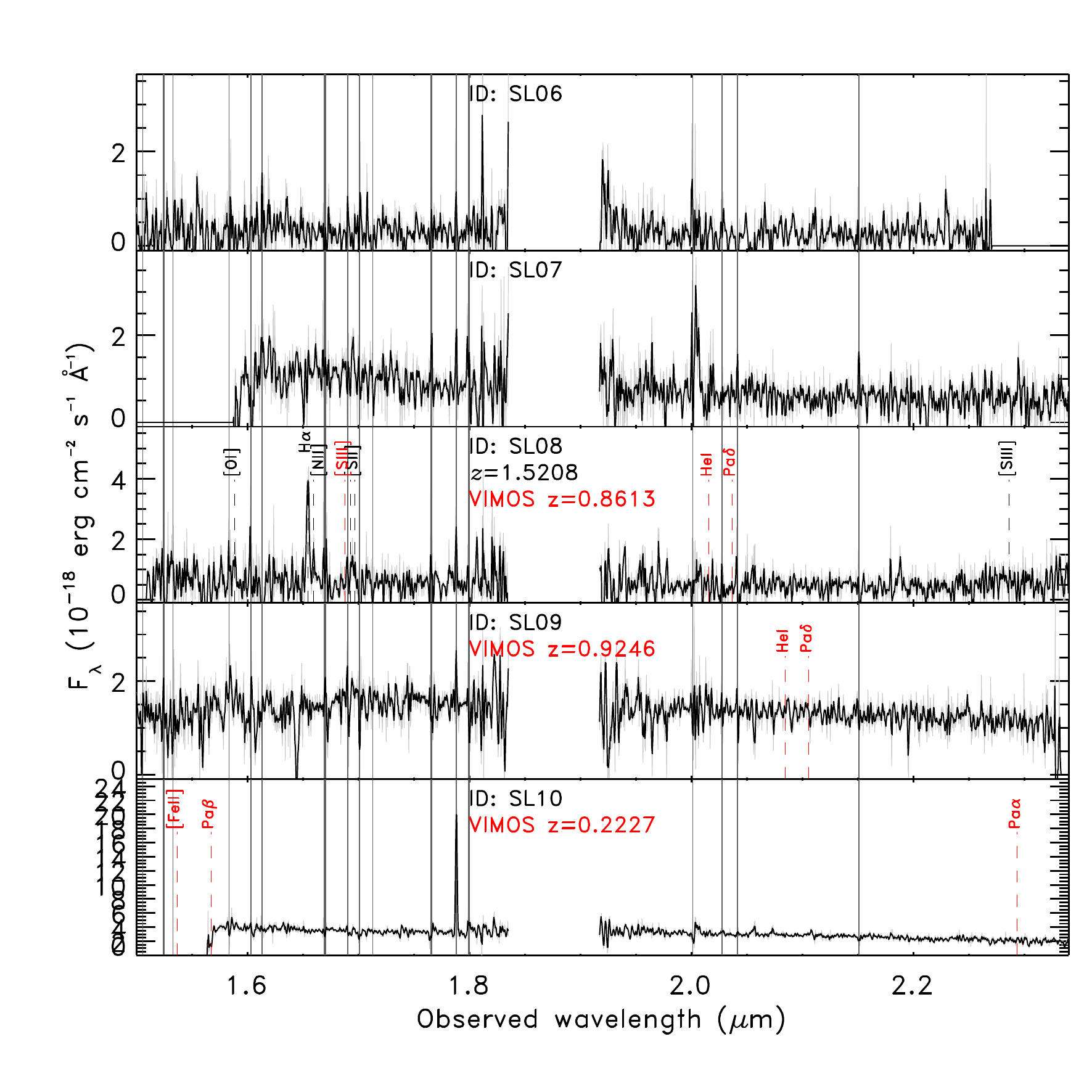}
\caption{{\small Continued.}}
\end{figure}
\begin{figure}\ContinuedFloat 
\centering
\includegraphics[width=\linewidth]{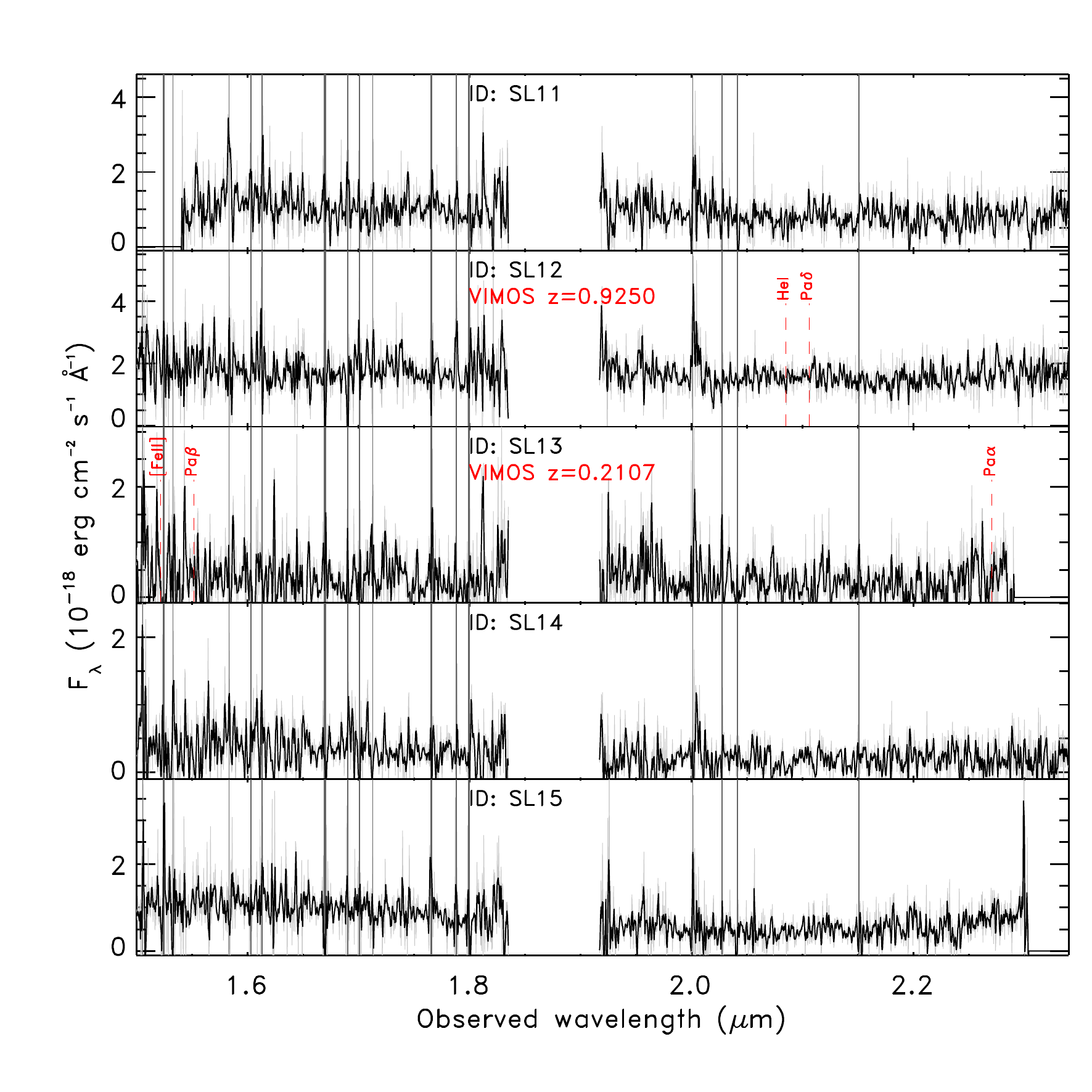}
\caption{{\small Continued.}}
\end{figure}

\begin{figure}
\centering
\includegraphics[width=\linewidth]{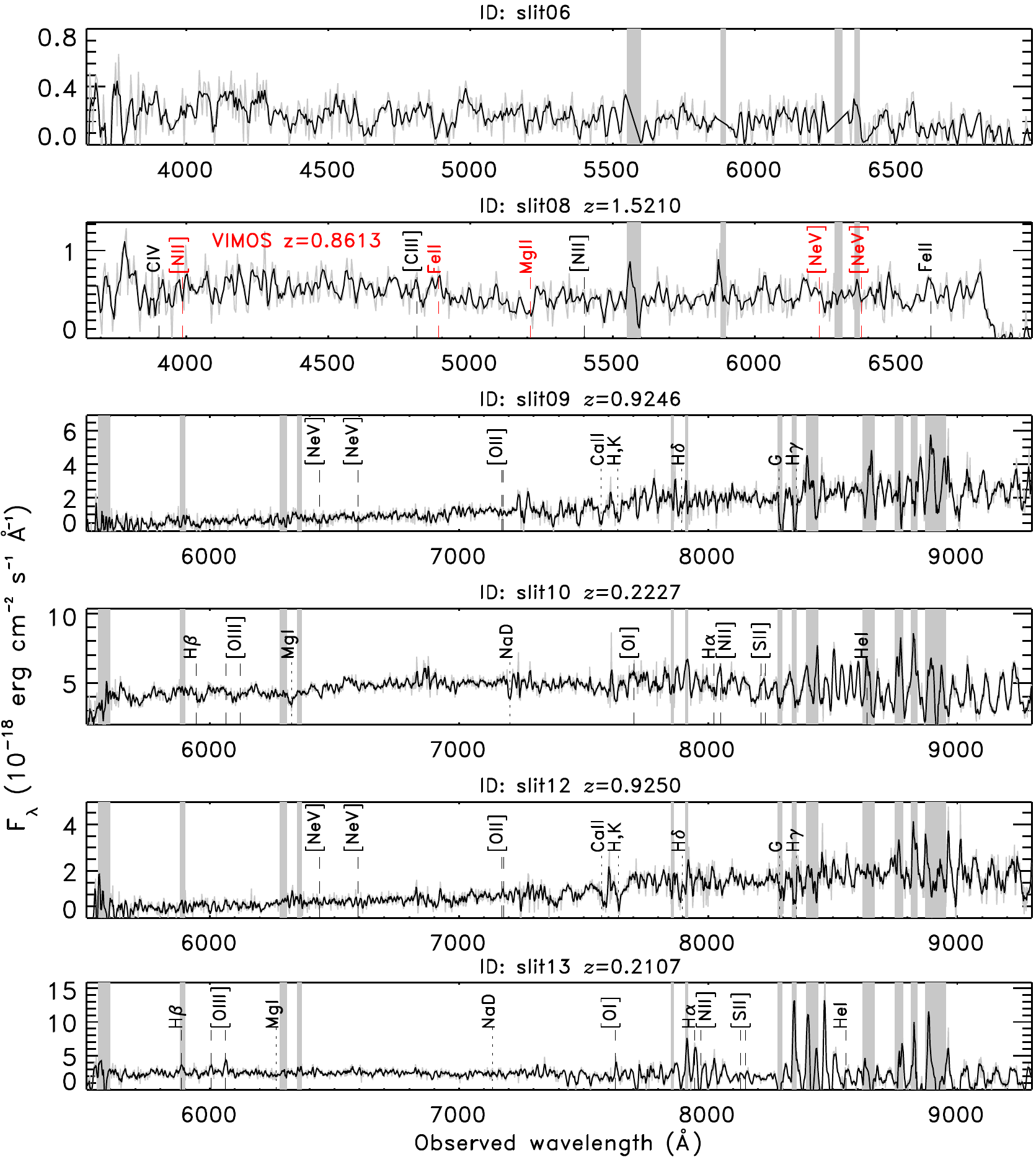}
\caption{{\small zCosmos-Deep VIMOS spectra of the sources observed with LBT/LUCI
(gray line, in black after rebinning).  Identifiers and spectroscopic redshifts
are annotated at the top of each panel.  The main emission and absorption lines are shown with
vertical dashed, or dotted lines, respectively. Spectral regions contaminated by bright sky lines are marked with
gray vertical bands. In case of ID slit08, we show the spectral features expected
at the LBT/LUCI redshift in black, and those at the VIMOS optical redshift
in red. The LUCI redshift is more reliable than the VIMOS one in this
case.}}
\label{fig:lbt_vimos_spectra}
\end{figure}

\clearpage
\section{VIMOS spectra}

In Fig.~\ref{fig:spectra_2p16}, we show the zCosmos-Deep/VIMOS spectra of
the sources at 2.15${\leq}z_{\mathrm spec}{<}$2.164, in Fig.~\ref{fig:spectra_2p19},
those 2.19${<}z_{\rm spec}{<}$2.20, and in Fig.~\ref{fig:spectra_int} those
at 2.164${<}z_{\rm spec}{<}$2.19.

\begin{figure}[h!]
\centering
\includegraphics[width=\linewidth]{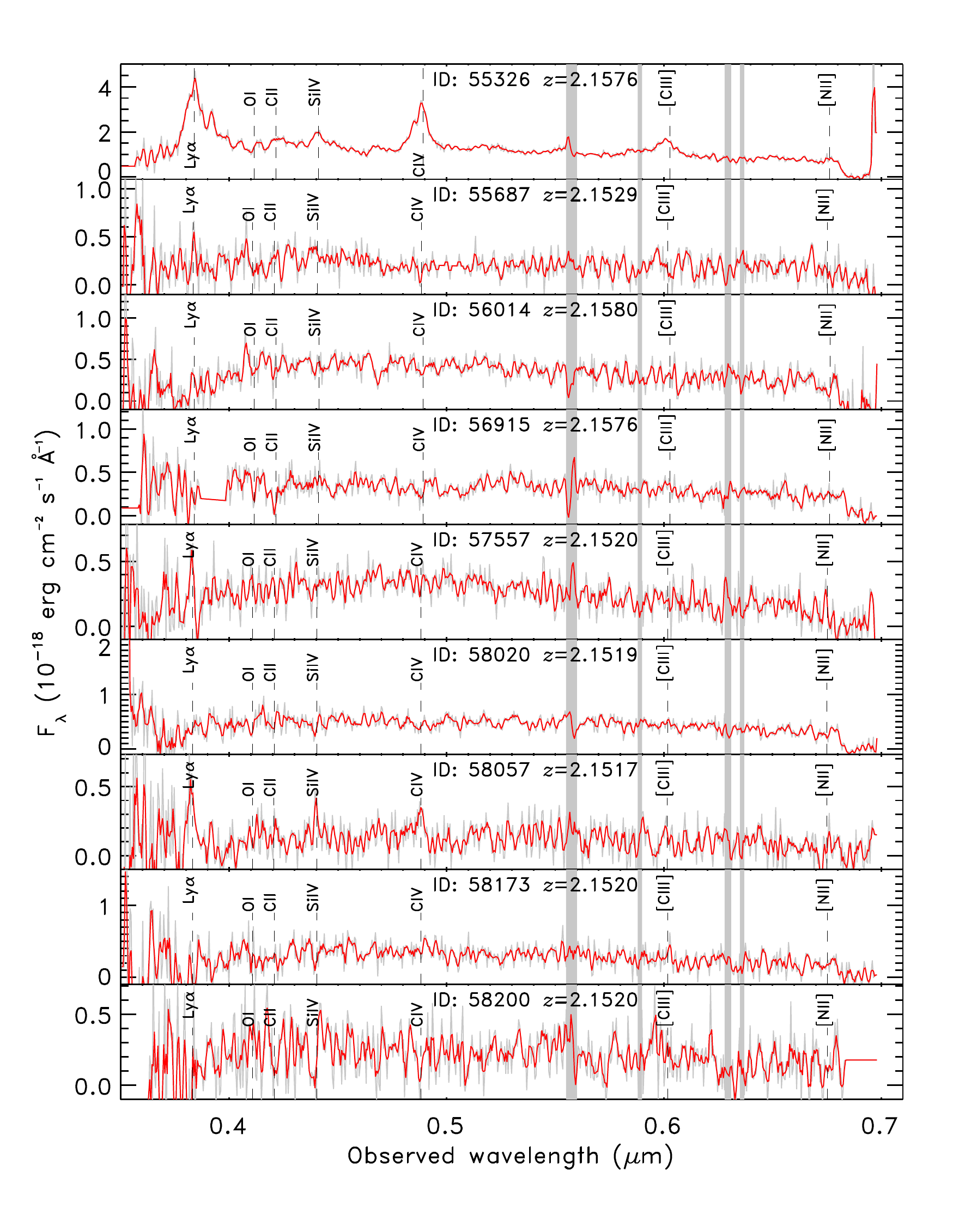}
\caption{{\small VIMOS (DEIMOS for L710971) spectra (gray line) of 18 member candidates with $z$\,=\,2.15--2.164
in the 10$^{\prime}\times12^{\prime}$ region centered at $\alpha$\,=\,150.465\deg, and
$\delta$\,=\,2.31\deg\ where G237 is located.  The red lines are smoothed
spectra. The vertical dashed lines represent the location of the main
spectral lines as annotated. Source IDs and spectroscopic redshifts are annotated in each
panel. Regions contaminated by bright sky lines are highlighted with vertical gray
bars.}}
\label{fig:spectra_2p16}
\end{figure}
\begin{figure}\ContinuedFloat 
\centering
\includegraphics[width=\linewidth]{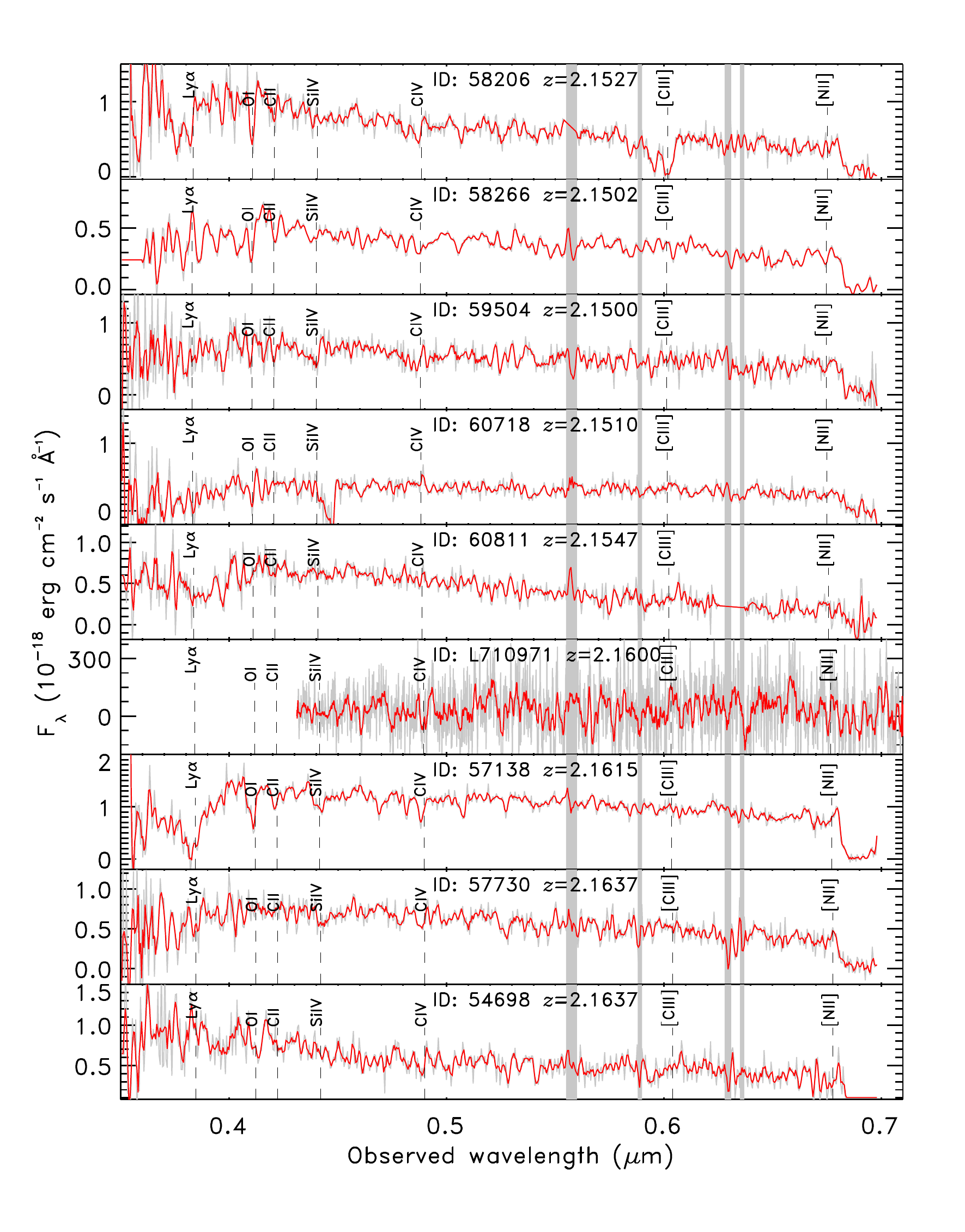}
\caption{{\small Continued.}}
\end{figure}

\begin{figure}[h!]
\centering
\includegraphics[width=\linewidth]{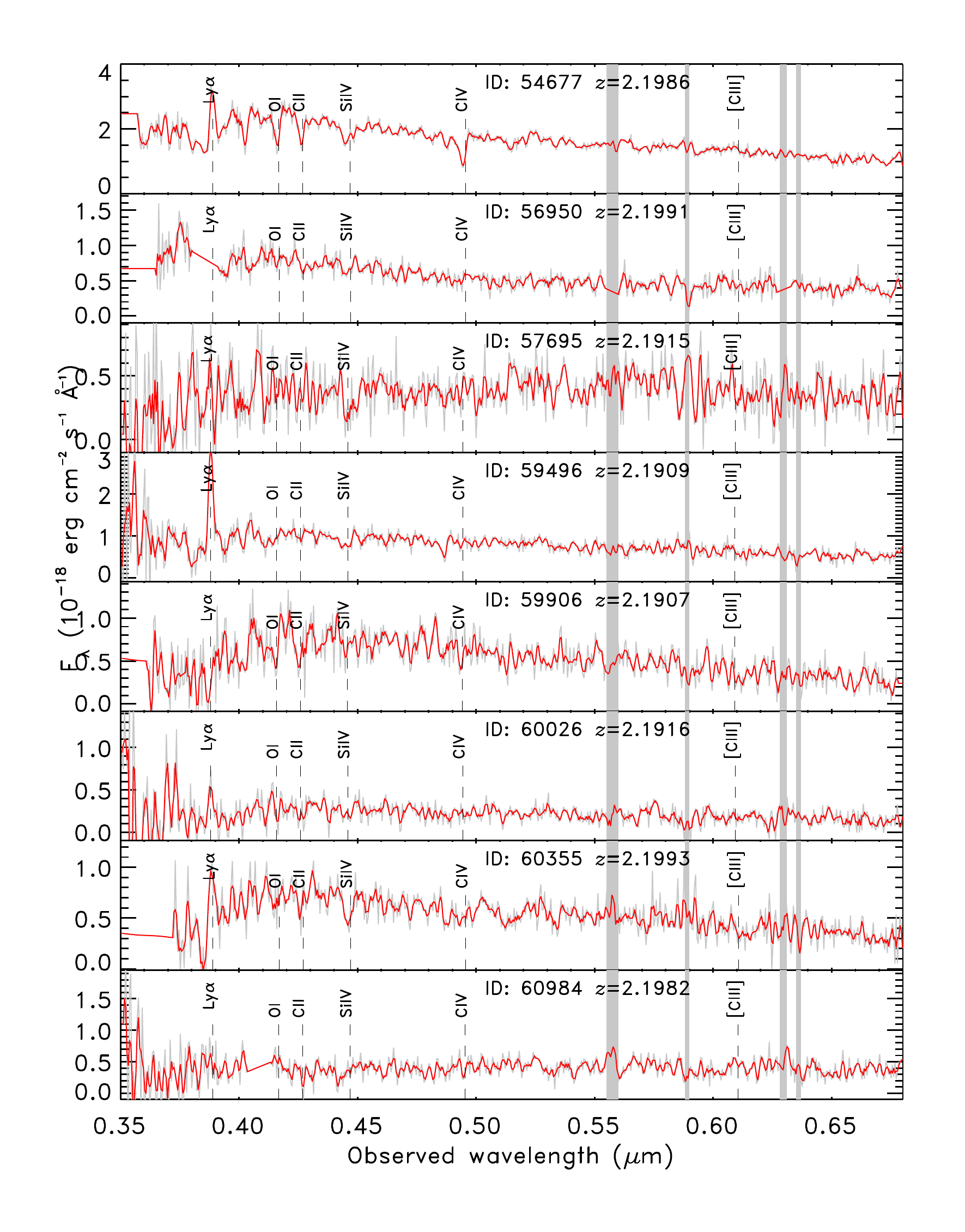}
\caption{{\small VIMOS spectra (gray line) of eight member candidates with
$z$\,=\,2.19--2.20 in the 10$^{\prime}\times12^{\prime}$ region centered at
$\alpha$\,=\,150.465\deg, and $\delta$\,=\,2.31\deg\ where G237 is located.  The red lines are smoothed
spectra. The vertical dashed lines represent the location of the main
spectral lines as annotated. Source IDs and spectroscopic redshifts are annotated in each
panel. Regions contaminated by bright sky lines are highlighted with vertical gray
bars.}}
\label{fig:spectra_2p19}
\end{figure}

\begin{figure}[h!]
\centering
\includegraphics[width=\linewidth]{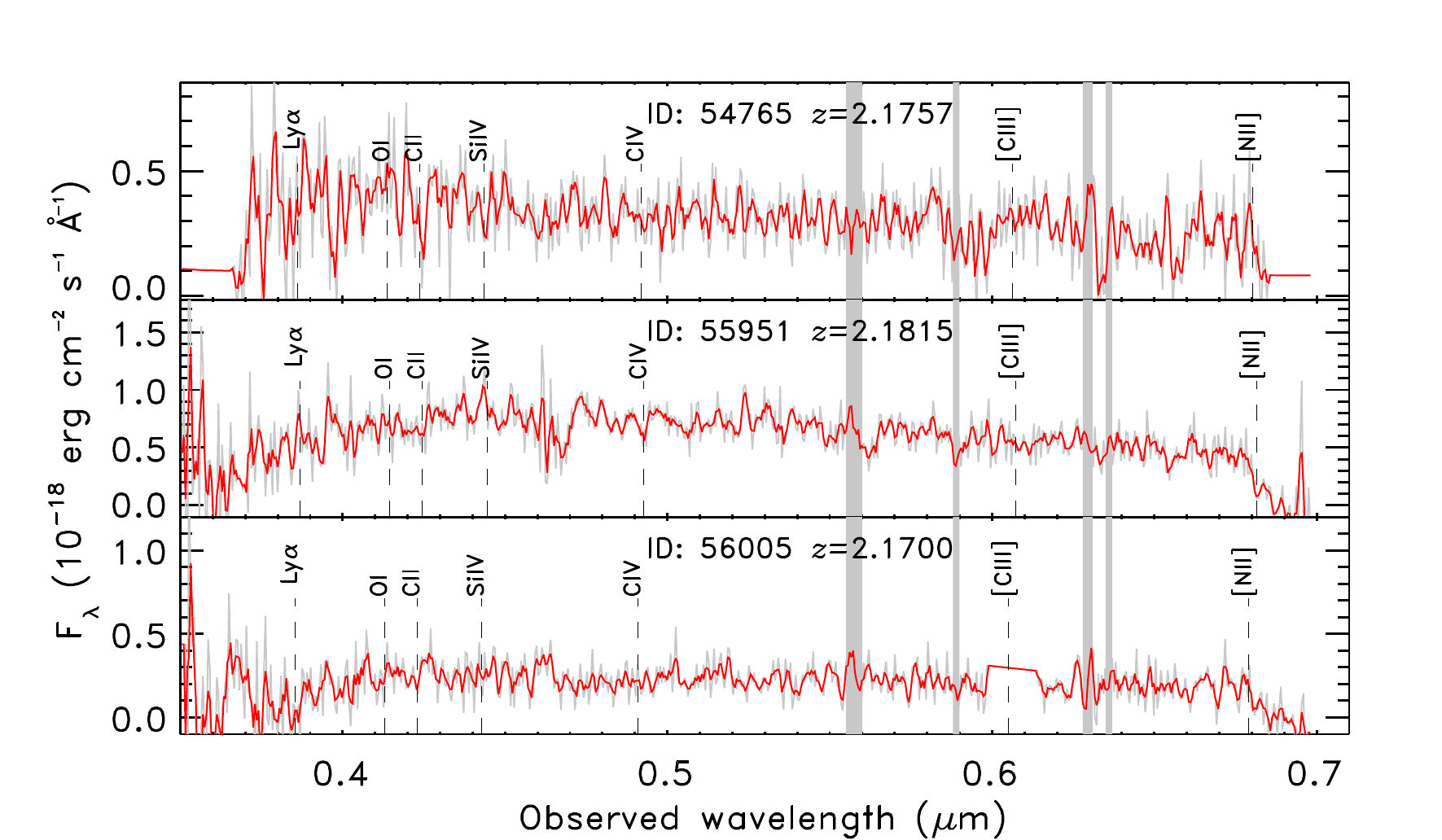}
\caption{{\small VIMOS spectra (gray line) of three member candidates with
$z$\,=\,2.164--2.19 in the 10$^{\prime}\times12^{\prime}$ region centered at
$\alpha$\,=\,150.465\deg, and $\delta$\,=\,2.31\deg\ where G237 is located.  The red lines are smoothed
spectra. The vertical dashed lines represent the location of the main
spectral lines as annotated. Source IDs and spectroscopic redshifts are annotated in each
panel. Regions contaminated by bright sky lines are highlighted with vertical gray
bars.}}
\label{fig:spectra_int}
\end{figure}

\clearpage
\section{Stacked spectrum assessment}\label{app:spec_comparison}

\begin{figure*}[h!]
\centering
\includegraphics[width=\linewidth]{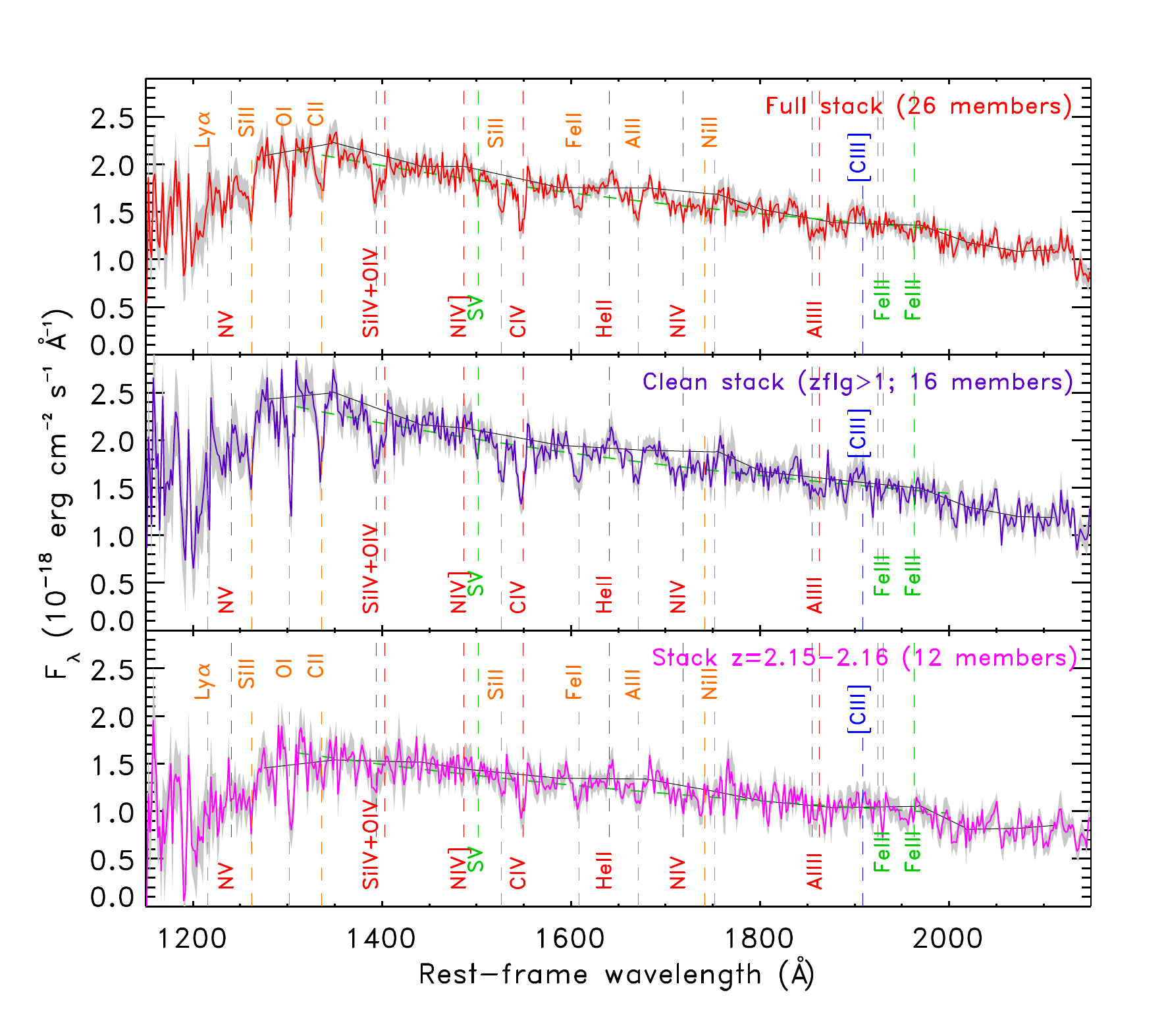}
\caption{{\small Stacked rest-frame VIMOS spectra of 26 member candidates with
$z$\,=\,2.15--2.20 (red solid line in {\it top panel}), of 16 members with
\textit{zflg}$\geq$2 (purple
solid line in {\it middle panel}), and of 12 members with $z$\,=\,2.15--2.16
(magenta solid line in {\it bottom panel}). In all panels, the gray filled region
represents the 1$\sigma$ uncertainty, the dashed green line is a linear fit obtained to
estimate the UV spectral slope $\beta$, where F$_{\lambda}\propto
\lambda^{\beta}$, and the solid black curve represents the pseudo-continuum
obtained by linearly interpolating the average flux density in the
"features-free" windows defined in Table~3 in~\citet{rix04}.
The vertical dashed lines represent the location of the main
spectral features as annotated (nebular: blue, ISM: orange, stellar 
photosphere: green, stellar winds: red).}}
\label{fig:spectra_stack_x3}
\end{figure*}

In Fig.~\ref{fig:spectra_stack_x3}, we show the
co-added spectra obtained with all 26 non AGN sources at $z$\,=2.15--2.20
(\textit{full} stack in the {\it top panel}), with the 16 galaxies with
\textit{zflg}$\geq$2 (\textit{clean} stack in the {\it middle panel}), and
with the 12 galaxies at $z$\,=2.15--2.16 ({\it bottom panel}).  
All three co-added spectra show numerous absorption features and are
consistent within 1$\sigma$, providing evidence that most redshift estimates
with low \textit{zflg} are correct.  The median S/N of the \textit{full}
stack is 14, that of the \textit{clean} stack is 12, and of the
\textit{z}\,=\,2.15--2.16 stack is 8.4.  In Sect.~\ref{sec:stack_spectrum}, we carry out the spectral
analysis of the ``full stack'' spectrum because has a higher S/N and it is
more complete and representative of the member population. 

\clearpage
\section{Morphological analysis}

\begin{figure*}[h!]
\centering
  \subfloat[55326: ET]{\includegraphics[height=2cm]{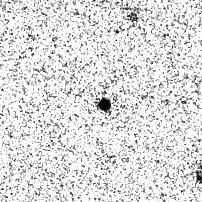} \includegraphics[height=2cm]{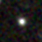}}
\;\subfloat[55687: Disk]{\includegraphics[height=2cm]{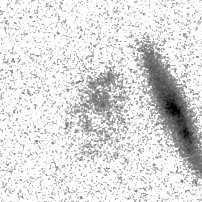} \includegraphics[height=2cm]{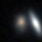}}
\;\subfloat[56014: Irr]{\includegraphics[height=2cm]{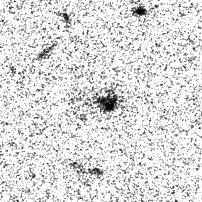}  \includegraphics[height=2cm]{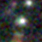}}
\;\subfloat[56915: Disk]{\includegraphics[height=2cm]{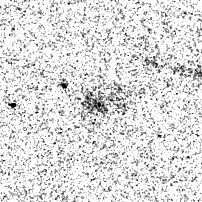} \includegraphics[height=2cm]{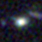}}\\
\subfloat[57557: Disk]{\includegraphics[height=2cm]{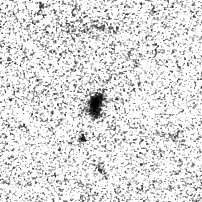} \includegraphics[height=2cm]{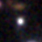}}
\;\subfloat[58020: Irr]{\includegraphics[height=2cm]{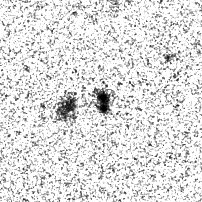} \includegraphics[height=2cm]{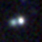}}
\; \subfloat[58057: Disk]{\includegraphics[height=2cm]{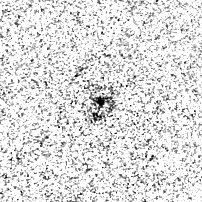} \includegraphics[height=2cm]{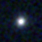}}
\;\subfloat[58173: Irr]{\includegraphics[height=2cm]{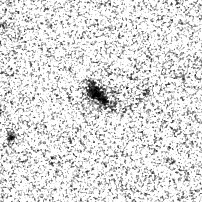} \includegraphics[height=2cm]{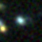}}\\
\subfloat[58200: Too faint]{\includegraphics[height=2cm]{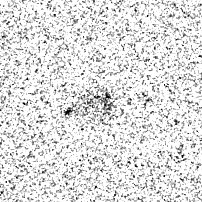} \includegraphics[height=2cm]{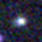}}
\;\subfloat[58206: Disk]{\includegraphics[height=2cm]{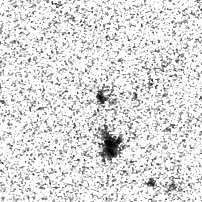} \includegraphics[height=2cm]{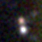}}
\;\subfloat[58266: Irr]{\includegraphics[height=2cm]{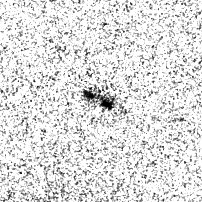} \includegraphics[height=2cm]{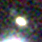}}
\;\subfloat[59504: Disk]{\includegraphics[height=2cm]{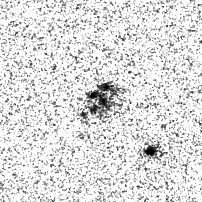} \includegraphics[height=2cm]{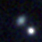}}\\
\subfloat[60718: Disk]{\includegraphics[height=2cm]{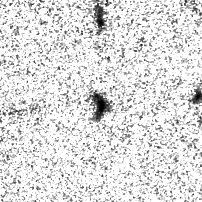} \includegraphics[height=2cm]{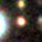}}
\;\subfloat[60811: Disk]{\includegraphics[height=2cm]{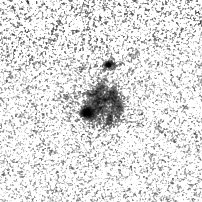} \includegraphics[height=2cm]{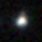}}
\;\subfloat[L710971: Too faint]{\includegraphics[height=2cm]{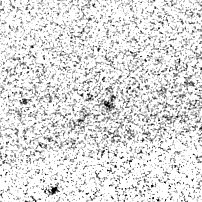} \includegraphics[height=2cm]{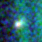}}
\;\subfloat[SL01: Too faint]{\includegraphics[height=2cm]{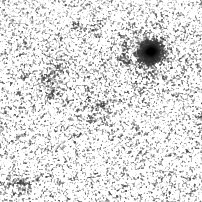}
\includegraphics[height=2cm]{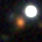}}\\
\subfloat[SL03: Disk]{\includegraphics[height=2cm]{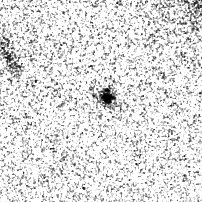}
\includegraphics[height=2cm]{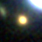}}
\;\subfloat[57138: Disk]{\includegraphics[height=2cm]{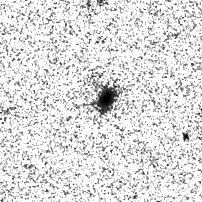}
\includegraphics[height=2cm]{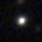}}
\;\subfloat[54698: Irr]{\includegraphics[height=2cm]{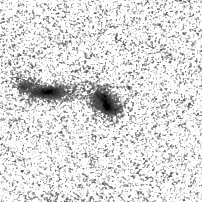} \includegraphics[height=2cm]{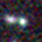}}
\;\subfloat[57730: Disk]{\includegraphics[height=2cm]{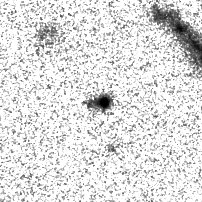} \includegraphics[height=2cm]{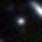}}\\
\caption{{\small 6$^{\prime\prime}\times 6^{\prime\prime}$ ACS I-band and UltraVISTA YJH cutouts of the sources at
2.15${\leq}z{<}$2.164. Source ID and morphological classification are reported
below the corresponding images.}}
\label{fig:cutouts1} 
\end{figure*}

\begin{figure*}
\centering
  \subfloat[54677: Irr]{\includegraphics[height=2cm]{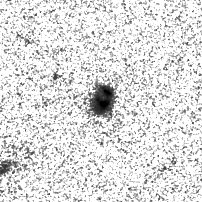} \includegraphics[height=2cm]{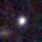}}
\;\subfloat[56950: Irr]{\includegraphics[height=2cm]{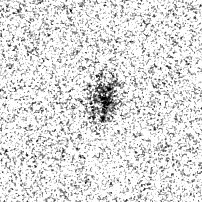} \includegraphics[height=2cm]{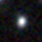}}
\;\subfloat[57695: Irr]{\includegraphics[height=2cm]{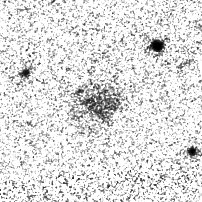} \includegraphics[height=2cm]{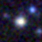}}
\;\subfloat[59496: Disk]{\includegraphics[height=2cm]{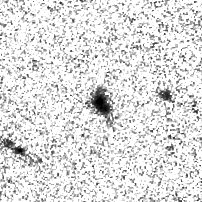} \includegraphics[height=2cm]{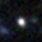}}\\
\;\subfloat[59906: Irr]{\includegraphics[height=2cm]{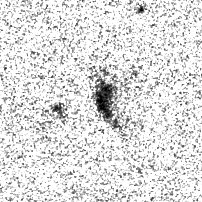} \includegraphics[height=2cm]{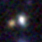}}
\;\subfloat[60026: Disk]{\includegraphics[height=2cm]{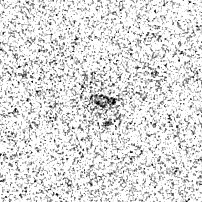} \includegraphics[height=2cm]{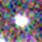}}
\;\subfloat[60355: Irr]{\includegraphics[height=2cm]{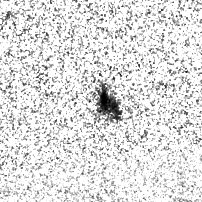} \includegraphics[height=2cm]{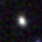}}
\;\subfloat[60984: Irr]{\includegraphics[height=2cm]{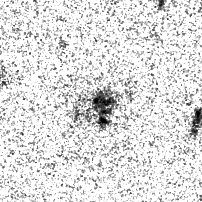} \includegraphics[height=2cm]{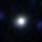}}\\
\caption{{\small 6$^{\prime\prime}\times 6^{\prime\prime}$ ACS I-band and UltraVISTA YJH cutouts of the sources at
$z{>}$2.19. Source ID and morphological classification are reported
below the corresponding images.}}
\label{fig:cutouts2} 
\end{figure*}

\begin{figure*}
\centering
\subfloat[54765: Disk]{\includegraphics[height=2cm]{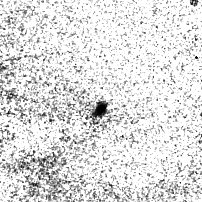} \includegraphics[height=2cm]{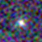}}
\;\subfloat[55951: Disk]{\includegraphics[height=2cm]{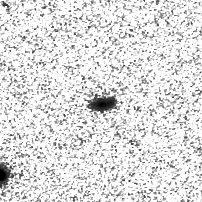} \includegraphics[height=2cm]{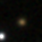}}
\;\subfloat[56005: Disk]{\includegraphics[height=2cm]{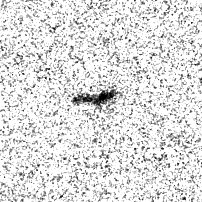} \includegraphics[height=2cm]{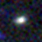}}\\
\caption{{\small 6$^{\prime\prime}\times 6^{\prime\prime}$ ACS I-band and UltraVISTA YJH cutouts of the sources at
2.164${<}z{<}$2.19.  Source ID and morphological classification are reported
below the corresponding images.}}
\label{fig:cutouts3} 
\end{figure*}

Postage stamps (6\arcsec$\times$6\arcsec) in the HST/ACS F814W-band and in the
UltraVISTA/YJH bands of all 31 spectroscopic members are shown in
Fig.~\ref{fig:cutouts1},~\ref{fig:cutouts2}, and ~\ref{fig:cutouts3}.
The identifier (ID) and the morphological classification are annotated below
each pair of images.

\clearpage
\section{Spectral energy distributions}

In Fig.~\ref{fig:cig_seds}, we show the SEDs of the 31 spectroscopic members
and the best fit models obtained with CIGALE (see~Sect.~\ref{sec:mstar_sfr}).

\begin{figure*}[h!] 
\centering
\includegraphics[width=17.6cm]{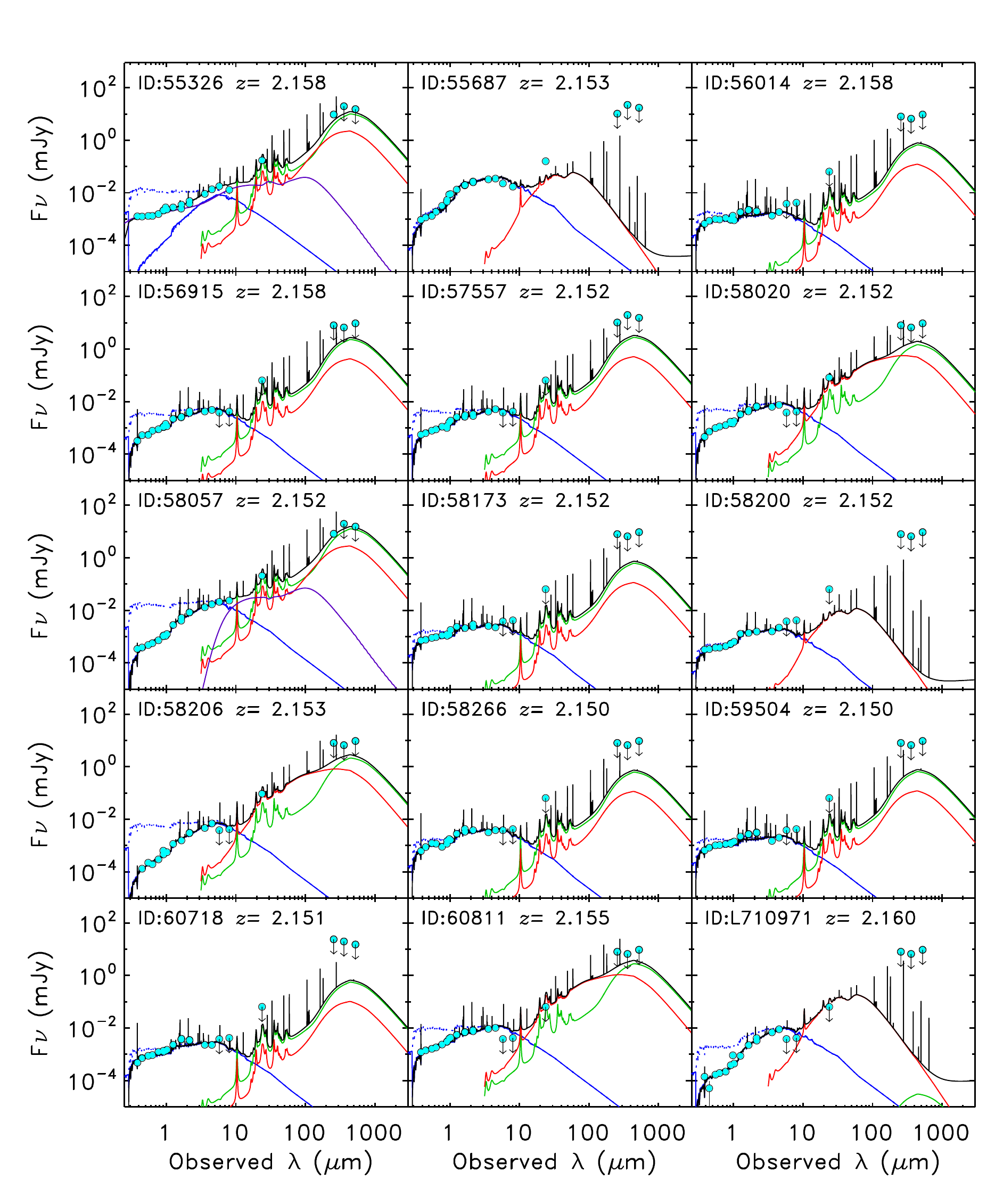}
\caption{{\small Observed spectral energy distributions (full cyan
circles) and best fit templates (black lines) obtained with CIGALE
at the spectroscopic redshift. Downward arrows are 5$\sigma$ upper limits. 
The blue solid and dotted lines represent the stellar component after and
before intrinsic extinction is applied. The green and red lines represent
emission from dust. The purple line represents an AGN component, present
only in IDs 55326, and 58057. Source ID and spectroscopic redshifts are
annotated.}}
\label{fig:cig_seds}
\end{figure*}
\begin{figure*}[h!]
\setcounter{figure}{0}
\centering
\includegraphics[width=17.6cm]{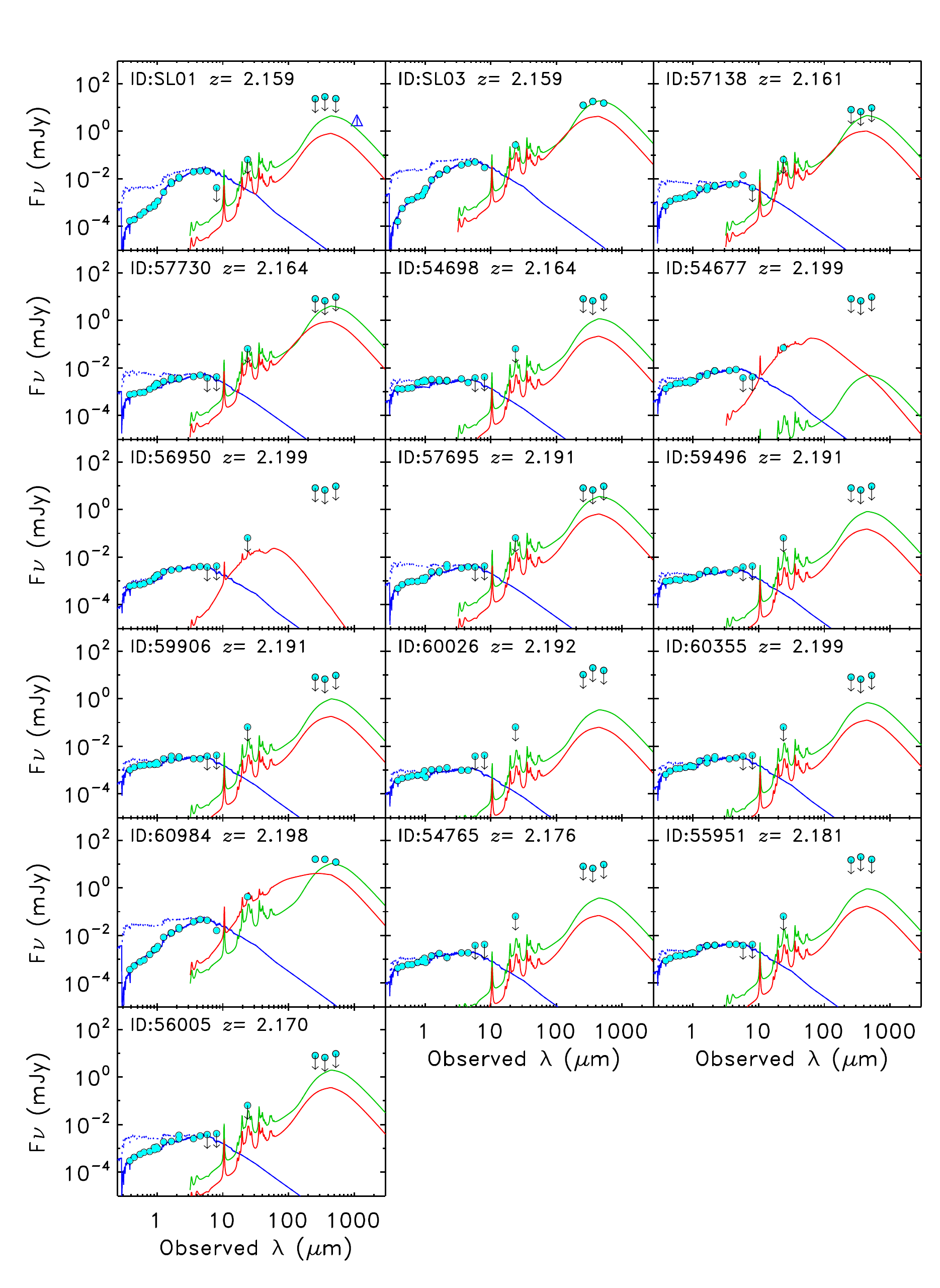}
\caption{{\small Continued.}}
\end{figure*}


\end{document}